\documentclass[a4paper,fleqn,usenatbib]{mnras}

\usepackage{mathptmx}
\usepackage[T1]{fontenc}
\usepackage{ae,aecompl}
\usepackage{supertabular}

\usepackage{graphicx}	% Including figure files
\usepackage{amsmath}	% Advanced maths commands
\usepackage{amssymb}	% Extra maths symbols
\usepackage{pdflscape}
\title[Stellar activity with LAMOST]{Stellar activity with LAMOST. I. Spot configuration in Pleiades}

\author[X.-S. Fang et al.]{
Xiang-Song Fang,\thanks{E-mail: xsfang@bao.ac.cn}
Gang Zhao,
Jing-Kun Zhao,
Yu-Qin Chen
and
Yerra Bharat Kumar
\\
Key Laboratory of Optical Astronomy, National Astronomical Observatories, Chinese Academy of Sciences, Beijing 100012, China\\
}

\date{Accepted XXX. Received YYY; in original form ZZZ}

\pubyear{2016}

\begin{document}
\label{firstpage}
\pagerange{\pageref{firstpage}--\pageref{lastpage}}
\maketitle

% Abstract of the paper
\begin{abstract}
We use the spectra of Pleiades and field stars from LAMOST DR2 archive to study
how spottedness and activity vary as a function of mass at young ages.
We obtained standard TiO band strength by measuring TiO bands near 7050~\AA~from LAMOST spectra (R$\approx$1800) for 
large sample of field GKM dwarfs with solar metallicity. Analysis show that active dwarfs, including late G- and early K-type, 
have extra TiO absorption compare to inactive counterparts, indicating the presence of cool spots on their surface.
Active late K- and M-dwarfs show deeper TiO2 and shallower TiO4 compare to inactive stars at a given TiO5, 
which could be partly explained through cool spots. We estimated cool spot fractional coverage for 304 Pleiades candidates 
by modelling their TiO2 (\&TiO5) band strength with respect to standard value. 
Results show that surface of large fraction of K- and M-type members have very large spot coverage ($\sim50\%$). 
We analysed a correlation between spot coverage, rotation and the amplitude of light variation, 
and found spot coverage on slow rotators ($R_{o} > 0.1$) increases with decreasing Rossby Number $R_{o}$.
Interestingly, we detected a saturation-like feature for spot coverage in fast rotators 
with a saturation level of $40\%-50\%$. In addition, spot distribution in hotter fast rotators show more symmetrical compare to slow rotators. 
More interestingly, we detected large spot coverage in many M type members with no or little light variation.
In bigger picture, these results provide important constraints for stellar dynamo on these cool active stars.

\end{abstract}

% Select between one and six entries from the list of approved keywords.
% Don't make up new ones.
\begin{keywords}
%stars: atmospheres -- stars: activity -- stars: starspots -- stars: late-type
stars: late-type -- stars: activity -- stars: starspots
\end{keywords}

%%%%%%%%%%%%%%%%%%%%%%%%%%%%%%%%%%%%%%%%%%%%%%%%%%

%%%%%%%%%%%%%%%%% BODY OF PAPER %%%%%%%%%%%%%%%%%%

\section{Introduction}
%This is a simple template for authors to write new MNRAS papers.
%See \texttt{mnras\_sample.tex} for a more complex example, and \texttt{mnras\_guide.tex}
%for a full user guide.
%All papers should start with an Introduction section, which sets the work
%in context, cites relevant earlier studies in the field by \citet{Others2013},
%and describes the problem the authors aim to solve \citep[e.g.][]{Author2012}.
Stellar spots, a ubiquitous manifestation of magnetic activity on late-type stars, 
are thought to be the fingerprints of magnetic lines on their 
photospheres. Spots size and distribution patterns on the surface are important constraints for
stellar dynamo mechanism \citep{stra2009}. Additionally, spots can inhibit exoplanet detection and characterization, 
since they can create noise in light curve \citep{aigr2004}, 
and radial velocity (RV) perturbations \citep{deso2007,meun2013}, e.g., the RV `jitter' \citep{ande2015}. 
The spot-induced noise in light curve or RV variation are also related to spot configuration (size, shape and temperature) on the photosphere.

Many observational studies showed larger radii of fast rotating and active K/M-type stars 
(thus lower effective temperatures) compare to model predictions, 
no matter whether they are binary members \citep[e.g.][]{torr2002,riba2006,lope2007} 
or single stars \citep[e.g.][]{mora2008}. Similar features were found in cool members of young open clusters, e.g., NGC 2516 
\citep{an++2007,jack2009} and Pleiades \citep{hart2010}. 
The discrepancies of the radii and effective temperatures between observations and models 
are thought to be due to presence of magnetic activities at high level \citep{torr2013},  e.g., 
existence of large and cool magnetic starspots \citep{chab2007}, and/or strong magnetic fields inhibiting convection 
\citep{mull2001,feid2012}. \citet{jack2014} found the validity of   
starspot model by comparing the loci of active cool members of NGC 2516 and Pleiades, 
and inactive field stars in color-magnitude diagrams (CMDs); 
if starspots are the sole cause for radius inflation detected in these highly magnetically active cool stars, 
it needs a very large spot coverage (e.g., 0.35-0.51).

Starspots can cause light variation due to its rotational modulation. 
However, some young low-mass stars with high levels of chromospheric activity show no spot modulation. 
\citet{jack2012} showed detail analysis on 210 low-mass stars with detected periods and 144 low-mass stars 
without detected periods in the young open cluster NGC 2516,
found no significant differences in their positions on CMDs, the distribution of their $v\sin i$, and 
their levels of chromospheric activity, suggested that the lack of spot modulation may be due to high 
axi-symmetric distributions of starspots being in very small angular length scale \citep[e.g. $\sim 3^{\circ}$,][]{jack2013},  
although the coverage fraction is probably very large, e.g., up to about 0.4 or even larger \citep{jack2014}.

Large spots may exist in late type active stars, mainly in fast rotating low-mass (K/M-type) stars, and probably locate randomly on the surface in very small length scale.
To understand this issue, the key is to derive the real spot coverage for active stars, 
especially for those stars showing absence or very small spot rotational modulation, 
since they may provide important clues to the spot distribution and hence dynamo mechanism \citep{jack2012}. 
 Traditionally, starspots are detected using the observational techniques, e.g., light-curve inversion \citep[e.g.][]{budd1977,sava2008} and 
Doppler imaging \citep[e.g.][]{coll1994,stra2002}, based on the fact that 
spot rotational modulation can result in photometric (broad-band fluxes) and spectroscopic
(line profiles) variability. 
However, these techniques are sensitive only to the asymmetric part of the starspot distribution, 
in other words, if the spots are highly axi-symmetric the techniques of light-curve and line profile 
modelling will underestimate the spot coverage.

Molecular band absorptions are the primary features in the spectra of M-type stars, 
e.g., the energy distributions of M dwarfs in optical spectra are entirely governed by TiO and CaH bands. 
TiO bands are sensitive to the temperature, and formed in the region where $T_{eff} < 4000$ K \citep{chab2000}, 
thus being used to classify M-dwarf spectral types \citep{kirk1991}.
\citet{vogt1979,vogt1981} and \citet{rams1980} are the first studies that gave an idea to use TiO bands as starspot indicators.
\citet{vogt1979} reported that the single-line spectroscopic binary HD 224085 (=II Peg; spectral type $\sim$K2 IV-V) 
showed VO and the $\gamma$ system TiO molecular absorption features in their spectrum. 
By comparing the ratio of spectra of II Peg obtained at minimum (spot in view) and maximum (spot out of view) 
with standards,
\citet{vogt1981} estimated the spot's equivalent spectral type as M6 or later. 
\citet{rams1980} observed enhanced absorption of TiO band near 8860 \AA\ in the RS CVn binary HR~1099 
(K1 IV + G5 V) during light minimum, and they concluded that the TiO feature is resulted from spotted regions 
which is about 1000 K cooler than the quiescent photosphere on the active K1-subtype component. 
More quantitative attempts towards deriving spot coverages from the TiO molecular absorption bands were carried out later.
\citet{huen1987} found the spot coverage of 35\%-50\% and 35\%-40\% from the analysis of TiO bands at  7127 \AA\ 
and 8860 \AA, respectively, for II~Peg. 
\citet{neff1995} made independent measurements of the area and the temperature of starspots 
for several stars using the TiO bands at 7055 \AA\ and 8860 \AA. In previous studies, 
the spot coverages of several active dwarfs and giants have been derived from 
the TiO molecular absorption bands \citep{onea1996, onea1998, onea2004}. 
As pointed out by \citet{onea2004}, molecular absorption bands such as TiO bands are used to study spots 
properties regardless of their distribution patterns, even on slowly rotating stars.

Up to date, the number of active stars whose spot coverages have been studied is still limited 
and a very few systematic studies about the properties of stellar activities on atmosphere, which could be 
due to lack of observational data for active stars, particularly the spectroscopic observations. 
LAMOST (Large sky Area Multi-Object fiber Spectroscopic Telescope), a quasi-meridian reflecting Schmidt telescope \citep{cui+2012}, 
survey on Galactic stars observed few million stellar spectra with signal-to-noise ratio $SNR > 10$ \citep{liu+2015},
allows us to initiate systematic study of magnetic activities in low mass stars.
By using the results from large photometric surveys (e.g., from HATNet and KEPLER) 
we aim to make full use of LAMOST spectral data to characterize stellar magnetic activities  
in cool member stars of open clusters and cool active field stars including spot configuration (e.g., spot coverage and distribution pattern), 
and chromospheric activity property, and further check for correlations between activity and stellar parameters. 
For this study we started with Pleiades, an open cluster, which is thought to be an analogue of NGC 2516 and 
a good laboratory to make systematic study of stellar activity for its proximity, richness, youth, and solar metallicity. 
In this paper, we estimated the spot fractional coverage for 304 Pleiades candidate members by modelling their TiO molecular 
bands near 7050~\AA. Further, we have characterized the spot coverages and spot distribution patterns, 
and showed the effect of stellar rotation on spot coverage with the help of results (rotation periods and amplitudes of light variation) 
from the HATNet survey data \citep[e.g.][]{hart2010}.

%%%%%%%%%%%%%%%%%%%%%%%%%%%%%%%%%%%%%
\section{Data and Target selection}
\subsection{LAMOST Data}
The LAMOST telescope (also called the Guo Shou Jing Telescope), characterized by both a wide field of view (5 deg in diameter) and a large aperture 
(effective aperture of $\sim$4 m),
is a reflecting Schmidt telescope located at the Xinglong station of NAOC, China. A total of 4000 fibers are mounted on its focal plane. 
The theoretical resolution of the spectrograph setup is $R\approx1000$ with the spectral wavelength coverage of $3700-9000$ \AA.
In practice, the resolution increased to
$R\approx1800$ ($\sim 3.9$ \AA~at 7100 \AA)
by narrowing the slit to 2/3 of the fiber width \citep{zhao2012,luo+2015}.
Two separate CCDs with 4K $\times$ 4K pixels are used to record blue and red part of the spectra, and each Angstrom has been sampled in two pixels in the original data. The final output  LAMOST spectra (blue and red channels are combined) have been re-sampled with the same difference
in wavelength between two adjacent 'pixels' ($\Delta \log(\lambda)=0.0001$, e.g., $\sim$1.6 \AA~at 7100 \AA) \citep[see][]{luo+2015}.

The LAMOST was planned to perform Galactic and extra-galactic surveys for first five years \citep{zhao2012}.
In the pilot survey (from autumn 2011 to summer 2012), both Galactic and extra-galactic targets were observed,
and then regular surveys were started from autumn 2012 which are mainly focused on Galactic stars survey \citep{luo+2015}.
By May 2015, the LAMOST collected $\sim$ 4.5 million stellar spectra with SNRs larger than 10 in the Galaxy \citep{liu+2015}.      
The LAMOST Date Release~2 (DR2) was made available to the Chinese astronomical community and international partners during January 2015,
containing more than 4 million spectra collected from autumn 2011 to summer 2014,
among them nearly 3.8 million are stellar spectra. 
LAMOST stellar parameter pipeline (LASP) provides the stellar parameters like effective temperature  ($T_{eff}$), 
surface gravity ($\log g$), metallicity ([Fe/H]) and radial velocity (RV)
for $\sim$ 2.2 million stellar spectra of AFGK stars whose spectra are relative flux calibrated and the signal-to-noise ratio criterion \citep{wu++2011,luo+2015},
Besides the AFGK star catalogue, DR2 also include a catalogue of 0.2 million M-type stars, whose stellar parameters are not derived by LASP. 
\subsection{Pleiades}
Pleiades is a famous and quintessential open cluster, due to its proximity, richness, youth, and solar metallicity. 
It has a distance in the range of $120-136$ pc \citep{pins1998,sode2005,vanl2009,meli2014}, 
and thus a very small reddening $E(B-V)=0.03-0.04$ \citep{stau1987,an++2007}, 
an age of 100-125 Myr \citep{meyn1993,stau1998,vanl2009}, a metallicity of $[Fe/H]=0.03^{+0.02}_{-0.05}$ 
\citep{sode2009}. A large number of stars have been identified as probable members of Pleiades in the literature, e.g., 
\citet{stau2007} compiled a catalogue of 1471 Pleiades member candidates (hereafter S07), \citet{lodi2012} 
identified $\sim$1000 Pleiades member candidates based on UKIDSS GCS survey data. More recently, new 812 stars were found to 
be probable members by \citet{bouy2015}, who compiled a catalogue of 2109 high-probability members (hereafter B15), 
which is the most completed census of the cluster to the date. In addition, \citet{hart2010} detected rotation periods for 
368 Pleiades stars using the HATNet transit survey data, and compiled a catalogue of rotation periods for 383 Pleiades members. 
Cross-matching S07 with LAMOST DR2 stellar catalogue, 
we identified 271 spectra with $SNRr>10$ of 238 probable members of Pleiades (hereafter sample-1). 
In addition, we identified 79 spectra of 66 probable Pleiades members which were not included in S07 but in B15 with 
probability more than 0.75, as a complementary sample (hereafter sample-2). 
In total, we selected 304 probable Pleiades members with 350 LAMOST spectra having $SNRr>10$ (a small fraction have multi-observations), 
most of which are K and M dwarfs, as listed in Table~\ref{tab:objects_list}.

\begin{table*}
\caption{Basic stellar parameters of sample Pleiades members observed by LAMOST. Full table is available online.}
\label{tab:objects_list}
\begin{tabular}{lccccccccccccccccccc}
   \hline
 Object name  & V   & $r^{a}$   & $I_{c}$ & $K_{s}$ &$T_{VI}$ &$T_{VK}$&$T_{rK}$&$T_{IK}$&Period&$A_{r}$ &$R_{o}^{b}$&B$^{c}$\\
                    & (mag)  & (mag)  & (mag)  & (mag)  &   (K)  &   (K)  &   (K)  &   (K)  &   (day)  &  (mag) &        &   \\
   \hline                                                                                             
     Melotte 22 174 & 11.620 & 11.358 &        &  9.374 &        &  5093  &  4982  &        & 0.474297 & 0.0645 & 0.0258 &  0  \\
  Melotte 22 HCG 39 & 12.857 & 12.307 & 11.618 & 10.134 &  4710  &  4687  &  4772  &  4664  & 6.461243 & 0.0169 & 0.2949 &  0  \\
  Melotte 22 DH 166 & 14.045 & 13.154 & 12.407 & 10.631 &  4190  &  4192  &  4166  &  4197  & 9.867441 & 0.0221 & 0.3765 &  0  \\
    Melotte 22 1100 & 12.250 & 11.851 & 10.970 &  9.400 &  4653  &  4597  &  4518  &  4539  & 7.502517 & 0.0233 & 0.3362 &  1  \\
  Melotte 22 DH 075 & 15.068 & 14.473 & 12.992 & 10.842 &  3716  &  3648  &  3658  &  3520  & 0.324622 & 0.0279 & 0.0097 &  0  \\
 Melotte 22 HHJ 295 & 17.600 &        & 14.500 & 12.218 &  3035  &  3130  &        &  3335  &          &        &        &  1  \\
J033754.79+252631.2 &        & 15.242 &        & 11.265 &        &        &  3467  &        &          &        &        &  0  \\
J040443.86+253333.0 &        & 14.906 &        & 11.715 &        &        &  3931  &        &          &        &        & -1  \\
    Melotte 22 1173 & 15.100 & 14.637 & 13.050 & 10.996 &  3739  &  3717  &  3653  &  3675  & 8.215426 & 0.0751 & 0.2476 &  0  \\
\hline
\multicolumn{13}{l}{$^a$ Sloan $r$-band magnitudes}\\
\multicolumn{13}{l}{$^b$ Rossby number}\\
\multicolumn{13}{l}{$^c$ Candidate single members (0); probable binary members (1); }\\
\multicolumn{13}{l}{ ~~ Fainter probable non-members (-1); brighter probable non-members (10). }\\
\end{tabular}
\end{table*}
%%%%
%%%%%%%%%%%%%%%%%%%%%%%%%%% 
\section{Methodology}
\subsection{Spectrum synthesis modelling technique}
\label{sec:maths}
The spectra of M-type stars are dominated by molecular bands, oxides (mainly TiO, VO), metal hydrides (e.g., CaH), and metal hydroxides (CaOH). 
\citet{reid1995} defined a number of narrow band flux ratios, so called indices (e.g., TiO2, TiO5, CaH2, CaH3) 
to measure the strength of molecular features in the spectra of M dwarfs. 
\citet{hawl1996} found that TiO band indices from magnetically active dMe stars differ systematically from old M dwarfs,
specifically stronger TiO2 band depths in active dMe stars compare to inactive counterparts. 
In this work, we showed that Pleiades low mass members have TiO2 (and TiO5) depths stronger than field stars of same temperature, 
and then used those data to infer spot fractional area coverage as a function of spectral type,  
based on the idea that TiO could be the indicator of photospheric temperature inhomogeneities \citep{vogt1979,rams1980}. 
Later papers by \citet{huen1987} and \citet{neff1995} expanded on this spectrum synthesis modelling technique. 
For illustrative purpose, here we give a simple description of this technique. 

The observed total flux from a star with inhomogeneous temperature photosphere 
(e.g., two-temperature structure: cool spotted region and hot quiescent region) can be expressed by
\begin{equation}
 f_{t}(\lambda)=(\frac{R}{D})^{2} \lbrack S_{q}(\lambda)(G_{\ast}-G_{s})+S_{s}(\lambda)G_{s} \rbrack ~,
\end{equation}
where $R$ is the stellar radius, $D$ is the distance from us, 
$S_{s}(\lambda)$ and $S_{q}(\lambda)$ are the specific intensity of spotted and quiescent photosphere, respectively.
$G$ is a geometrical quantity depending upon the limb-darkening effect (equivalently flux weighted area), 
and for a homogeneous hemisphere 
$G_{\ast}=\iint(1-\beta+\beta \cos \gamma)\cos \gamma \sin \theta d\theta d\phi=\pi(1-\beta/3)$ \citep{huen1987}, 
assuming a limb darkening law with the form of $1-\beta+\beta \cos\gamma$; $G_{s}$ means the flux weighted area of spotted region. 
Correspondingly, the observed relative flux (with nearby pseudo continuum band $\lambda_{c}$) can be predicted as
\begin{equation}
F_{t} = \frac{f_{t}(\lambda)}{f_{t}(\lambda_{c})}= 
\frac{f_{s}R_{\lambda_{c}}F_{s}+(1-f_{s})F_{q}}{f_{s}R_{\lambda_{c}}+(1-f_{s})} ~, 
\label{equ:Ftotal}
\end{equation}
assuming that $G$ is independent of wavelength, where $F_{s}=S_{s}(\lambda) / S_{s}(\lambda_{c})$ and 
$F_{q}=S_{q}(\lambda) / S_{q}(\lambda_{c})$ are the relative strength of spotted and quiescent photosphere, respectively, 
and $R_{\lambda_{c}} = S_{s}(\lambda_{c}) / S_{q}(\lambda_{c})$ denotes the contrast of continuum surface flux 
between the spotted and quiescent photosphere.
$f_{s}=G_{s}/G_{\ast}$ is the fractional coverage (projected area) of spotted region on the observed hemisphere (also called spot filling factor). 
So, based on the idea of spectrum synthesis we can derive the spot filling factor $f_{s}$ for a spotted star 
via modelling the observed relative flux spectrum as the weighted sum of the reference spectra of suitable inactive stars with 
$T_{eff}=T_{q}$ and $T_{eff}=T_{s}$ and similar luminosity class. For example, if an active K dwarf star with quiescent 
photosphere having $T_{eff}=4800$ K is covered by a spotted region with $T_{eff}=3500$ K, we can 
use the observed TiO2 band strength $F_{t}$ to derive the spot filling factor $f_{s}$, 
based on the TiO2 strength $F_{q}$ and $F_{s}$ of an inactive K dwarf star ($T_{eff}=4800$ K) 
and an inactive M dwarf star ($T_{eff}=3500$ K), respectively, 
and the continuum surface flux contrast $R_{\lambda_{c}}$ between these two stars using the following expression,  
\begin{equation} 
f_{s}=\frac{1}{1+r_{\lambda}R_{\lambda_{c}}} , ~~~r_{\lambda}=\frac{F_{t}-F_{s}}{F_{q}-F_{t}}.
\label{equ:fs}
\end{equation}

In this case, determination of spot coverage by modelling the observed TiO2 strength, $F_{t}$, 
of a spotted star needs standard values of $F_{q}$ and $F_{s}$, 
the TiO2 strength of the spotted and quiescent photosphere, respectively. 
Also required is corresponding continuum surface flux contrast $R_{\lambda_{c}}$ between the spotted and quiescent photosphere. 

%%%%%%%%%%%%%%%%%%%%%%
\subsection{Estimation of continuum surface flux contrast}
\label{sec:contrast}
\begin{figure}
\centering
\includegraphics[width=\columnwidth]{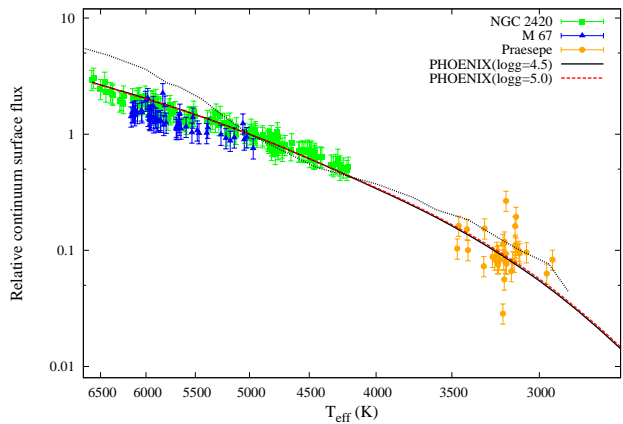}
\caption{Measurements of relative continuum surface flux over the bandpass of 7042-7048 \AA\ for the sample and model spectra is shown. 
Note the flux is normalized at $T_{eff}=5000$ K. 
Black solid line and red dashed line denotes the PHOENIX spectra with [Fe/H]=0.0 \& $\log g=4.5$, and [Fe/H]=0.0 \& $\log g=5.0$, respectively. 
Squares, triangles and circles represent members of NGC 2420, M 67 and Praesepe, respectively. 
Black dotted line indicates the R-band surface flux.}
\label{fig:contrast}
\end{figure}

PHOENIX stellar atmosphere models \citep{huss2013} are used to estimate continuum surface flux contrast ($R_{\lambda_{c}}$).
We have selected model spectra with $log$g=4.5, [Fe/H]=0.0, and effective temperatures from 2600 K to 6500 K,
and obtained the corresponding average continuum surface flux over the wavelength region of 7042 to 7048 \AA.
Fig.~\ref{fig:contrast} shows relative continuum surface flux (normalized by setting continuum surface flux to be 1 at $T_{eff}=5000$ K), 
equivalently being the contrast between a given star and the star with $T_{eff}=5000$ K. 
To check the agreement of these contrasts from models with those from the observations, 
we used the flux calibrated SDSS spectra of the members of NGC~2420, M~67 and Praesepe open clusters to estimate continuum surface flux, 
 by measuring the average observed flux $f_{t}(\lambda_{c})$ over the bandpass of 7042-7048~\AA, 
together with their radii and distances (e.g., $S(\lambda_{c}) \propto (R/D)^{-2} f_{t}(\lambda_{c})$).
140 members of NGC~2420 with $\log g >4.0$ and 52 dwarfs in M~67 were selected 
from the catalogue of \citet{lee+2008}. 
We estimated their radii using the temperatures from \citet{lee+2008}, 
based on PARSEC models \citep{chen2014}. 
 For NGC 2420, we adopted the PARSEC model with a metallicity of [Fe/H]$\sim-0.2$ \citep{jaco2011} 
and an age of $\log t \sim9.0$ (WEBDA, http://www.univie.ac.at/webda/); 
for M 67, the model with a solar metallicity \citep{jaco2011} and an age of $\log t \sim9.4$ (WEBDA) was adopted.  
In Praesepe, we selected 36 probable members and collected SDSS spectra of good quality ($SNR>10$) 
by cross-matching the catalogue of Praesepe candidate members \citep{doug2014} with SDSS DR12 SSPP output catalog \citep{lee+2008,smol2011}.
We estimated their effective temperatures using the colors $r-J$ and $r-K_{s}$, and then derived corresponding 
radii using PARSEC model adopting a metallicity of [Fe/H]$\sim+0.16$ \citep{carr2011} and an age of about 600 Myr \citep{foss2008}. 
The distances adopted from WEBDA for NGC 2420, M 67 and Praesepe are 3085, 908 and 187 $pc$, respectively. 
Based on the SDSS spectra of these members of open clusters and corresponding radii and distances, 
we corrected the effect of radius and distance on observed continuum flux and obtained 
the average continuum surface flux in the bandpass of 7042-7048~\AA. 
Using the values of NGC 2420 members with $T_{eff}$ around 5000 K as reference, 
we normalized the derived continuum surface fluxes, as shown in Fig.~\ref{fig:contrast}, 
wherein the error bar is equivalent to its uncertainty due to 10\% error of distance. 
One can see that there exists slight offset between members in NGC 2420 and M 67 at the same effective temperatures, 
which might be due to their uncertainties of distance (e.g., this offset is equivalent to $\sim10\%$ distance error for M 67), 
and/or the difference of basic physical parameters (e.g., metallicity and age) between these two clusters. 
As a supplementary check, we derived the R-band surface flux at a given $T_{eff}$ using the empirically determined Barnes-Evans relationship 
between visual surface-brightness parameter and the $V-R$ color index \citep{barn1978}, 
as shown by black dotted line in Fig.~\ref{fig:contrast}, 
where the $T_{eff}$ was estimated from the corresponding color $V-R$ 
adopting the relationship between $T_{eff}$ and $V-R$ color of \citet{john1966}.  
From Fig.~\ref{fig:contrast} we noticed that continuum surface fluxes in the bandpass of 7042-7048~\AA\ derived from 
SDSS spectra are in good agreement with predicted values from model spectra and the R-band surface fluxes of stars cooler than $\sim5200$ K.

\subsection{Scaling TiO2 band strength}
In principle, we could get the relative strength of any TiO band based on the spectra of stellar atmosphere models.
In practice, probably due to incompleteness of the spectral line lists and/or
the performance of LAMOST spectral data, there exist discrepancies between model spectrum and LAMOST
spectrum on some molecular bands (see Fig.~\ref{fig:teff_ewha} and Fig.~\ref{fig:tio2_tio5}).
Thus, we have taken a large sample of inactive dwarfs with solar metallicities as reference stars,
and used their LAMOST spectra to establish the ``standard'' scales of TiO2 band strength over a wide effective temperature range.

\subsubsection{Sample of reference stars} 
In this study, main targets are the spectra of 304 candidate members of Pleiades from LAMOST DR2.  
As analogues of the spotted and quiescent photosphere of an active star in Pleiades, we chose the reference 
stars with similar metallicities and surface gravities: 
the spectra of reference stars (FGK) are selected based on following criteria, 
$\log g > 4.0$, -0.1 < [Fe/H] < 0.1, and $T_{eff}$ in range of 4000-6500 K.
Since the number of hotter dwarfs is much larger than cooler ones in LAMOST DR2 archive, 
to accommodate similar number of stars in each $T_{eff}$ bin (e.g., bin-width of 50 K), 
we used hotter stars with relatively higher SNR.  
Thus our selected spectra in different temperature region have different SNR in r-band (hereafter SNRr) 
over the range of 30-100, e.g., for stars in the range of $4500\leq T_{eff}<5000$ K  and $5000\leq T_{eff}< 5300$ K have $SNRr > 35$, 
and $SNRr > 40$, respectively. 
Finally, the total sample of reference stars is about $132600$ with effective temperature range from 4000 K to 6500 K.

The situation is different for M-type stars. 
The energy distribution of M-type stars are strongly deviate from blackbody due to absorption of molecular bands, 
and is essentially shaped by the opacities of those molecules \citep{alla2000}. 
Thus, it is not straight forward to obtain their atmospheric parameters from the low resolution spectra. 
Fortunately, TiO is know to be sensitive to temperature and metallicity but weakly depends on gravity \citep{wool2006,lepi2007}, 
while CaH is sensitive to temperature and gravity, 
but less sensitive to metallicity \citep{mann2012}, making the combination of TiO and CaH be a good indicator of gravity 
and metallicity. \citet{lepi2007} found that the ratio between TiO and CaH is a function of both gravity and metallicity 
for M-type stars, however, this ratio is believed to be mostly affected by metallicity for 
the higher gravity stars such as M dwarfs/subdwarfs. Besides CaH, Na~{\sc i} doublet (8172-8197~\AA) are usually  
used as luminosity classification indicators \citep[e.g.][]{mann2012}, since they are quite shallow in 
giants and relatively deep in dwarfs. 
In the current work, we selected M dwarfs with solar metallicities using the TiO and CaH molecular bands, 
and Na~{\sc i} doublet.

\begin{table}
\centering
\caption{Definition of spectral indices}
\label{tab:index_defin}
\begin{tabular}{lcccccc}
   \hline
Index       &  Numerator   &   Denominator & Reference \\
   \hline
CaH2        &  6814-6846   &   7042-7046   & \citet{reid1995}\\
CaH3        &  6960-6990   &   7042-7046   & \citet{reid1995}\\
TiO5        &  7126-7135   &   7042-7046   & \citet{reid1995}\\
TiO2n       &  7057-7064   &   7042-7048   & this work\\
TiO3n       &  7090-7097   &   7077-7083   & this work\\
TiO4n       &  7126-7135   &   7115-7120   & this work\\
TiO5n       &  7126-7135   &   7042-7048   & this work\\
   \hline
\end{tabular}
\end{table}

We initially picked the spectra of about 110900 M-type stars with $SNRr >10$ from LAMOST DR2 M stars catalogue, 
and measured the strength of molecular bands (CaH2, CaH3, and TiO5), Na~{\sc i} doublet, and $H_{\alpha}$. 
The strengths of molecular bands are measured using spectral band indices defined by \citet{reid1995} 
 (See Table~\ref{tab:index_defin} for the indices and their definitions).
We measured the equivalent widths of Na~{\sc i} doublet and $H_{\alpha}$ using the following formula,
\begin{equation} 
EW = \int \frac{f(\lambda)-f(\lambda_{c})}{f(\lambda_{c})}d\lambda, 
\end{equation}
where $f(\lambda_{c})$ denotes the nearby pseudo continuum flux. Note that the equivalent width of an emission line measured 
using above formula has a positive value. To measure the equivalent width of Na~{\sc i} doublet ($EW_{Na}$), 
we integrated the line flux from 8176 \AA~to 8202 \AA, and the continuum flux $f(\lambda_{c})$ was estimated based on the 
flux between 8165-8175~\AA~and 8225-8235~\AA. The error propagation in each measurement was from the 
flux uncertainty per pixel of the spectrum. Since the wavelength region for these indices measurements is relatively narrow 
(e.g., the continuum of TiO5), it is important to correct all spectra for any significant shift in wavelengths. 
Therefore, all spectral measurements for M-type stare were carried out using the RV-corrected spectra. 
RVs were measured by cross-correlating each spectrum with the \citet{boch2007} 
M dwarf template of the best matched subtype (see Appendix~\ref{sec:rv_m}). 

To further select M-type stars with solar metallicity, 
we used the metallicity-dependent parameter $\zeta$, a ratio of CaH and TiO indices defined by \citet{lepi2007},
\begin{equation}
\zeta = \frac{1-TiO5}{1-[TiO5]_{Z_{\odot}}}~,
\label{equ:zeta_defin}
\end{equation}
where $[TiO5]_{Z_{\odot}}$ is a function of [CaH] (= CaH2 + CaH3), representing the expected value of the 
TiO5 index in stars with solar metallicities for a given value of [CaH]. 
Using their newly corrected spectral index values 
(the measurements from the Palomar-MSU spectroscopic survey of \citet{reid1995} are used as standards), 
\citet{lepi2013} re-calibrated $[TiO5]_{Z_{\odot}}$ as 
\begin{equation}
[TiO5]_{Z_{\odot}}=0.622-1.906[CaH]+2.211([CaH])^{2}-0.588([CaH])^{3}.
\label{equ:zeta_scale}
\end{equation}

\begin{figure}
\centering
\includegraphics[width=\columnwidth]{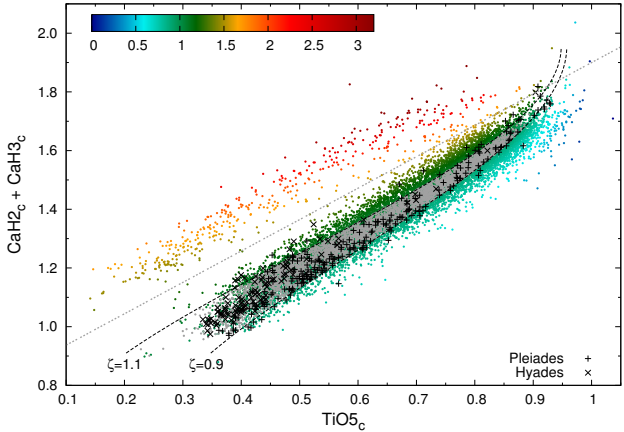}
\caption{Corrected indices (CaH2$_{c}$, CaH3$_{c}$ and TiO5$_{c}$) are shown for M-type stars (dot), Hyades (cross), 
and Pleiades (plus). Grey dotted line separates M-dwarfs (below the line) and M-giants (above the line). 
Black dashed lines represent model values for $\zeta$ = 0.9 \& 1.1. 
Color gradient indicates the values of $\zeta$. Grey colour points indicate selected sample of M dwarfs with solar metallicity. }
\label{fig:zeta_m}
\end{figure}

We compared our initial measurements of TiO5, CaH2 and CaH3 from LAMOST spectra 
with the corrected (standard) values for the same stars provided by \citet{lepi2013}, 
and found small but evident systematic differences. We then corrected our measurements of these indices 
using the correspondingly derived corrections (see Appendix~\ref{sec:correct_m}).     
Based on new $\zeta$ from corrected indices (TiO5$_{c}$, CaH2$_{c}$ and CaH3$_{c}$), 
we then selected $69600$ spectra of M-type stars with $0.9\leq \zeta \leq1.1$. 
The CaH bands and Na~{\sc i} doublet are especially sensitive to gravity and thus are used as luminosity class 
indicator \citep{mann2012,gaid2014}, e.g., the M dwarfs and giants lie in 
two branches in the [CaH] versus TiO5 diagram, although  
for late K- and early M-type stars the locus of dwarfs and giants tend to overlap, creating a region of ambiguity (see Fig.~\ref{fig:zeta_m}). 
As shown in Fig.~\ref{fig:zeta_m}, though the region between $\zeta=0.9$ and $\zeta=1.1$ always includes the majority of the M dwarfs, 
to further exclude the M giants among the stars with $0.9\leq \zeta \leq1.1$, 
we culled objects with larger [CaH] values (e.g., above the straight line displayed in Fig.~\ref{fig:zeta_m}) 
and shallower Na~{\sc i} doublet (e.g., $EW_{Na}>-1.0$). 
Moreover, we refined the sample by selecting higher SNRr (15-30) for early sub-type M dwarfs 
(because the LAMOST archive includes a large number of such stars). 
Finally, we got 32200 spectra of M dwarfs (with spectral type earlier than M6) having solar metallicities.  
We then estimated their effective temperatures for these M dwarfs from their spectral 
types using a correlation between $T_{eff}$ and spectral sub-types (see Appendix~\ref{sec:teff_m}). 
In total, we selected the spectra of about 164800 reference dwarfs (FGKM) with solar metallicities over the temperature range of 3000-6500 K.

To remove active stars from the selected reference dwarfs sample, we considered $H_{\alpha}$ line as the activity indicator and 
measured the equivalent width of $H_{\alpha}$ line ($EW_{H_{\alpha}}$), where the line was centered at 6563~\AA~over 12~\AA~bandpasses, 
the continuum flux was taken to be the average flux between 6547-6557~\AA~and 6570-6580~\AA.
We binned the stars on $T_{eff}$ with a bin-width of 50 K, and then got mean value of $EW_{H_{\alpha}}$ in each bin, 
as well as corresponding standard deviation ($\sigma$), by fitting the distribution (histogram) of $EW_{H_{\alpha}}$ with a Gaussian function.  
We then culled sample stars with slight loose limits: 
about 10000 stars with $EW_{H_{\alpha}}$ larger than $3.5\sigma$ were identified to be active stars, 
several hundreds ($\sim$800) spectra with $EW_{H_{\alpha}}$ lower than $4.0\sigma$ (might be due to bad measurements). 
In this study, we used $\sim154000$ spectra of FGKM dwarfs as inactive reference stars, shown with grey dots in Fig.~\ref{fig:teff_ewha}.
%
%%%%%%%%%%%%%%%%%%%%%
\subsubsection{Standard TiO2 strength}  
\begin{figure*}
\centering
\includegraphics[width=\columnwidth]{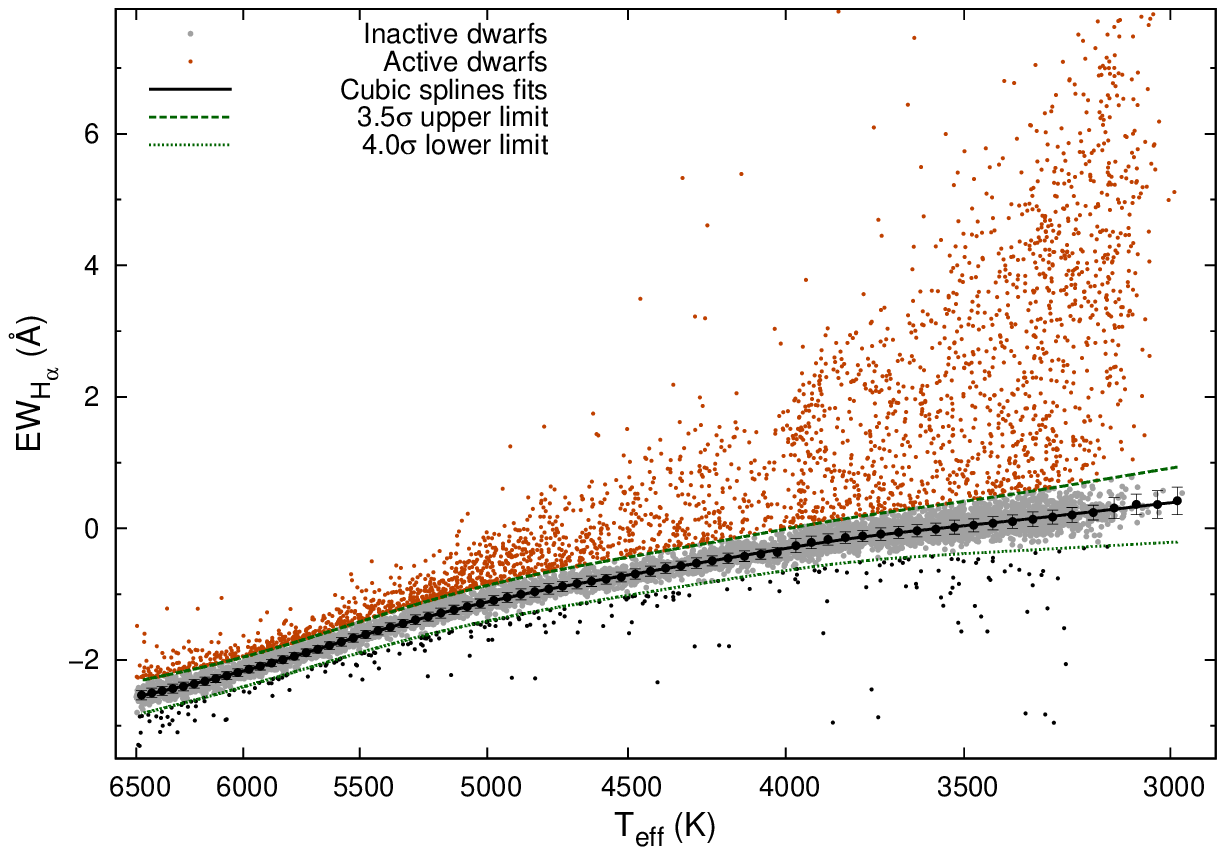} 
 \includegraphics[width=\columnwidth]{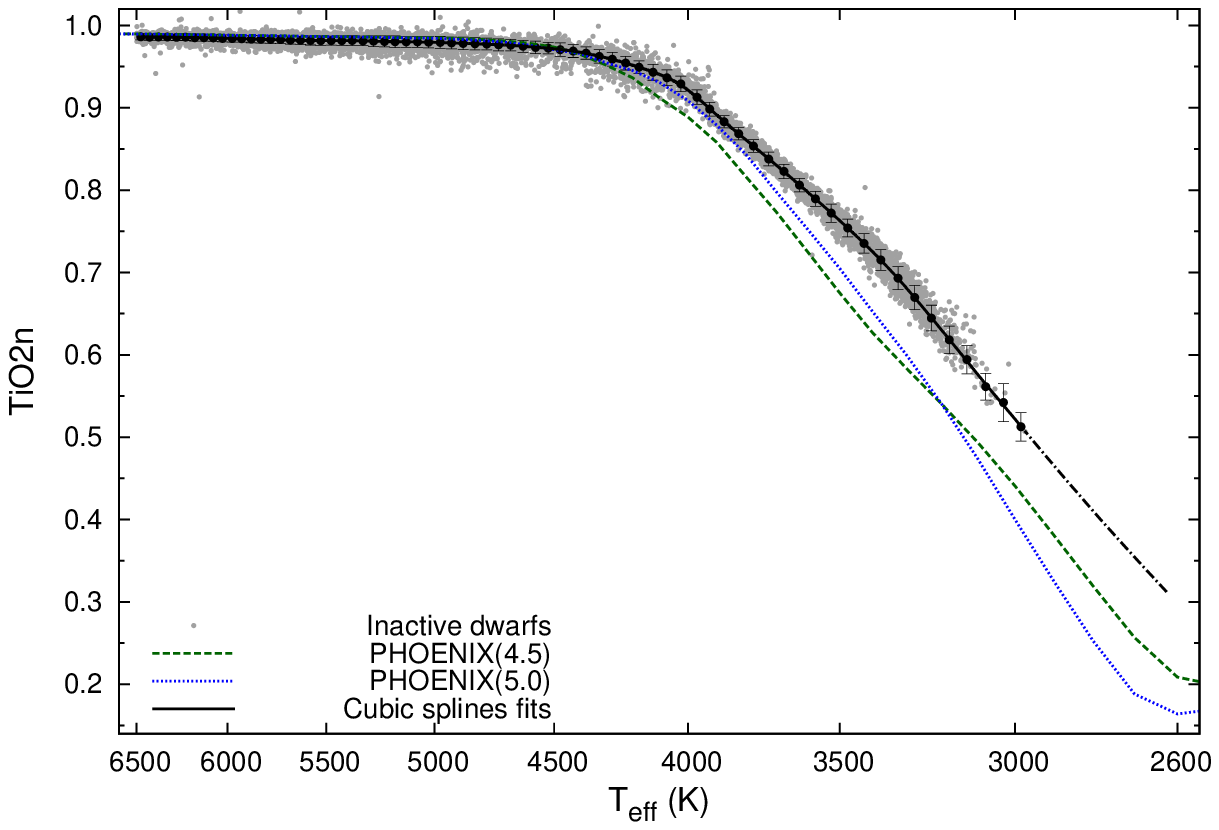}
\caption{Left panel shows the distribution of $EW_{H_{\alpha}}$ of reference sample dwarfs. 
Black dots with error bars are the mean values with standard deviations 
of Gaussian $EW_{H_{\alpha}}$ distributions in individual $T_{eff}$ bins. 
Black solid line represents the smoothed fit to the binned mean values. 
Upper $3.5\sigma$ and lower $4.0\sigma$ to the mean value are shown with green dashed and dotted lines, respectively. 
Note final inactive reference sample selected between these two lines are shown in grey points. 
Right panel shows derived standard relation between TiO2n and $T_{eff}$ for these inactive reference sample. 
Black solid line represents the smoothed fitting to the binned mean values of TiO2n, and black dot-dashed line 
represents a linear extrapolation of TiO2n with $2600<T_{eff}< 3000$ K. 
Green dashed and blue dotted lines indicate relations from PHOENIX models with $\log g$ =4.5\&5.0, respectively. }
\label{fig:teff_ewha}
\end{figure*}

In this analysis, all the measurements were carried out using RV corrected spectra. 
For FGK dwarfs, we adopted RV values derived by LASP, while for M dwarfs we measured their 
RVs in the manner described previously (but see also Appendix~\ref{sec:rv_m}).
In order to reduce the uncertainty of measurements due to RV corrections, 
we slightly broadened the continuum and molecular band region, 
as listed in Table~\ref{tab:index_defin} (added ``n'' to the index name to designate modified indices). 
Based on those selected inactive reference stars, we then got the binned mean TiO2n by Gaussian fitting to the distribution in each $T_{eff}$ bin (50 K),  
and finally got the standard relation of TiO2n versus $T_{eff}$, 
as shown by large black dots with error bars and further a smoothed black line in the right panel of Fig.~\ref{fig:teff_ewha}. 
As a comparison, we also plotted the scales of TiO2n predicted by PHOENIX model 
spectra with solar metallicity and $\log g=4.5-5.0)$, 
after degrading their high spectral resolutions by convolving a Gaussian profile to match LAMOST spectra.

%%%%%%%%%%%%
\subsection{Determination of spot filling factor }
In fact, besides the cool spots, there might also be hot spots (e.g., faculae) on the surface of an active star, 
thus the contribution from hot spots to the total flux should be considered as well. 
For simplicity, in this work, we assumed the photometric properties of hot spots are similar to those of quiescent photosphere.   
In this case, we could derive the cool spot filling factor $f_{s}$ with a two-temperature model using the modelling technique mentioned in Sec.~\ref{sec:maths}. 
This technique need independent determination of the quiescent photosphere temperature $T_{q}$ for active stars. 
We estimated $T_{q}$ for Pleiades candidates using $V-I$ color index, 
which is least influenced by spottedness \citep{stau2003}, 
or adopted a corrected $T_{eff}$ value from other colors (see Appendix~\ref{sec:teffq} for more details). 
Besides $T_{q}$, determining the spot filling factor also need an estimate of the cool spot temperature $T_{s}$. 
Our current understanding about cool spot 
temperature includes several techniques, e.g., simultaneous modelling of brightness and color variations,
Doppler imaging, modelling of molecular bands, and modelling of atomic line-depth ratios \citep[see a review by][]{berd2005}. 
However, due to various limitations in these current methods, only few measurements of absolute spot temperatures are available, 
which show method dependency \citep{stra2009}. 
Now we only know that the spot temperature contrast with respect to the quiescent photospheric temperature 
($\Delta T = T_{q} -T_{s}$) seems to be a function of $T_{q}$, e.g., $\Delta T$ have values near 2000 K 
for early G stars and drops to 200 K for the late M stars \citep{berd2005}. In principle, 
we could simultaneously determine the spot coverage and spot temperature by fitting two TiO bands with different temperature sensitivity, 
e.g., \citet{neff1995} estimated spot filling factor and spot temperature by fitting TiO bands near 7055 \AA\ and 8860 \AA. 
In practice, we found such method would result in large uncertainties, 
mainly due to large uncertainties in observed continuum flux and weak strength of TiO near 8860 \AA~band that falls in the red end of LAMOST spectra, 
and partly due to weak dependency of spot coverage on the spot temperature near 3500-3600 K 
(thought to be the spot temperature for late G- and K-type stars). 
In this study, for simplicity, we fixed the quiescent photosphere temperature $T_{q}$ (thus the $F_{q}$ of quiescent reference star), 
and then modelled TiO2n band strength $F_{t}$ of a target star using $F_{s}$ arrays of spot reference stars 
spanning the range, 2600 K $\leq T_{s} < T_{q}$ (e.g., equivalently $T_{s}/T_{q}$ $\sim 0.5-1.0$). 
Among these derived $f_{s}$, we adopted the minimum value as the ``real'' spot fractional coverage. 
Further analysis showed that such procedure would not result in large deviations (see Sec.~\ref{sec:uncertain}). 
In this analysis, for stars with larger TiO2n compare to their inactive counterparts, 
we simply set their cool spot filling factors to be zero.
%%%%%%%
%
\begin{table*}
\caption{Spectroscopic measurements of sample Pleiades members. Full table is available online.}
\label{tab:objects_results}
\begin{tabular}{lccccccccccccccccccc}
   \hline
   Object name   &$T_{spec}$& RV&$EW_{H_{\alpha}}$&TiO2n&TiO5n &$T_{q}$ &$f_{s1}^{a}$ &$T_{s1}^{a}$ &$f_{s2}^{b}$ & $T_{s2}^{b}$ & Flag$^{c}$\\
                 & (K)  &(km\,s$^{-1}$)& (\AA)&        &       & (K)  &       &  (K)  &        &  (K)  &      \\
   \hline
     Melotte 22 174 & 5022 &  -4.50 &  0.0090 & 0.9763 & 0.9432 & 5152 & 0.110 & 3525 & 0.126 & 3595 & 1-1-2-0-0 \\
  Melotte 22 HCG 39 & 4708 &   3.14 & -0.2760 & 0.9766 & 0.9417 & 4710 & 0.000 &      & 0.059 & 3595 & 1-1-1-0-0 \\
  Melotte 22 DH 166 & 4044 &  12.19 &  0.0291 & 0.9361 & 0.8773 & 4190 & 0.179 & 3390 & 0.198 & 3500 & 1-1-1-1-1 \\
    Melotte 22 1100 & 4513 &   1.90 &  0.3080 & 0.9615 & 0.9125 & 3653 & 0.236 & 3480 & 0.309 & 3555 & 1-1-1-0-0 \\
  Melotte 22 DH 075 & 3731 &  11.48 &  3.8819 & 0.8143 & 0.7054 & 3716 & 0.221 & 3120 & 0.000 &      & 1-1-1-1-1 \\
 Melotte 22 HHJ 295 & 3145 & -11.83 &  9.2937 & 0.5357 & 0.4000 & 3145 & 0.573 & 2740 & 0.062 & 2835 & 1-1-6-1-1 \\
J033754.79+252631.2 & 3496 &   4.12 &  2.0233 & 0.7240 & 0.5824 & 3479 & 0.308 & 3000 & 0.000 &      & 2-1-3-1-1 \\
J040443.86+253333.0 & 3980 & -32.82 & -0.4746 & 0.9068 & 0.8356 & 3977 & 0.100 & 3280 & 0.003 & 3445 & 2-1-3-1-1 \\
    Melotte 22 1173 & 3672 &   8.48 &  2.1087 & 0.8093 & 0.6712 & 3739 & 0.318 & 3115 & 0.287 & 3212 & 1-3-1-1-1 \\
\hline
\multicolumn{12}{l}{$^a$ Spot filling factor ($f_{s1}$) and spot temperature ($T_{s1}$) from TiO2n }\\ 
\multicolumn{12}{l}{$^b$ Spot filling factor ($f_{s2}$) and spot temperature ($T_{s2}$) from TiO5n }\\ 
\multicolumn{12}{l}{$^c$ Flag N1-N2-N3-N4-N5:  }\\ 
\multicolumn{12}{l}{~~~ N1: 1 = sample-1, 2 = sample-2;}\\
\multicolumn{12}{l}{~~~ N2: number of LAMOST observations; }\\
\multicolumn{12}{l}{~~~~N3: $T_{q}$ from $T_{color}$: 1 = $T_{VI}$, 2 = corrected $T_{VK}$, 3 = corrected $T_{rK}$, 4 = corrected $T_{IK}$;}\\
\multicolumn{12}{l}{~~~~~~~ N3 = original N3 + 5 for $T_{color}-T_{spec} < -100$ K;}\\
\multicolumn{12}{l}{~~~ N4: $T_{spec}$ from LAMOST pipeline (0) or our measurement (1); }\\ 
\multicolumn{12}{l}{~~~ N5: RV from LAMOST pipeline (0) or our measurement (1) }\\ 
\end{tabular}
\end{table*}

%%%%%%%%%%%  
%%%%%%%%%%%%%%%%%%%%%%%%%%%%%%%%%%%%%%%%%%%%%%%%%
\section{Limitations and uncertainties}
\label{sec:uncertain}
\begin{figure*}
\centering
\includegraphics[width=\columnwidth]{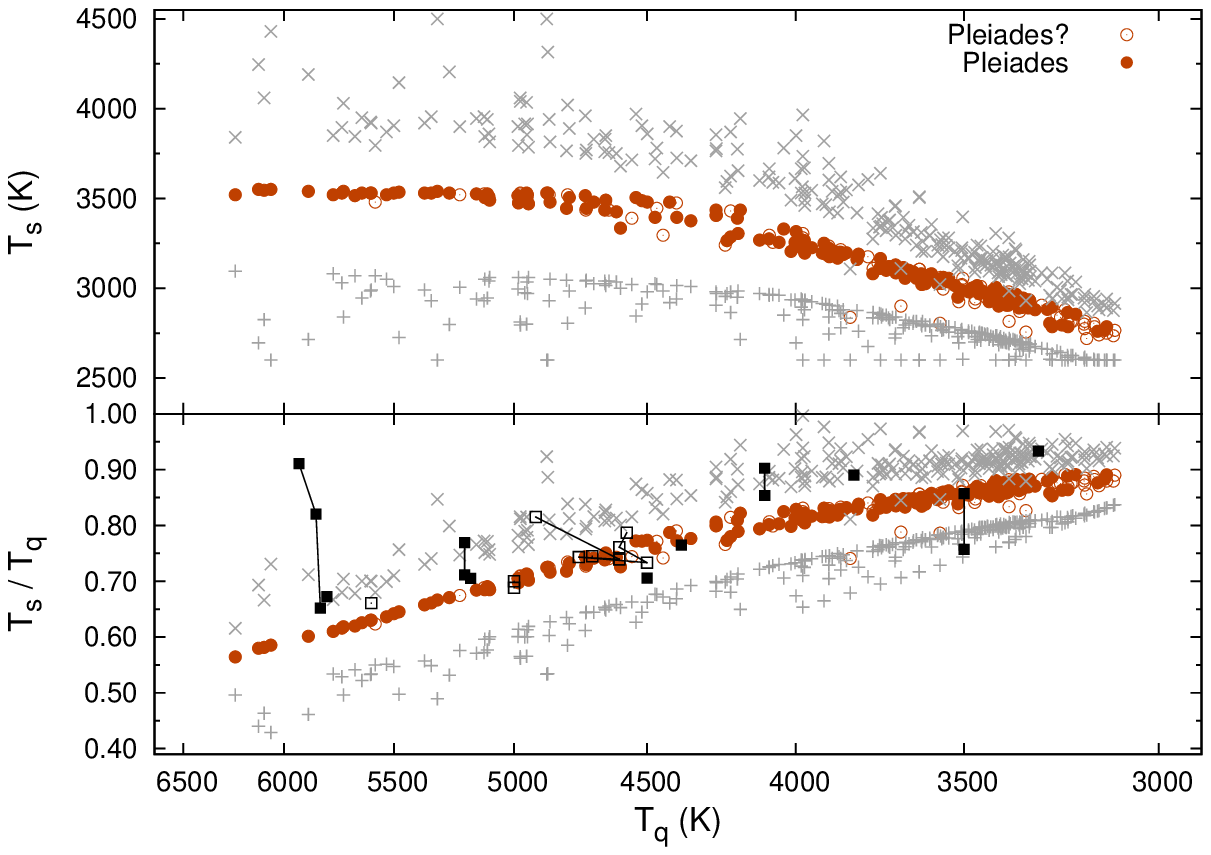}
\includegraphics[width=\columnwidth]{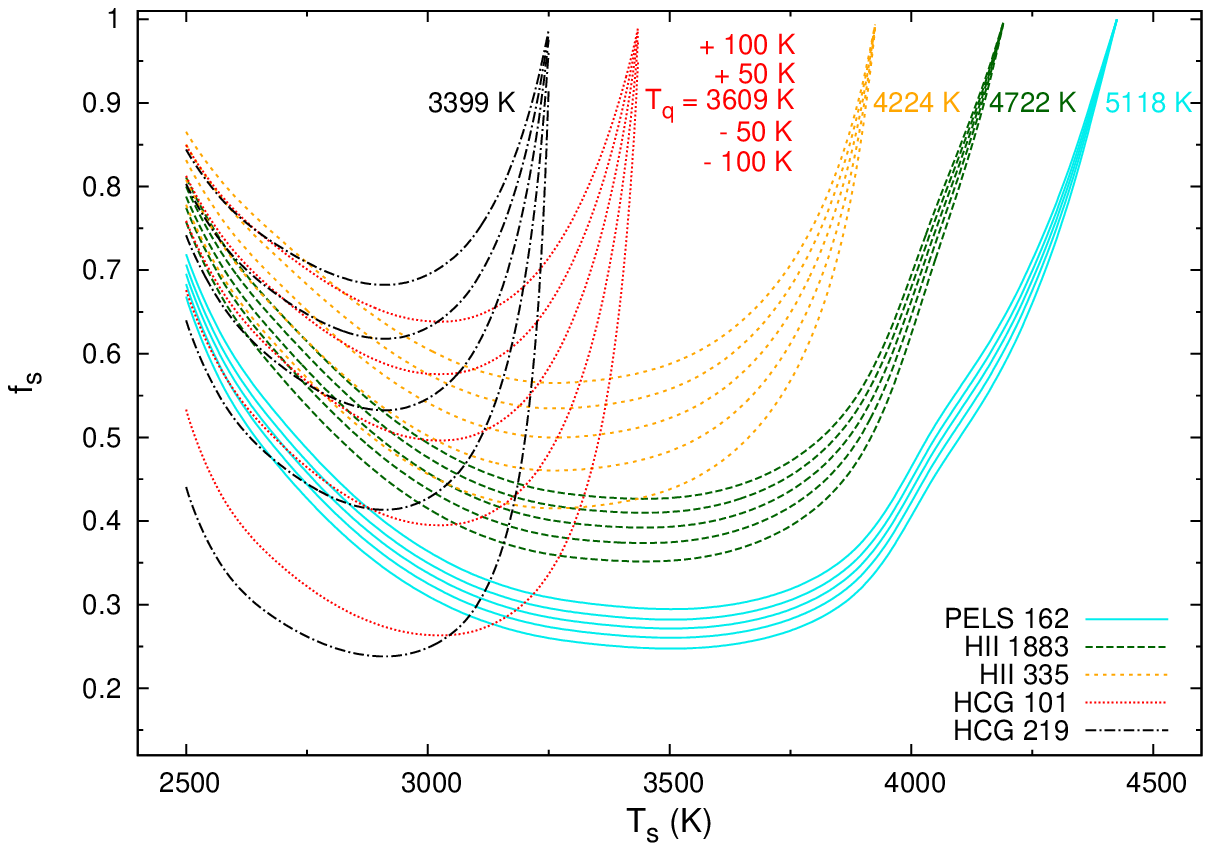}
\caption{ Left panels show the relations between $T_{q}$ and adopted $T_{s}$ (corresponding to $f_{s}$ minima) (top),
and ratio of $T_{s}$ to $T_{q}$ (bottom). 
Lower and higher spot temperatures corresponding to spot filling factors 5\% larger than minimum value are shown in grey plus and cross symbols, 
respectively. 
Filled circles represent probable single members and open circles with dots denote probable binary members or non-members.
Filled and open squares with dots indicate the spot temperatures from previous studies for dwarfs and sub-giants, respectively. 
Black solid lines are drawn connecting symbols of same star. 
Right panel presents five examples showing the dependence of $f_{s}$ on spot temperature $T_{s}$.
For each object, we show five cases corresponding to five quiescent photosphere temperatures: $T_{q}$, $T_{q}\pm50$ K, $T_{q}\pm100$ K.}
\label{fig:sfs_ts_tq}
\end{figure*}

In this work, we modeled TiO2 band strength for an active star using different pairs of
spot/quiescent reference stars over a wide range of spot temperatures,  
and finally adopted the results with minimum spot filling factors as the ``real" spot fractional coverage. 
We displayed the spot temperatures corresponding 
to the $f_{s}$ minima in Fig.~\ref{fig:sfs_ts_tq}, showing that the spot temperatures $T_{s}$ 
fall in the range of 2800-3600 K, equivalent spot temperature contrast ratios ($T_{s}$/$T_{q}$) are in the range of 0.6-0.9. 
Previous studies showed $T_{s}$/$T_{q}$ is around 0.8 \citep[see][and reference therein]{berd2005,stra2009}, 
shown in the lower left panel of Fig.~\ref{fig:sfs_ts_tq}, 
wherein we present the spot temperature contrast ratios for several active dwarfs and sub-giants that were derived based on different techniques 
and collected by \citet{berd2005} from literature. 
We found that the spot temperatures corresponding to $f_{s}$ minima are similar to those values previously derived for spotted stars, 
just being slightly cooler for earlier type stars. To check how the spot temperature affect the derived spot coverage, 
we showed five examples in right panel of Fig.~\ref{fig:sfs_ts_tq}, 
where we displayed behaviour of $f_{s}$ with different spot temperatures $T_{s}$ for five Pleiades candidate members 
with spectral types from late G- to medium M-type. 
We noticed $f_{s}$ varies very slowly over a wide $T_{s}$ range around the minimum of $f_{s}$ for K-type stars,  
indicating  $f_{s}$ is insensitive to $T_{s}$ in the temperature region around adopted $T_{s}$. 
e.g., for PELS 162 spot coverage have nearly constant values over $T_{s}$ of 3200-3800 K 
(equivalent $T_{s}$/$T_{q}$ $\sim$0.63-0.75). 
In Fig.~\ref{fig:sfs_ts_tq} the plus and cross symbols, respectively, represent the cooler and hotter 
spot temperatures corresponding to the spot filling factor 5\% larger than the adopted minima, 
defining a $T_{s}$ region with $f_{s}$ uncertainty of 5\%. The spot temperatures of active dwarfs and sub-giants, derived by previous studies, are
found almost within this region, which indicates that our adopted minima of spot filling factors could be a good indicator of real spot coverages. 
\begin{figure*}
\centering
\includegraphics[width=\columnwidth]{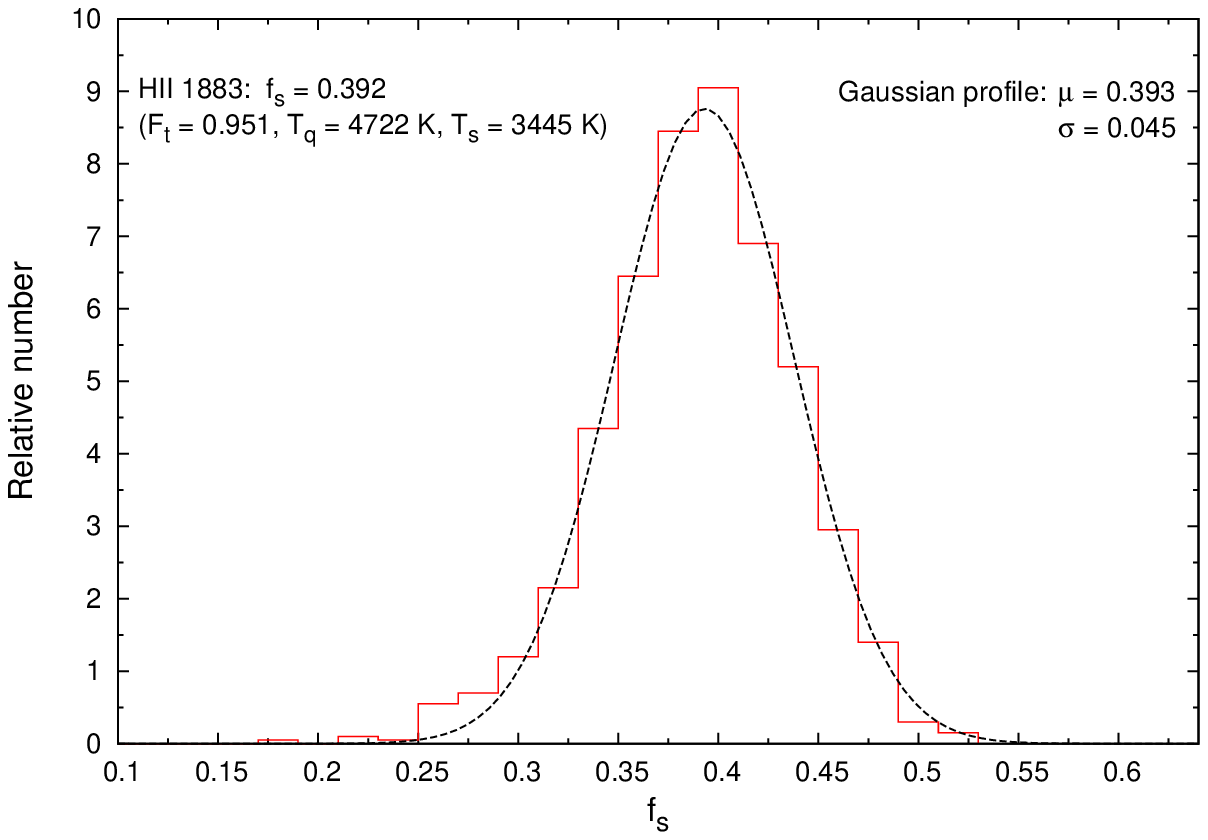}
\includegraphics[width=\columnwidth]{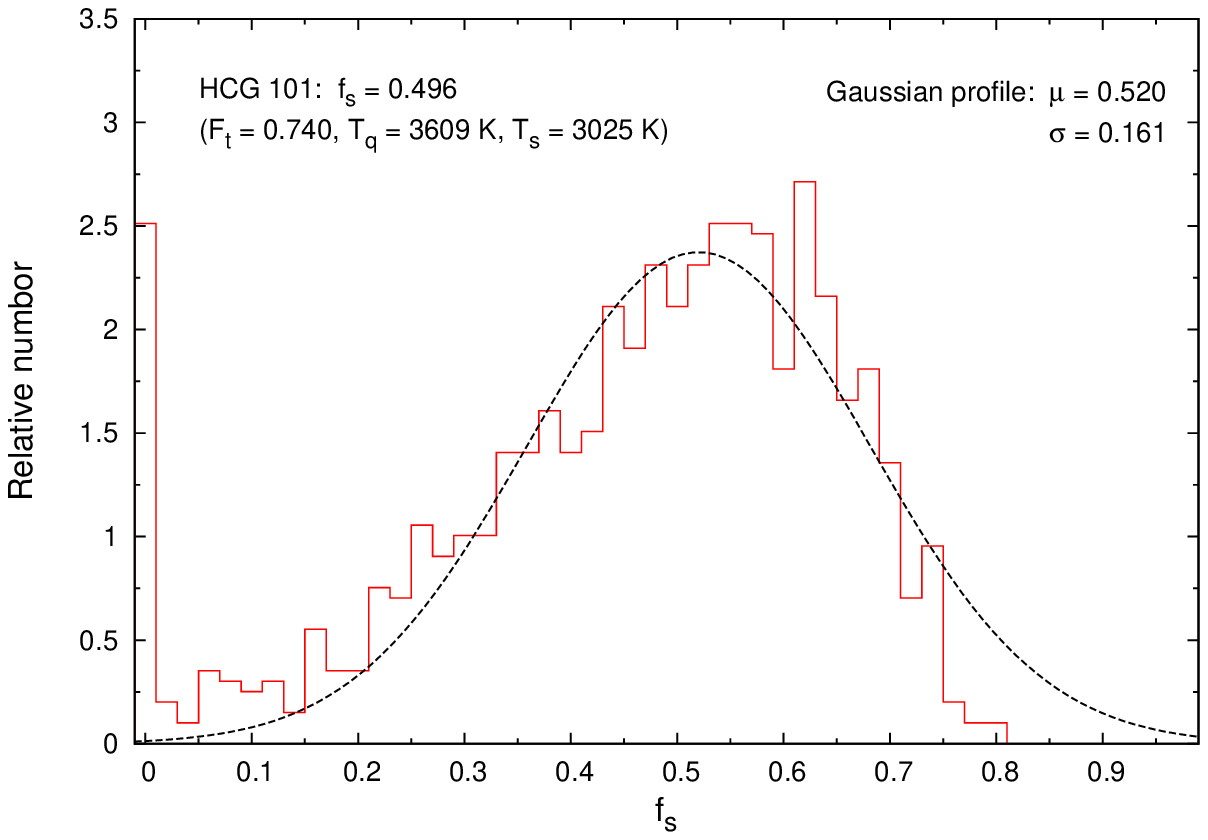}
\caption{Distribution of spot filling factors ($f_{s}$) for Pleiades members HII 1883 and HCG 101 are shown in left and right panel, respectively.
Dashed line represents the Gaussian fit to the $f_{s}$ distribution. Resulting mean ($\mu$) and standard deviations ($\sigma$)
are given in respective panels. }
\label{fig:sfs_mc_example}
\end{figure*}
%

%{\bf 
The uncertainties of spot filling factors measurements mainly propagates from the observed spectra 
(errors of observed TiO band flux, uncertainties in RV, and rotational broadening),
the models (continuum surface flux ratio $R_{\lambda_{c}}$, and TiO2 scaling based on reference dwarfs),
and the estimates of quiescent photosphere temperature. we discussed these issues as follows. 

\begin{figure}
\centering
\includegraphics[width=\columnwidth]{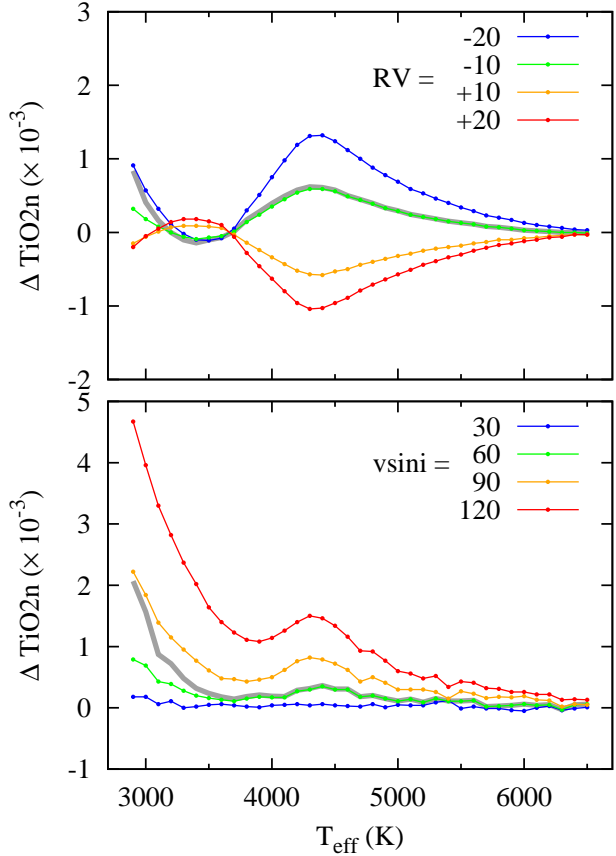}
\caption{Estimated differential TiO2n values due to wavelength shifts, RV$=\pm 10$, $\pm 20$ km\,s$^{-1}$ (upper panel), 
and rotational broadenings, $v\sin i=30,~60,~90,~120$ km\,s$^{-1}$ (lower panel). 
Note grey color solid line represent ratio of $\Delta$TiO2n to TiO2n 
at shift of +10 km\,s$^{-1}$ (upper panel) and broadening of 60 km\,s$^{-1}$ (lower panel).}   
\label{fig:diff_rv_rot}
\end{figure}

The spectral resolution of LAMOST spectra (R$\sim$1800), 
and the large rotational broadening are two factors that limits the accuracy of RV measurements for Pleiades candidates. 
To check how RV uncertainties affect TiO2n measurements, we have taken PHOENIX spectra (R$\sim$500000) 
and degraded the resolution to 1800 by convolving a Gaussian profile, 
and found the differences of TiO2n ($\Delta$TiO2n) due to RV shifts of $\pm$ 20 km\,s$^{-1}$ 
are less than 0.002, shown in Fig.~\ref{fig:diff_rv_rot}. 
In addition, many Pleiades K and M dwarfs are found to be rapid rotators, e.g., HII 2208 have a $v\sin i$ value of 73 km\,s$^{-1}$ \citep{stau1987}, 
and HII 1883, the almost extreme case, have a $v\sin i$ of 140 km\,s$^{-1}$ \citep{stau1984}. 
We thus evaluated the affects of rotational broadening to TiO2n measurements, as shown in the lower panel of Fig.~\ref{fig:diff_rv_rot}, 
and found the difference of TiO2n due to $v\sin i$ of $\sim100$ km\,s$^{-1}$ are typically less than 0.002, 
which is comparable with errors due to RV uncertainties, and even less than typical errors of TiO2n measurements 
propagated from the uncertainties of observed flux (For GK members, the typical value of uncertainty of TiO2n is less than 0.005, 
in case of fainter members it is slight larger than 0.005). 
Our analysis showed that the error of 0.002~in TiO2n would not result in large deviation of spot filling factors 
(see Table~\ref{tab:results_mc}).   

The spot coverage depends strongly on the quiescent photosphere temperature, particularly for very cool stars, 
as shown in the right panel of Fig.~\ref{fig:sfs_ts_tq}. The errors of colors used in this paper are typically less than 0.05 mag,
which corresponds to the uncertainties of $T_{eff}$ around 100 K.
For early K-type stars, 100 K deviation in $T_{q}$ will lead to a spot coverage difference of about $2\%-3\%$, while
for a medium M-type star such a temperature deviation results in very large ($15\%-20\%$) variation,
since TiO2 is strongly temperature-sensitive, as shown in Fig.~\ref{fig:teff_ewha}.
To estimate the uncertainties on the determinations of $f_{s}$, a Monte Carlo approach was used.
For simplicity, we performed the Monte Carlo simulation 1000 times for five example stars using the target
TiO2n strength $F_{t}$ with standard deviation corresponds to measurement error (typically 0.005),
and quiescent photosphere temperature $T_{q}$ with standard deviation of 100 K, as listed in Table~\ref{tab:results_mc}.
Figure \ref{fig:sfs_mc_example} gives two examples of $f_{s}$ simulations (HII 1883 and HCG 101). 
From the Table~\ref{tab:results_mc}, 
one can see that the uncertainties of $f_{s}$ of hotter stars, compare to K dwarfs, are mainly comes from the errors of $F_{t}$  
because of the sharp continuum flux contrast between spot and quiescent photosphere 
($R_{\lambda_{c}}$ is smaller, the same error of $F_{t}$ result in larger $f_{s}$ uncertainty). 
For cooler stars (e.g., M stars), the uncertainty in quiescent photosphere temperature dominates the $f_{s}$ uncertainty, 
since the TiO2 strength have strong temperature-sensitivity when $T_{eff}<4000$ K. 
In other words, comparing hotter and cooler stars, K dwarfs have smaller uncertainties in spot coverage.   

For M dwarf stars, we estimated their effective temperatures from spectral characteristics. 
By comparing the temperatures from spectroscopy with those from photometry for M dwarf reference stars, 
we found that the effective temperatures from spectral features are underestimated about 50 K (see Appendix~\ref{sec:teff_m}), 
which would overestimate the spot filling factors, especially for very cool stars.
To evaluate the deviation of $f_{s}$ due to any systematic deviation in the standard relation of TiO2n vs. $T_{eff}$, 
we derived spot filling factors for five example members using the standard relation of TiO2n vs. $T_{eff}$ but with shifted $T_{eff}$ of 50 K, 
and found that the resultant differences of $f_{s}$ 
are about 0.02, 0.03, 0.04, 0.11 and 0.13 for PELS 162, HII 1883, HII 335, HCG 101 and HCG 219, respectively. 
Therefore, for GK-type stars, the underestimation of $T_{eff}$ by about 50 K in TiO2n vs. $T_{eff}$ standard relation 
would affect spot filling factors marginally, 
but for M-type members, especially medium and late M-type members, 
the derived spot filling factors might be overestimated (up to 0.1 or above).
%}
%
\begin{table}
\caption{Estimated parameters from Monte Carlo simulations for 5 Pleiades members}
\label{tab:results_mc}
\begin{tabular}{lccccccccccccccccccc}
   \hline
Object &$T_{q}$ &$\sigma(T_{q})$ &$F_{t}$ &$\sigma(F_{t})$& $\sigma_{1}^a$&$\sigma_{2}^b$&$\sigma_{3}^c$\\
%            & (K)    & (K)            &      &               &                &              &        \\
   \hline
PELS 162        &5118  &100          & 0.969 &    0.005  &    0.024   & 0.086           & 0.093  \\
HII 1883        &4722  &100          & 0.951 &    0.003  &    0.037   & 0.030           & 0.045  \\
HII 335         &4224  &100          & 0.900 &    0.003  &    0.071   & 0.020           & 0.077  \\
HCG 101         &3609  &100          & 0.740 &    0.004  &    0.159   & 0.022           & 0.161  \\
HCG 291         &3399  &100          & 0.658 &    0.010  &    0.160   & 0.051           & 0.160  \\
\hline
\multicolumn{8}{l}{$^a$ standard deviation of $f_{s}$ due to $\sigma(T_{q})$ }\\ 
\multicolumn{8}{l}{$^b$ standard deviation of $f_{s}$ due to $\sigma(F_{t})$ }\\ 
\multicolumn{8}{l}{$^c$ standard deviation of $f_{s}$ due to both $\sigma(F_{q})$ \& $\sigma(F_{t})$ }\\ 
\end{tabular}
\end{table}

It is important to point out that our derived standard TiO2n scales cut off at $T_{eff}\sim3000$ K, 
because of the paucity of LAMOST spectra of M dwarfs later than M6. 
However, in the procedure of searching for minimum of spot coverage, 
we simply linearly extrapolated the TiO2n when $T_{s} < 3000$ K, as shown by black dot-dashed line in Fig.~\ref{fig:teff_ewha}. 
We noticed such a linear extrapolation would not affect the final results significantly, since the PHOENIX models predict that TiO2 of M dwarfs later 
than M6 still decreases with temperature in a similar way like hotter M dwarfs, 
until $T_{eff}\sim2600$ K at which the TiO2 overturns (probably due to the formation of grains
and to the veiling effect by dust scattering). 
Finally, we note that there might exist discrepancies between the derived spot coverages 
and the real values for very cool Pleiades members (approximately mass < 0.5$M_{\sun}$ or $T_{eff}<3500$ K), 
since these stars in Pleiades with an age of $\sim120$ Myr are still 
in their pre-main-sequence phase, gravity effect would affect the TiO2 strength and thus the final results. 
Any features shown for these very cool stars in this paper should be only qualitative in nature. 

%%%%%%%%%%%%%%%%%% 
\section{Results and Discussion}
\subsection{Manifestation of activity on TiO band strengths}

\begin{figure*}
\centering
\includegraphics[width=\columnwidth]{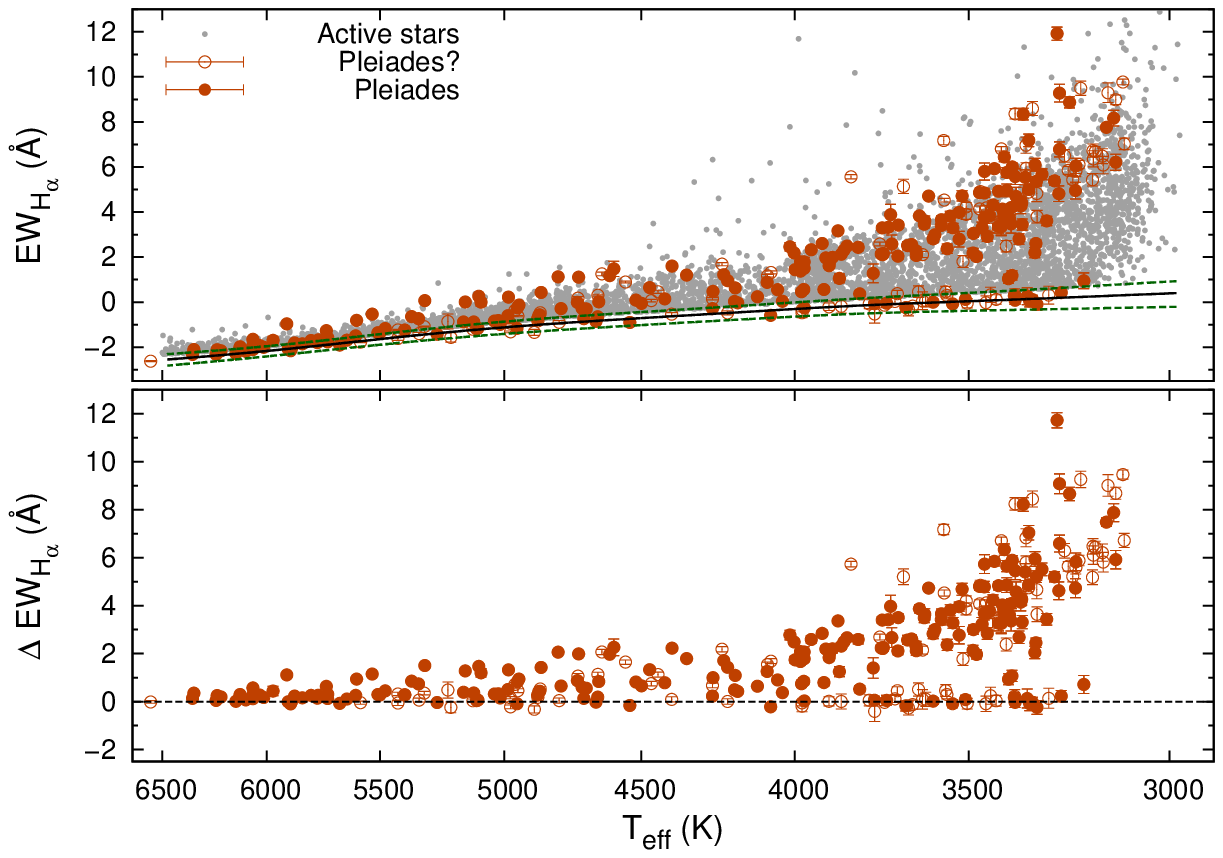}
\includegraphics[width=\columnwidth]{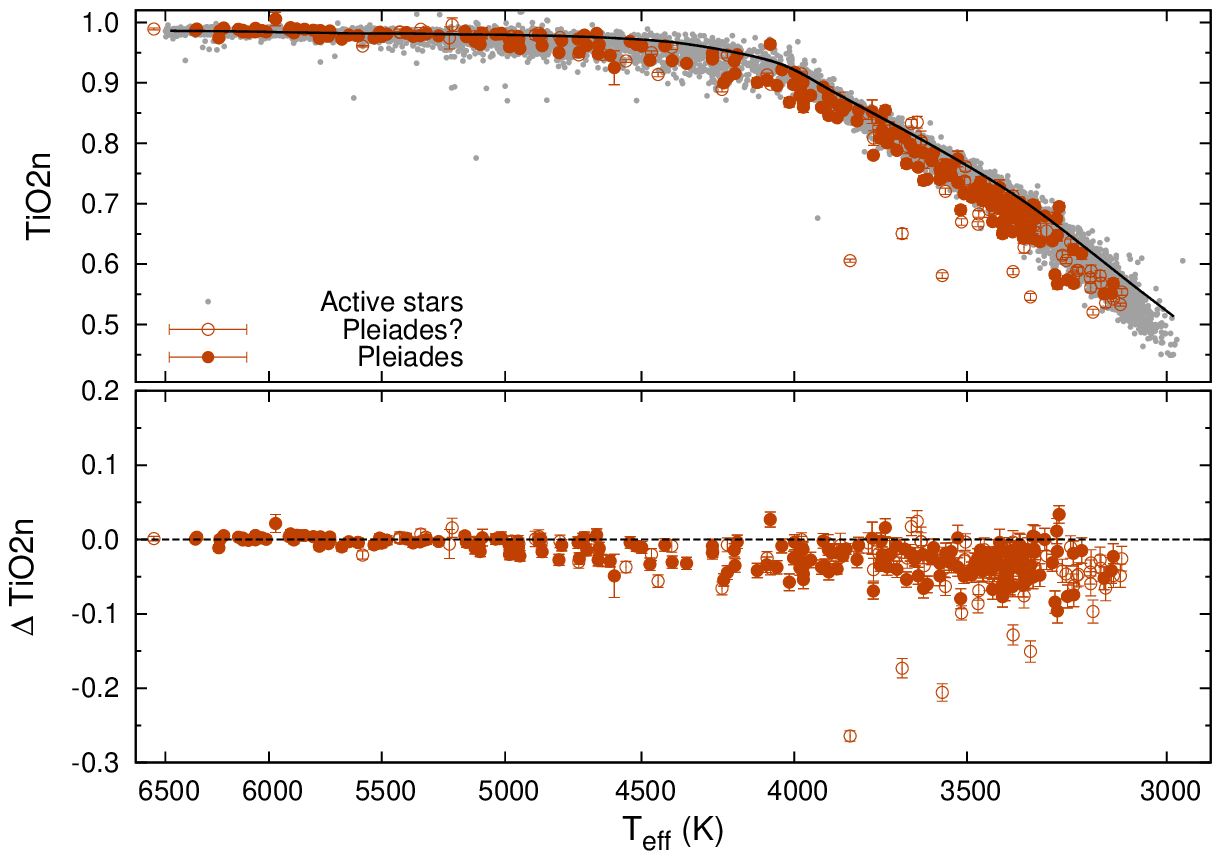}
\caption{Measurements of $EW_{H_{\alpha}}$ and TiO2n and their corresponding 
deviations ($\Delta EW_{H_{\alpha}}$ and $\Delta$TiO2n) to the mean values for entire sample stars are shown in left and right panel, respectively. 
Black solid lines show the mean relations for inactive counterparts. 
In the upper left panel, upper $3.5\sigma$ and lower $4\sigma$ to mean $EW_{H_{\alpha}}$ are shown in green dashed lines. 
Note $T_{eff}$ for Pleiades members are quiescent photosphere temperature. For members having LAMOST multi-observations, 
we displayed their average values rather than individual observational values.}
\label{fig:ewha_tio2_active}
\end{figure*}

$H_{\alpha}$ line is the strongest and widely studied activity indicator for low mass dwarf stars \citep[e.g.][]{west2004}.
We measured $EW_{H_{\alpha}}$ for Pleiades members, in which most of them show extra $H_{\alpha}$ emission features 
($\Delta EW_{H_{\alpha}} > 0$) relative to their inactive counterparts, 
indicating they are chromospherically active (see left panels of Fig.~\ref{fig:ewha_tio2_active}).  
The corresponding TiO2n measurements for Pleiades members along with previously selected $\sim 10000$ active dwarfs, 
from the right panels of Fig.~\ref{fig:ewha_tio2_active}, 
showed that the active stars and Pleiades candidate members including hotter stars (e.g., early K-type dwarfs) 
are found to have deeper TiO2 absorption ($\Delta$TiO2n < 0) compare to their inactive counterparts, indicating the presence of cool spots.

\begin{figure*}
\centering
\includegraphics[width=\columnwidth]{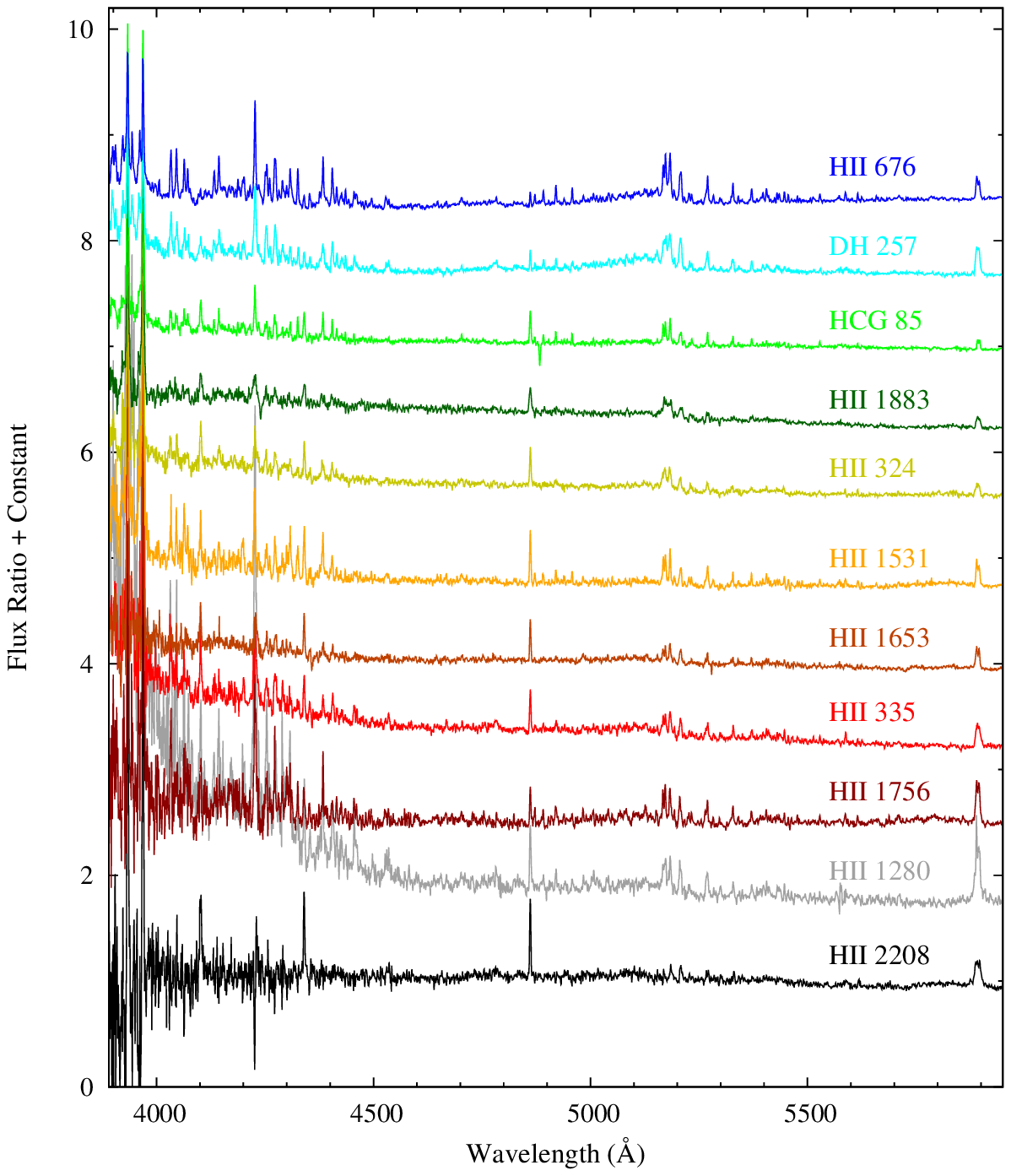}
\includegraphics[width=\columnwidth]{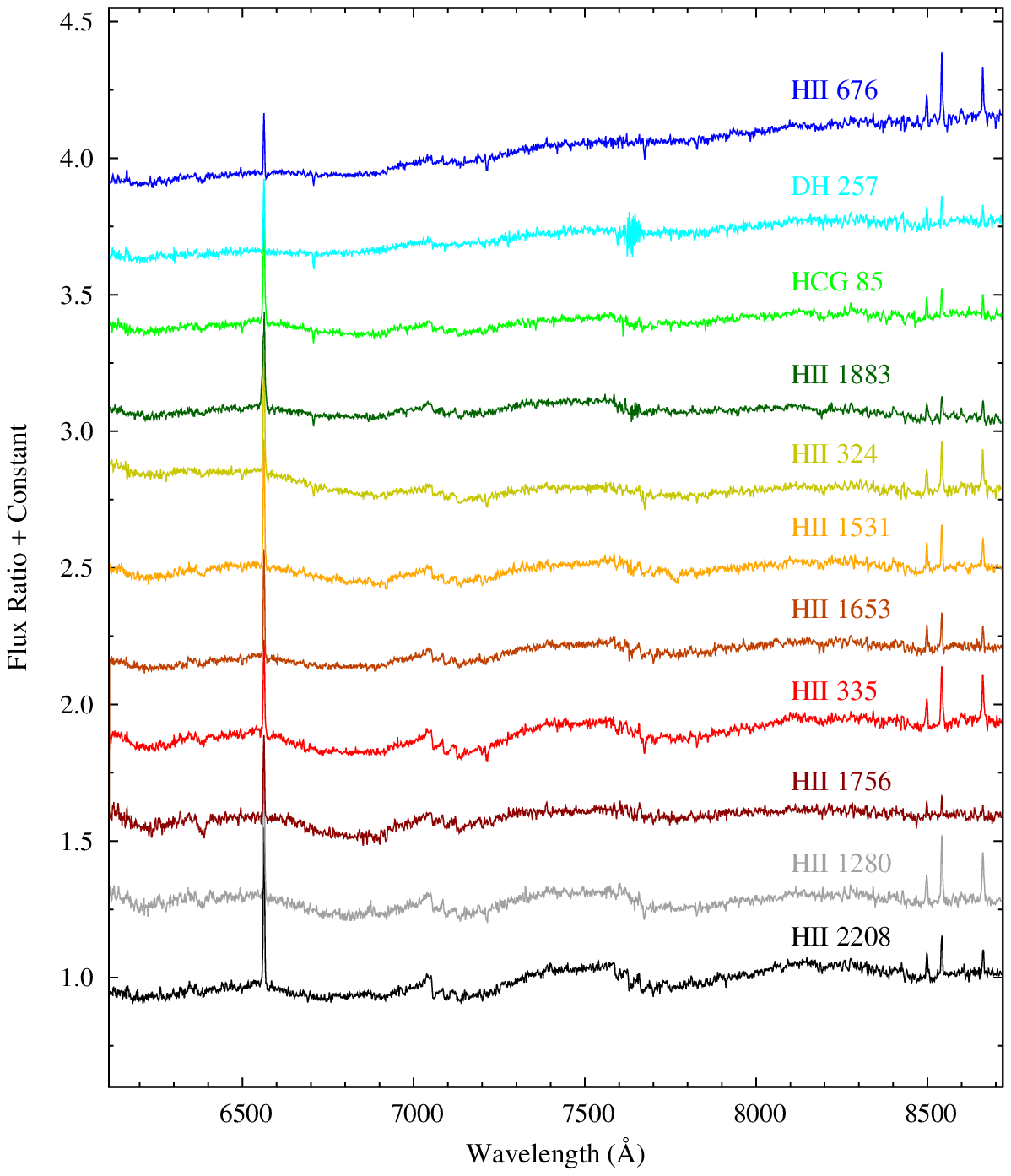}
\caption{Ratio spectra in blue region (left panel) and red region (right panel) of 
11 Pleiades cool members with spectral type from early-K (blue) to early-M (black) is shown. 
Object names are marked for reference. Note the residual emission lines of atomic species 
and absorption of molecular species (see the text for details).}
\label{fig:ratio_pleiades}
\end{figure*}

In fact, besides TiO2, large cool spots could produce measurable signals of many molecular bands in observed spectra. 
To illustrate, we computed the flux ratio spectra of 11 rapidly rotating K and early M Pleiades members (potentially single stars). 
For these 11 stars, we divided each spectra by a range of inactive template spectra 
and found the best matched template through visual check in several wavelength regions with relatively free of molecular bands. 
Note during matching the spectra with the templates, the templates were wavelength-shifted and rotationally broadened. 
To assemble the inactive template spectra, 
high quality spectra of a subsample of reference stars with similar temperatures (e.g., about 100 stars in each 50 K bin) 
were first shifted to a zero-velocity rest frame, then normalized at band of 7040-7050~\AA\ and co-added. 
The ratio spectra, in Fig.~\ref{fig:ratio_pleiades}, 
shows blue continuum rising to shorter wavelength and slight rise of red continuum towards red, 
confirming the finding of \citet{stau2003} about anomaly of spectral energy distributions (SEDs) of Pleiades K dwarfs. 
Also, the spectra show residual emission lines  of Balmer lines, Ca~{\sc ii} H\&K, and Ca~{\sc ii} IRT lines ($\lambda$8498, 8542 and 8662), 
and several neutral metal quasi-emission lines that are strongly temperature-sensitive 
such as Ca~{\sc i} $\lambda 4226$, Mg~{\sc i} b triplet lines, Fe~{\sc i} $\lambda 5208$,  
and Na~{\sc i} D lines, indicates the presence of chromospheric activity in these stars. Particularly, these spectra  
show well-defined molecular absorption features, e.g., CaOH ($\sim$6230~\AA),
CaH ($\sim$6380, 6800 and 6900~\AA), and TiO ($\sim$7055, 7088, 7126 and 7600~\AA), 
in combine with anomalous SEDs and neutral metal quasi-emission lines indicate 
that these Pleiades K stars are hotter than inactive counterparts 
in the blue band which may be due to the presence of hot spots like faculae, 
but cooler than inactive counterparts in the red that provides unambiguous proof of the presence of cool spots.  

The spectral division process could give estimate of quiescent temperatures, e.g., 
the temperature of best-matched inactive template spectrum. In these 11 Pleiades stars, 
we found the difference between the quiescent temperatures from division process and 
the temperatures derived based on broad-band colors are in range of $-$93 to 105 K. 
On an average, the temperature from division process is slight hotter.
In addition, different molecular bands have different temperature-sensitivity, 
thus the relative strengths and overall appearances of different band features could lead to possible estimates of spot temperatures.   
However, we discuss these issues elsewhere (Fang et al, in preparation), 
including detailed spectral division procedure, methods to estimate quiescent temperature and spot temperature, 
and the schemes of estimating spot filling factors, e.g., instead of simply fitting the strength of one band, we try to make simultaneous fits  
to multiple spectral regions containing molecular bands using inactive template spectra .
 
\begin{figure*}
\centering
\includegraphics[width=\columnwidth]{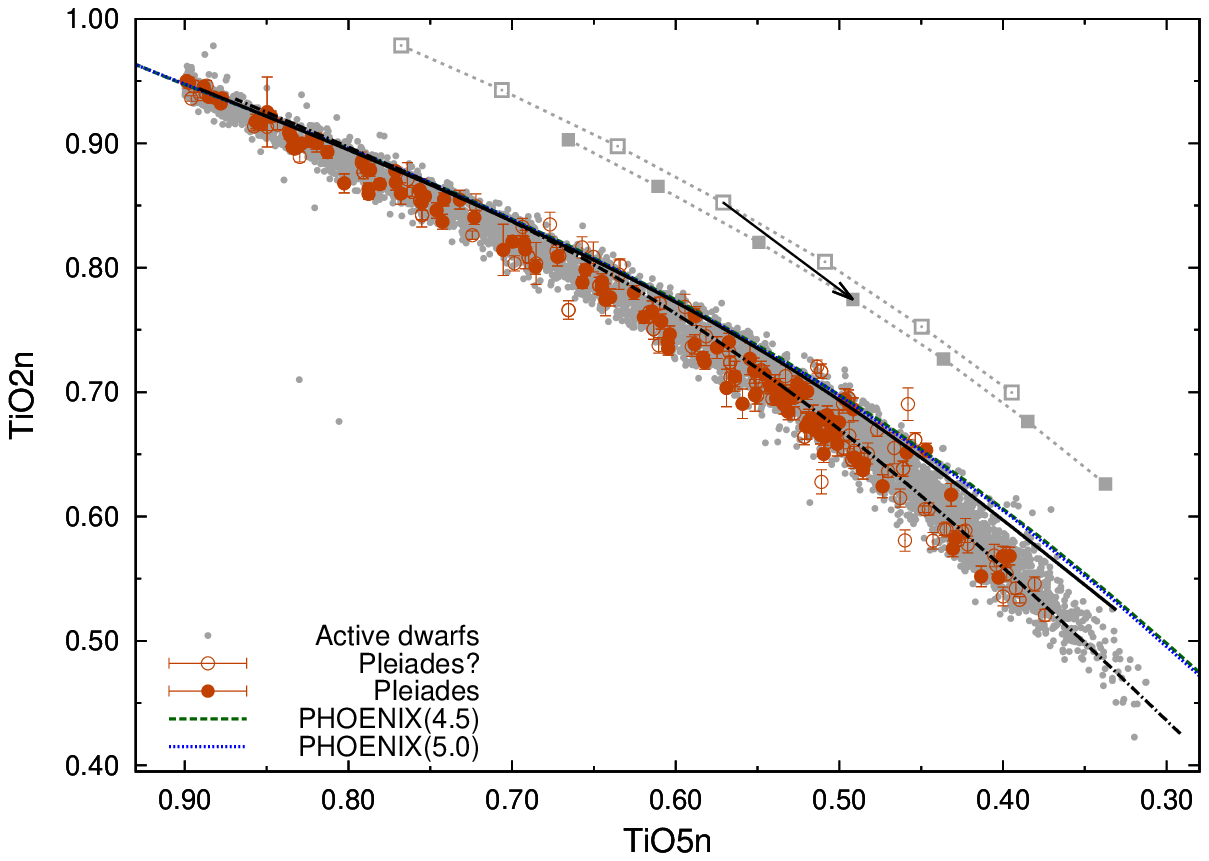}
\includegraphics[width=\columnwidth]{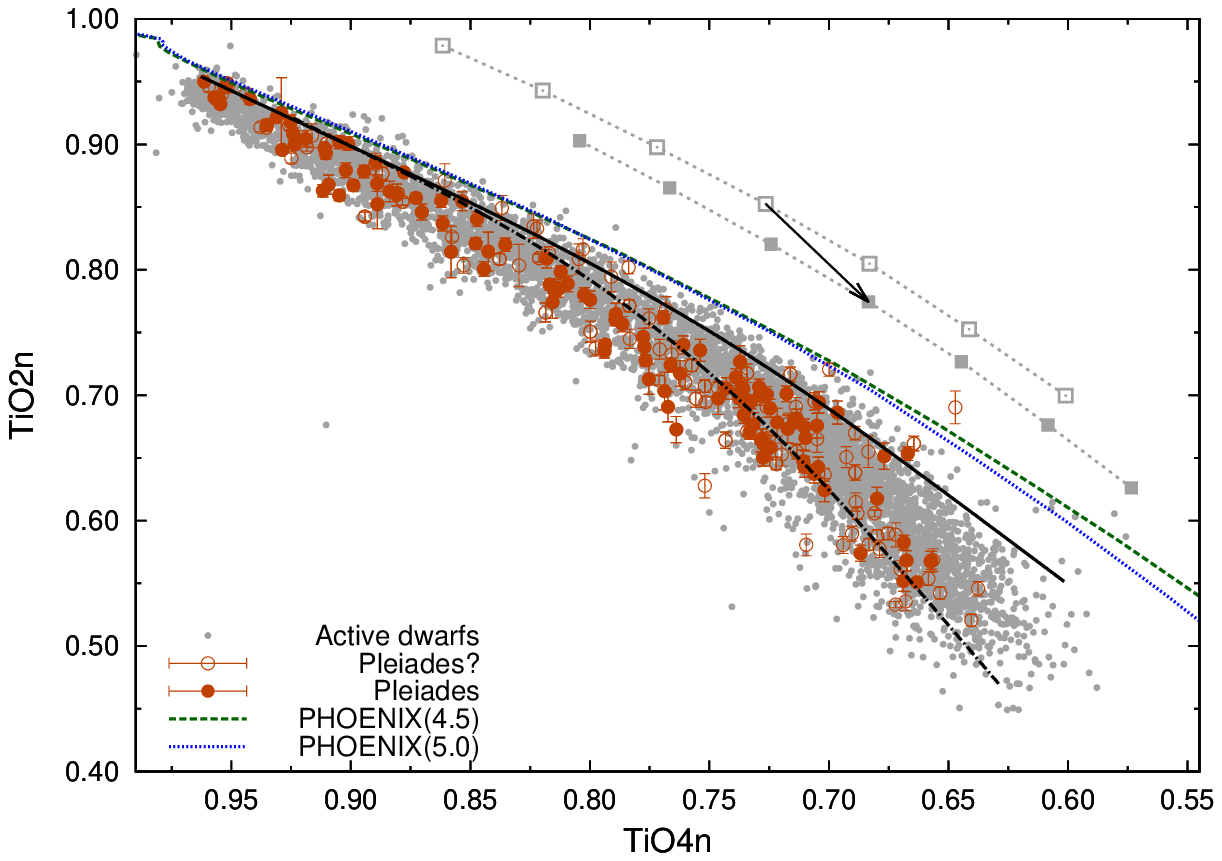}
\includegraphics[width=\columnwidth]{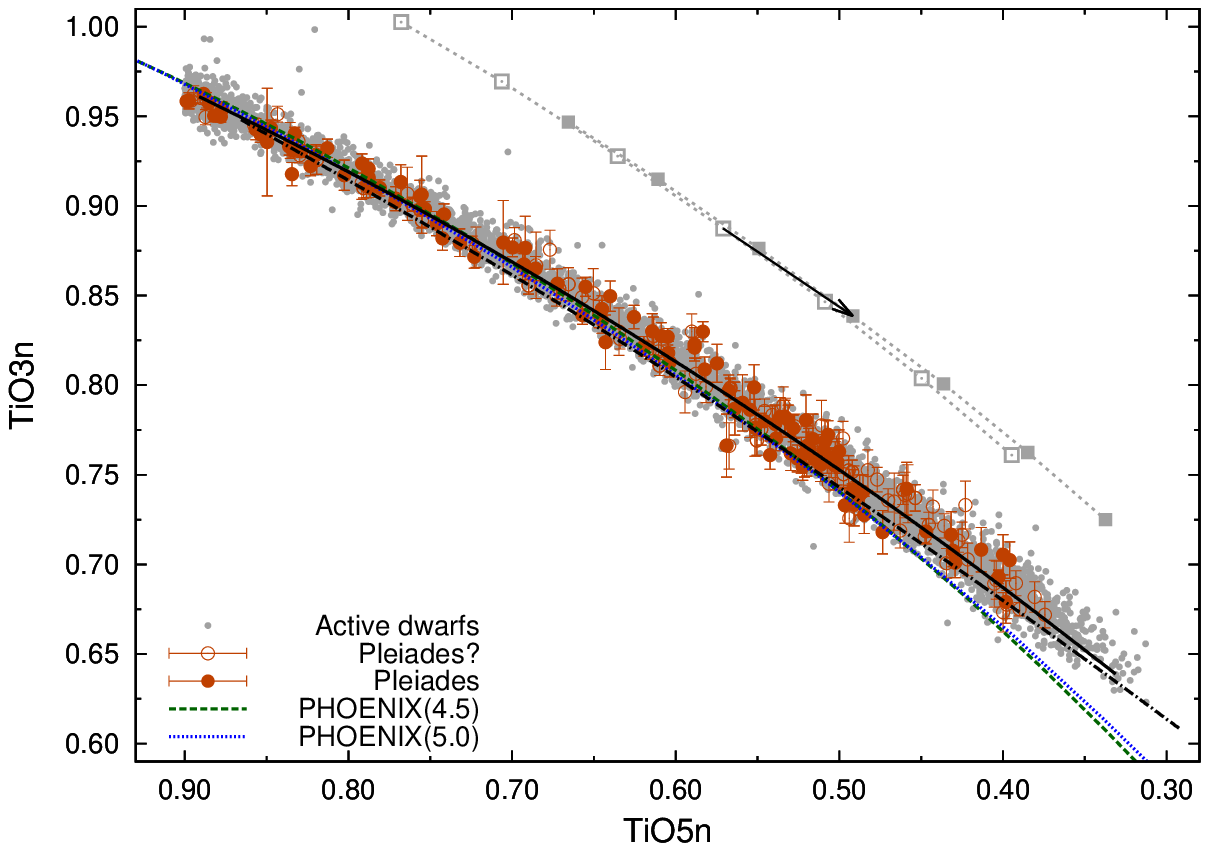}
\includegraphics[width=\columnwidth]{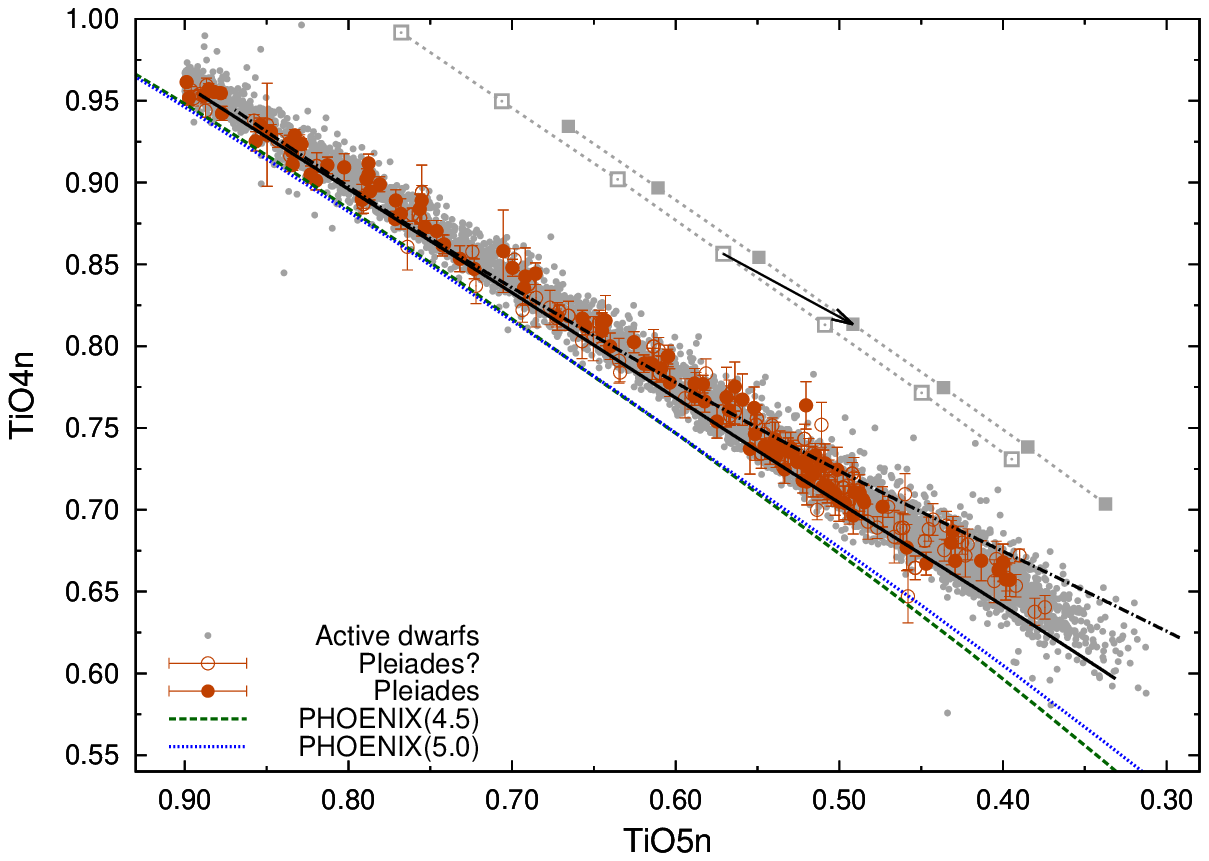}
\caption{Relations between various TiO band indices are shown. 
Black solid and dot-dashed lines in each panel indicates the mean relations for inactive M dwarfs and giants, respectively. 
``Theoretical'' relations from PHOENIX models with $\log g=4.5$ and $\log g=5.0$ are shown in green dashed and blue dotted lines, respectively. 
Filled squares represent spotted stars with $f_{s}=50\%$ and $T_{s}=0.85T_{q}$, 
and open squares represent corresponding quiescent stars with $T_{eff}=T_{q}$. 
Arrow indicates the displacement of TiO bands from quiescent star with $T_{eff}=3600$ K to spotted star with $T_{q}=3600$ K. 
Note the dotted lines connecting filled and open squares are shifted up by adding the same constants for better display.} 
\label{fig:tio2_tio5}
\end{figure*}

In Fig.~\ref{fig:tio2_tio5}, 
we showed the correlations among TiO2n, TiO3n, TiO4n and TiO5n 
for active M-type stars and Pleiades M-type candidates, comparing with the mean relation for inactive counterparts. 
We noticed that these active dwarfs have stronger (deeper) TiO2  and weaker TiO4 at a given TiO5 value, 
showing opposite behaviour. 
The large offset of TiO2 at a given TiO4 value for active stars, in upper right panel of Fig.~\ref{fig:tio2_tio5}, 
confirmed the anomalous behaviour of TiO4 found by \citet{hawl1996} and \citet{boch2007}.  
However, unlike TiO2 and TiO4, there exist no evident difference in TiO3 between active stars and inactive counterparts 
(though slightly weaker for medium- and late-M type active stars than inactive counterparts). 
These differences may arise partly due to combined effect of presence of cool spots 
and their different temperature-sensitivities. 
Our preliminary check of these indices using PHOENIX models ($\log g\sim4.5-5.5$) show that they have different temperature-sensitivities, 
namely, different strength gradient as a function of temperature, 
and different saturate onset or overturn temperature, e.g., 
TiO4 tends to overturn at $T_{eff}\sim2800$ K, TiO3 overturns at slight lower temperature, 
TiO2 shows a reversal relation at $T_{eff}\sim2600$ K. 
TiO5, measured full depth of these three subbands, have the combinational characteristics, e.g., \citet{reid1995} found that 
TiO5 become weaker as effective temperature decreases, showing a reversal relation from M7. 
One can imagine that the same cool spot would lead to different amount of extra absorptions for different TiO bands because of their 
different temperature-sensitivities. To check this effect, 
we computed the resulting TiO bands due to large cool spots with $f_{s}=50\%$ 
for several example of M dwarfs with $3300\leq T_{q}\leq3900$ K and $T_{s}=0.85T_{q}$, using PHOENIX model spectra (solar metallicity, $\log g=5.0$), 
shown as solid squares in the Fig.~\ref{fig:tio2_tio5}, and values for corresponding quiescent stars with $T_{eff}=T_{q}$ are shown as open squares. 
We see that the displacements of these TiO bands of spotted stars compare to corresponding quiescent stars show similar trends that of active M stars in these diagrams.  
In addition, to check potential gravity effect on these indices for M-type stars, 
we show the mean relations for M giants that are derived based on a sample of $\sim$2100 M giants 
(selected with $\zeta>1.6$ and $EW_{H_{\alpha}}<0$ from M-type stars sample those lie in M giants branch, see Fig.~\ref{fig:zeta_m}). 
We can see that the M giants agree well with inactive dwarfs of earlier M-type, 
but for later M-type stars M giants tend to follow the sequence of active M stars, 
which gives a clue of probable gravity effect on TiO bands in these very cool active stars. 
However, the underlying physical mechanisms for above issues are not well understood, thus need further study.
%
   
%%%%%%%%%%%%%%%%%%%%%%%%%%%%% 
\subsection{Spot configuration}
Based on the idea that the extra TiO2 (\&TiO5) band absorption in an active dwarf indicates the presence of cool spots, 
we derived spot filling factors for 304 probable members of Pleiades by modelling their TiO2 (\&TiO5) band strengths, 
and shown in the left panel of Fig.~\ref{fig:teffph_sfs} 
(only the results from TiO2n are displayed, see Appendix~\ref{sec:best_tio} for results from TiO5n), 
and listed in Table~\ref{tab:objects_results}. 
The results show overall picture that the upper envelope 
of $f_{s}$ increases slightly as stars become cooler. G-type stars have spot filling factors of $f_{s}<30\%$ whereas
a large fraction of later type stars have values of $f_{s}>30\%$ in the range of 30\%-50\%. 
Many studies indicate that the radii of young low-mass active stars tended to be larger 
than predicted values \citep[e.g.][]{stau2007,jack2009}, and modelling results show that cool spot model can solely 
cause the detected radius inflation, however, this spot model need a very large spot coverage \citep{jack2014}. 
Our results found large spot coverages for few member stars that may provide indirect support to the spot model for explaining the radius inflation.  
\begin{figure*}
\centering
\includegraphics[width=\columnwidth]{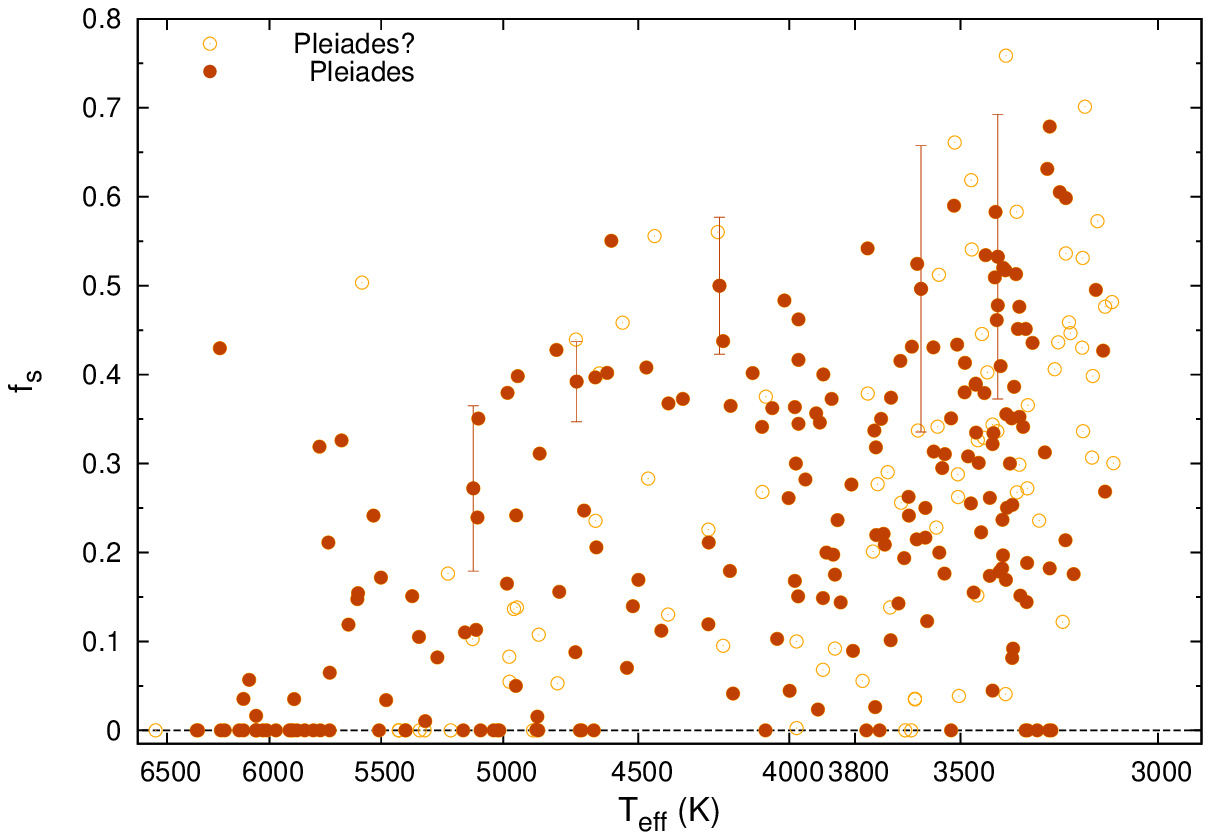}
\includegraphics[width=\columnwidth]{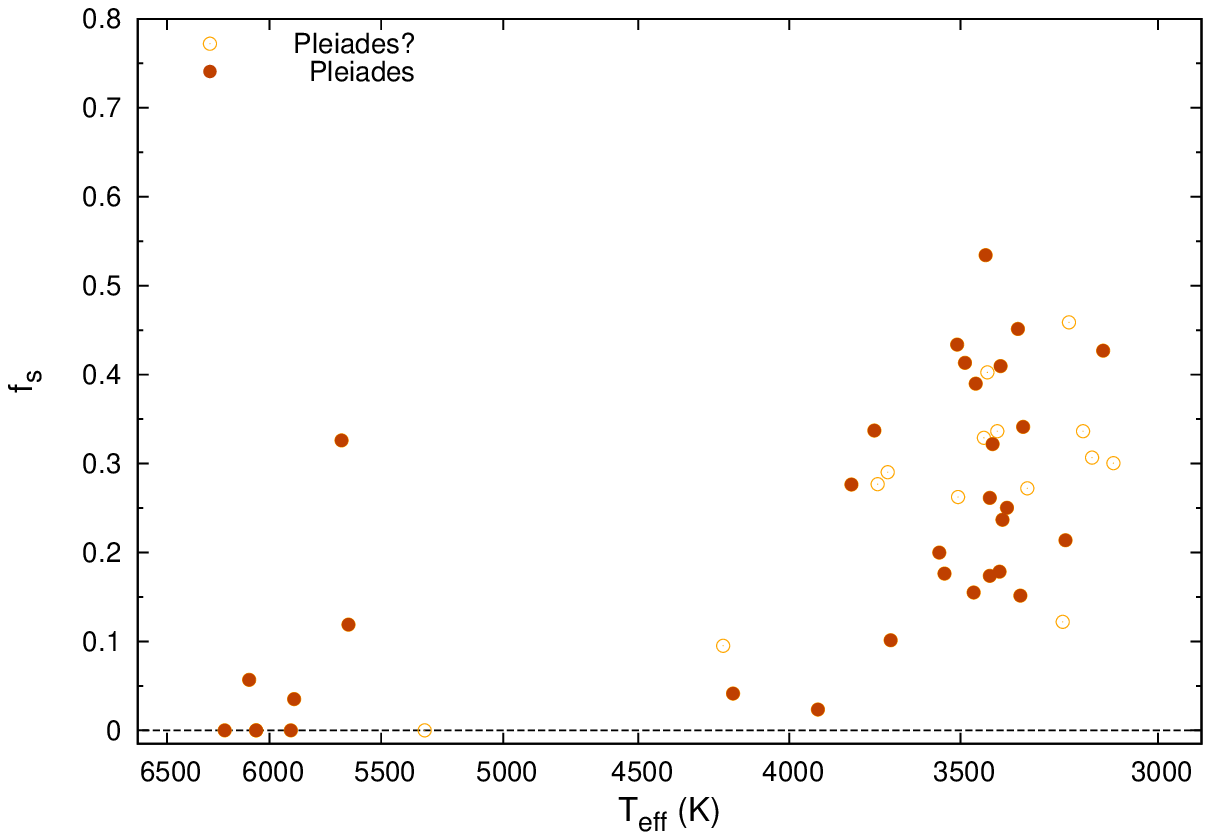}
\caption{Distribution of derived spot filling factors of Pleiades candidate members are in left panel. 
Error bars of spot filling factors are marked for five typical members with spectral type from late-G to medium M (see Table~\ref{tab:results_mc}).
Spot filling factors for Pleiades members without detected periods (showing absence or very small spot rotational modulation) are in right panel.}
\label{fig:teffph_sfs}
\end{figure*}

\begin{figure*}
\centering
\includegraphics[width=\columnwidth]{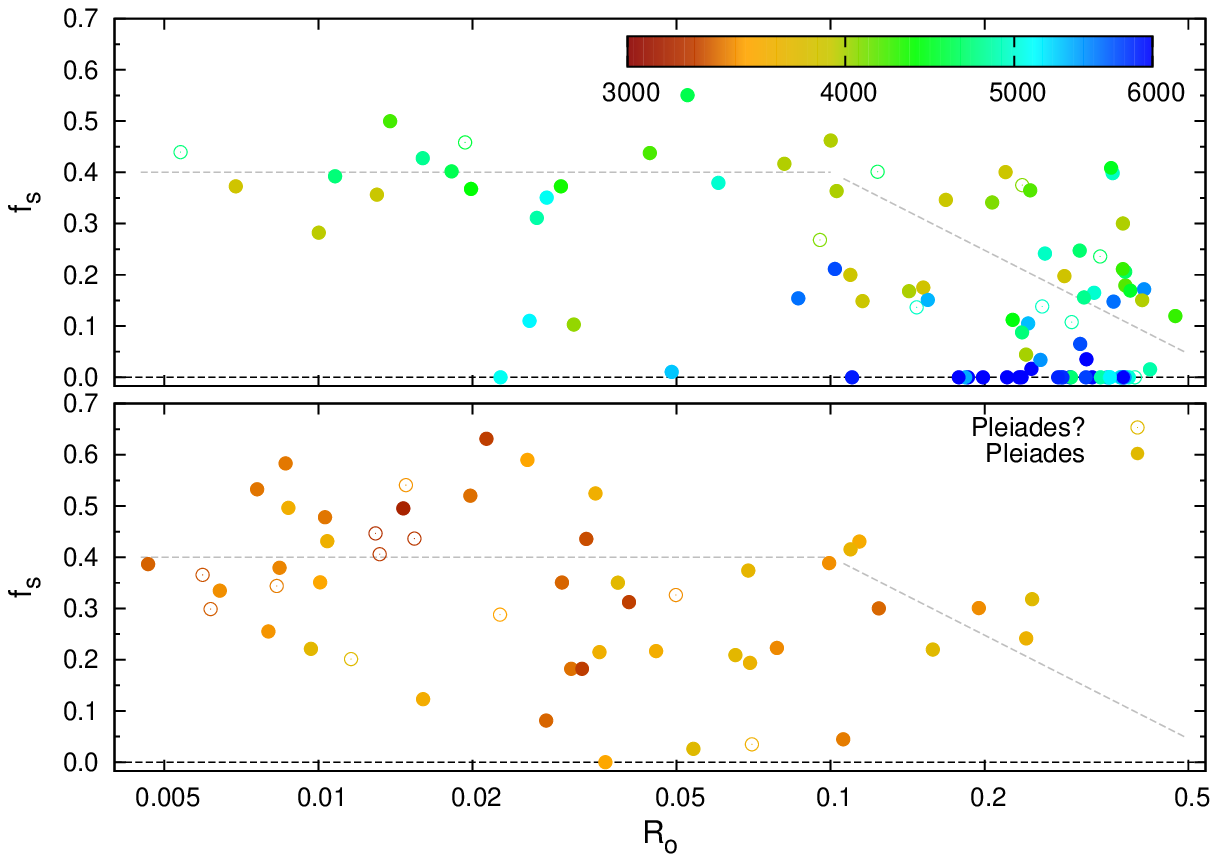}
\includegraphics[width=\columnwidth]{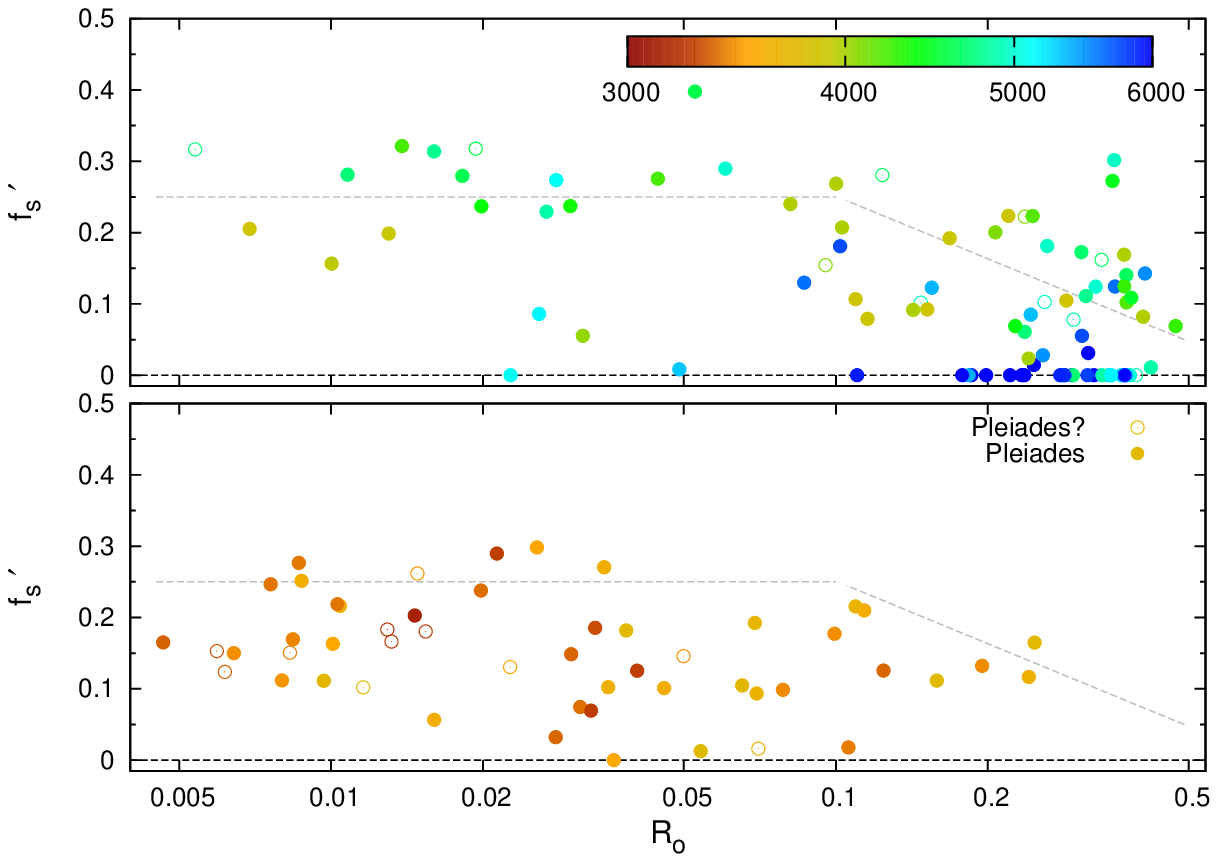}
\caption{Left: $f_{s}$ vs. $R_{o}$ for Pleiades candidate members with $T_{eff}>3800$ K (top panel) 
and $T_{eff}\leq3800$ K (bottom panel). 
Dotted line is drawn to indicate the general trend for slow rotating members and show saturated spot coverage level for fast rotators. 
Color gradient represent different effective temperatures. 
Right: same as left panels, but for $f'_{s}$ vs. $R_{o}$.}
\label{fig:sfs_rossby}
\end{figure*}

Fig.~\ref{fig:teffph_sfs} shows that spot filling factors of Pleiades members with similar colors spread a wide range, 
in particular for K- and M-type stars, indicating possible affect from other parameters to stellar spot coverage. 
More interestingly, we found a plateau of spot coverage around $f_{s}\sim 40\%$ over the $T_{eff}$ interval from $\sim$3800 K to $\sim$5000 K, showing a saturation-like phase, a phenomenon detected on fast rotating stars like 
chromospheric activities of the NGC 2516 members \citep{jack2010} 
and the coronal activities of fast rotating late-type stars \citep[e.g.][]{wrig2011}. 
To further demonstrate this feature, in Fig.~\ref{fig:sfs_rossby}, we showed the derived spot filling factors versus 
Rossby numbers $R_{o}$ ($R_{o}=P/\tau$, where $P$, $\tau$ denote the stellar rotation period and convective overturn time, respectively) 
for stars whose periods are available. To calculate Rossby numbers, rotation periods were adopted from the catalogue of \citet{hart2010}, 
and convective overturn times were estimated from stellar masses  (masses were converted from quiescent photosphere temperature using PARSEC models) 
using the correlation between convective overturn time and mass by \citet{wrig2011}. 
Fig.~\ref{fig:sfs_rossby} shows the existence of a trend that 
rapidly rotating K-type stars have larger spot coverage compare to slowly rotating counterparts. 
However, the trend is non-monotonous. For slowly rotating stars (Approx. $R_{o}>0.1$), 
spot coverage increases with decreasing $R_{o}$, 
while for the rapidly rotating stars ($R_{o}$ $<$ $\sim0.1$) the $f_{s}$ do not increase any more 
with decrement in $R_{o}$ showing a saturated phase, but it seems that members with different spectral types have different saturation levels.
Most of the M-type stars in our sample have $R_{o}<0.1$ and their spot coverage reached saturated phase, 
despite there exist large scatter (partly due to large uncertainties of spot coverage determination for these stars). 
Furthermore, we introduced a new parameter, namely, $f'_{s} = f_{s}\times(1.0- (\frac{T_{s}}{T_{q}})^4)$, 
representing the ratio of ``lost" flux due to appearance of cool spotted region to expected flux of a quiescent photosphere 
(may be an indicator of photometric activity level). We plotted $f'_{s}$ with $R_{o}$ in the right panel of Fig.~\ref{fig:sfs_rossby}, 
and noticed very similar features like $f_{s}$ versus $R_{o}$. 
In fact, we found that there also exists saturation of chromospheric activity levels for these members of Pleiades; 
a detail discussion about this aspect will appear in subsequent paper (Fang et al., in preparation).

\begin{figure*}
\centering
\includegraphics[width=140mm,height=96mm]{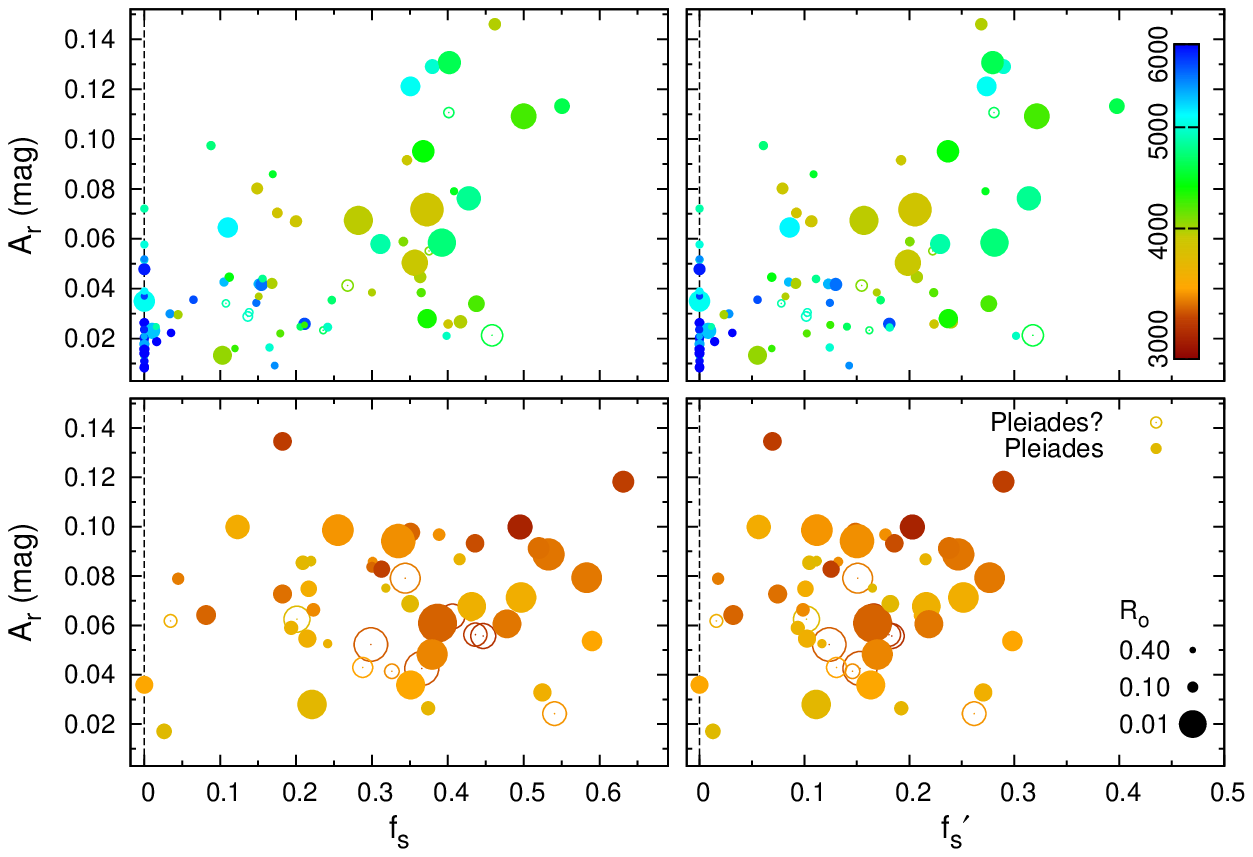}
\caption{ $A_{r}$ against $f_{s}$ (left panels), and $f'_{s}$ (right panels)  
for Pleiades candidate members with $T_{eff}>3800$ K (upper panels) and with $T_{eff}\leq3800$ K (lower panels). 
Note the size of each symbol is scaled with $R{o}$ value, and color gradient denotes its effective temperature.}
\label{fig:sfs_amp}
\end{figure*}

To characterize the distribution pattern of spots, 
we first focused on the correlation between derived spot coverage and amplitude of light variation due to spot rotational modulation.  
In Fig.~\ref{fig:sfs_amp}, we showed the Sloan r-band amplitude variation 
(the peak-to-peak amplitudes of the best-fitting sinusoid to the light curves, hereafter $A_{r}$) of Pleiades members against 
derived $f_{s}$  and $f'_{s}$ , where $A_{r}$ were derived by \citet{hart2010} 
based on data from the Hungarian-made Automated Telescope Network (HATNet) survey for transiting exoplanets. 
As shown in the upper left panel, there exists an overall trend of increasing $A_{r}$  with the increment of $f_{s}$, 
an expected relationship between spot coverage and corresponding light variation due to rotational modulation, 
assuming the spots are located asymmetrically on the surface. However, the scatter is very large, e.g., 
the stars having similar $A_{r}$ span wide regions in the diagram (the faster rotators tend to have larger $f_{s}$), 
especially for stars with $A_{r} < 0.10$ mag.   
Of course, these large scatter may be partly due to uncertainties in the measurements of spot filling factors, 
but we believe that these scatters mainly comes from the different spot distribution patterns on their photosphere.  
To illustrate, we assume the derived total spot coverage consist of two parts, namely,
the asymmetric part $f_{s}(asym)$ resulting in light variation and the symmetric part $f_{s}(sym)$ 
having no contribution to light variation (e.g., polar-like spots and numerous, axi-symmetrically distributed spots).  
In this case, comparing with slow rotators showing similar $A_{r}$ the faster rotators have similar $f_{s}(asym)$ (thus similar $A_{r}$), 
but larger $f_{s}(sym)$, indicates spots become more symmetrical. 
If such a case is real, then indicates these two parts of spot coverage may have different dependence on rotation, e.g., 
the $f_{s}(sym)$ may get more benefits from the increase of stellar rotation rate until saturates. 
As shown by lower left panel, no evident trend is found in M-type stars, 
although there exists a weak gradual increase in $A_{r}$ with $f_{s}$. 
However, considering large uncertainties in the spot coverages for these very cool stars, 
it is very hard to say these features in M members are real or not. 
Furthermore, we checked the derived spot filling factors for the members without detected periods in the HATNet survey,
which are displayed in the right panel of Fig.~\ref{fig:teffph_sfs}.
Among these stars most of them are M-type stars which have large spot coverages.
Since very few G- and K-type members in our sample are without detected periods,
it is very hard to conclude about this issue with earlier type stars. 
However, the results show that some M-type stars with high spot coverage showing no or very low light variations which indicates their photosphere may be covered by many small spotted regions, 
providing partly support for the assumption that large spot coverage might be
made up of many small and randomly located spots on the stellar photosphere \citep{jack2012,jack2014}.

%%%%%%%%%%%%%%%%%%%%%%%%%%%
\section{Conclusion}
We have measured the strength of TiO bands in a large sample of LAMOST spectra of inactive dwarfs with solar metallicity over 
a wide temperature range from 3000 K to 6500 K, and obtained the standard values. 
We found extra TiO absorption features appear on the spectra of active stars, 
even hotter stars like G- and early K-type stars, unambiguously indicating the presence of cool starspots on their photospheres. 
In addition, active cool stars like late K- and M-type stars have stronger (deeper) TiO2, 
but weaker (shallower) TiO4 than inactive stars at a given TiO5 value. 
We believe that the anomalies in these TiO indices mainly result from the presence of cool spots, 
but for very young M-type stars the gravity effect may also be accounted.
 
We have estimated the spot fractional coverages for 304 Pleiades candidate members 
by modelling their TiO2n (\&TiO5n) band strengths using standard values obtained from reference stars.
The results show that the spot coverage of G-type members are generally less than 30\%, 
while a large fraction of K- and M-type members appear to be covered by very large cool spots with filling factors over the range of $30\%-50\%$.

We have obtained the correlations between spot coverage and rotation, and the amplitude of light variation.
We found faster rotating G- and K-type stars being covered by larger spots and slowly rotating G- and K-type members 
show good dependency of spot coverage on stellar rotation. 
More interestingly, we detected a saturation-like feature of spot coverage among fast rotators 
($R_{o} < \sim0.1$) with a saturation level of $\sim40-50\%$. In G- and K-type members 
we noticed that the fraction of symmetrically distributed spots to the total spots among faster rotators is larger, 
suggests the symmetric part of spot coverage benefits more from the increment of stellar rotation rate. 
In addition, we found large spot coverages were detected in many M type members showing no or little light variation, 
all of them give strong constraints for stellar dynamo mechanism in these very active members. 

In subsequent publications, we present results of our extended analysis 
from LAMOST spectra of members of various open clusters in comparison with Pleiades, and active stars in the Kepler field. 
Also, present detail study towards the characteristics of magnetic activities on stellar photosphere, chromosphere and coronal region, 
and the correlations between these activity levels and stellar parameters: activity, rotation,
the amplitude of light variation, and flux-flux relation  (e.g., chromospheric-coronal flux-flux relation).
\section*{Acknowledgements}
We are thankful to the referee, John Stauffer, for his observant comments and constructive suggestions that helped to improve our manuscript. 
This study is supported by the National Key Basic Research Program of China (973 program, Nos. 2014CB845701 and 2014CB845703), 
and the National Natural Science Fundation of China (Grant No. U1431106). 
Y.B.K is thankful to the support from Chinese Academy of Sciences Visiting Fellowship for Researchers from Developing Countries, 
Grant No. 2013FFJB0008. 
This work has made use of LAMOST data. The Guo Shou Jing Telescope (the Large sky Area Multi-Object fiber Spectroscopic Telescope, LAMOST)
is a National Major Scientific Project built by the Chinese Academy of Sciences.
Funding for the project has been provided by the National Development and Reform Commission.
LAMOST is operated and managed by National Astronomical Observatories, Chinese Academy of Sciences. 

%%%%%%%%%%%%%%%%%%%%%%%%%%%%%%%%%%%%%%%%%%%%%%%%%%

%%%%%%%%%%%%%%%%%%%% REFERENCES %%%%%%%%%%%%%%%%%%

% The best way to enter references is to use BibTeX:

%\bibliographystyle{mnras}
%\bibliography{example} % if your bibtex file is called example.bib

\begin{thebibliography}{99}

\bibitem[\protect\citeauthoryear{Ahn et al.}{2012}]{ahn+2012}
Ahn, C. P.,  Alexandroff, R., Allende Prieto, C., et al. 2012, ApJS, 203, 21

\bibitem[\protect\citeauthoryear{Aigrain \& Irwin}{2004}]{aigr2004}
Aigrain, S., \& Irwin, M. 2004, MNRAS, 350, 331
\bibitem[\protect\citeauthoryear{Allard et al.}{2000}]{alla2000}
Allard, F., Hauschildt, P. H., \& Schweitzer, A. 2000, ApJ, 539, 366

\bibitem[\protect\citeauthoryear{An et al.}{2007}]{an++2007}
An, D., Terndrup, D. M., et al. 2007, ApJ, 655, 233

\bibitem[\protect\citeauthoryear{Andersen \& Korhonen}{2015}]{ande2015}
Andersen, J. M., \& Korhonen, H. 2015, MNRAS, 448, 3053

\bibitem[\protect\citeauthoryear{Baraffe et al.}{1998}]{bara1998}
Baraffe, I., Chabrier, G., Allard, F., and Hauschildt, P. H. 1998, A\&A, 337, 403
\bibitem[\protect\citeauthoryear{Barnes et al.}{1978}]{barn1978}
Barnes, T. G., Evans, D. S., \& Moffett, T. J. 1978, MNRAS, 183, 285
\bibitem[\protect\citeauthoryear{Berdyugina}{2005}]{berd2005}
Berdyugina, S. V. 2005, Living Rev Solar Phys, 2, 8
\bibitem[\protect\citeauthoryear{Bessell}{2005}]{bess2005}
Bessell, M. S. 2005, Annu. Rev. Astro. Astrophys., 43, 293
\bibitem[\protect\citeauthoryear{Bochanski et al.}{2007}]{boch2007}
Bochanski, J. J., West, A. A., Hawley, S. L., \& Covey, K. R. 2007, AJ, 133, 531
\bibitem[\protect\citeauthoryear{Bouy et al.}{2015}]{bouy2015}
Bouy, H., Bertin, E., Sarro, L. M., et al. 2015, A\&A, 577, 148
\bibitem[\protect\citeauthoryear{Boyajian et al.}{2012}]{boya2012}
Boyajian, T. S., von Braun, K., et al. 2012, ApJ, 757, 112
\bibitem[\protect\citeauthoryear{Budding}{1977}]{budd1977}
Budding, E. 1977, Ap\&SS, 48, 207

\bibitem[\protect\citeauthoryear{Cardelli et al.}{1989}]{card1989}
Cardelli, J. A., Clayton, G. C., \& Mathis, J. S. 1989, ApJ, 345,245
\bibitem[\protect\citeauthoryear{Carpenter}{2001}]{carp2001}
Carpenter, J. M. 2001, AJ, 121, 2851
\bibitem[\protect\citeauthoryear{Carrera \& Pancino}{2011}]{carr2011}
Carrera, R., \& Pancino, E. 2011, A\&A, 535, 30
\bibitem[\protect\citeauthoryear{Casagrande et al.}{2008}]{casa2008}
Casagrande, L., Flynn, C., \& Bessell, M. 2008, MNRAS, 389, 585
\bibitem[\protect\citeauthoryear{Chen et al.}{2014}]{chen2014}
Chen, Y., Girardi, L., et al. 2014, MNRAS, 444, 2525
\bibitem[\protect\citeauthoryear{Chabrier et al.}{2007}]{chab2007}
Chabrier, G., Gallardo, J., \& Baraffe, I. 2007, A\&A, 472, L17
\bibitem[\protect\citeauthoryear{Chabrier \& Baraffe}{2000}]{chab2000}
Chabrier, G., \& Baraffe, I. 2000, Annu. Rev. Astron. Astrophys, 38, 337
\bibitem[\protect\citeauthoryear{Collier Cameron \& Unruh}{1994}]{coll1994}
Collier Cameron, A., \& Unruh, Y.C. 1994, MNRAS, 269, 814

\bibitem[\protect\citeauthoryear{Cui et al.}{2012}]{cui+2012}
Cui, X.-Q, Zhao, Y.-H., Chu, Y.-Q., et al. 2012, RAA (Research in Astronomy and Astrophysics), 12, 1197

\bibitem[\protect\citeauthoryear{Cutri et al.}{2003}]{cutr2003}
Cutri, R. M., Skrutskie, M. F., van Dyk, S., et al. 2003, VizieR Online Data Catalog, 2246, 0

\bibitem[\protect\citeauthoryear{Desort et al.}{2007}]{deso2007}
Desort, M., Lagrange, A.-M., Galland, F., et al. 2007, A\&A, 473, 983
\bibitem[\protect\citeauthoryear{Douglas et al.}{2014}]{doug2014}
Douglas, S. T., Agueros, M. A., et al. 2011, ApJ, 795, 161
\bibitem[\protect\citeauthoryear{Feiden \& Chaboyer}{2012}]{feid2012}
Feiden, G. A. \& Chaboyer, B. 2012, ApJ, 761, 30
\bibitem[\protect\citeauthoryear{Fossati et al.}{2008}]{foss2008}
Fossati, L., Bagnulo, S., et al. 2008, A\&A, 483, 891
\bibitem[\protect\citeauthoryear{Gaidos et al.}{2014}]{gaid2014}
Gaidos, E., Mann, A. W., et al. 2014, MNRAS, 443, 2561
\bibitem[\protect\citeauthoryear{Hartman et al.}{2010}]{hart2010}
Hartman, J. D., Bakos, G. A., Kovács, G., Noyes, R. W., 2010, MNRAS, 408, 475
\bibitem[\protect\citeauthoryear{Hawley et al.}{1996}]{hawl1996}
Hawley, S. L., Gizis, J. E., \& Reid, I. N. 1996, AJ, 112, 2799

\bibitem[\protect\citeauthoryear{Henden et al.}{2016}]{hend2016}
Henden, A.~A, Templeton, M., et al. 2016, VizieR Online Data Catalog, 2336, 0

\bibitem[\protect\citeauthoryear{Huenemoerder \& Ramsey}{1987}]{huen1987}
Huenemoerder, D. P., \& Ramsey, L. W. 1987, ApJ, 319, 392
\bibitem[\protect\citeauthoryear{Husser et al.}{2013}]{huss2013}
Husser, T.-O., Wendevon Berg, S., \& Dreizler, S., et al. 2013, A\&A, 553, A6
\bibitem[\protect\citeauthoryear{Jackson et al.}{2009}]{jack2009}
Jackson, R. J., Jeffries, R. D., Maxted, P. F. L. 2009, MNRAS, 399, L89
\bibitem[\protect\citeauthoryear{Jackson \& Jeffries}{2010}]{jack2010}
Jackson, R. J., \& Jeffries, R. D. 2010, MNRAS, 407, 465
\bibitem[\protect\citeauthoryear{Jackson \& Jeffries}{2012}]{jack2012}
Jackson, R. J., \& Jeffries, R. D. 2012, MNRAS, 423, 2966
\bibitem[\protect\citeauthoryear{Jackson \& Jeffries}{2013}]{jack2013}
Jackson, R. J., \& Jeffries, R. D. 2013, MNRAS, 431, 1883
\bibitem[\protect\citeauthoryear{Jackson \& Jeffries}{2014}]{jack2014}
Jackson, R. J., \& Jeffries, R. D. 2014, MNRAS, 441, 2111
\bibitem[\protect\citeauthoryear{Jacobson et al.}{2011}]{jaco2011}
Jacobson, H. R., Pilachowski, C. A., \& Friel, E. D. 2011, AJ, 142, 59
\bibitem[\protect\citeauthoryear{Johnson}{1966}]{john1966}
Johnson, H. L. 1966, Ann Rev. Astr. Ap., 4, 193
\bibitem[\protect\citeauthoryear{Kamai et al.}{2014}]{kama2014}
Kamai, B. L., Vrba, F. J., et al. 2014, AJ, 148, 30
\bibitem[\protect\citeauthoryear{Kirkpatrick et al.}{1991}]{kirk1991}
Kirkpatrick, J. D., Henry, T. J., McCarthy, D. W. 1991, ApJSuppl, 77, 417
\bibitem[\protect\citeauthoryear{Lee et al.}{2008}]{lee+2008}
Lee, Y. S., Beers, T. C., \& Sivarani, T., et al. 2008, AJ, 136, 2050
\bibitem[\protect\citeauthoryear{Lepine et al.}{2007}]{lepi2007}
Lepine, S., Rich, R. M., \& Shara, M. M. 2007, ApJ, 669, 1235
\bibitem[\protect\citeauthoryear{Lepine et al.}{2013}]{lepi2013}
Lepine, S., Hilton, E. J., et al. 2013, AJ, 145, 102
\bibitem[\protect\citeauthoryear{Liu et al.}{2015}]{liu+2015}
Liu, X.-W., Zhao, G., \& Hou, J.-L. 2015, RAA (Research in Astronomy and Astrophysics), 15, 1089
\bibitem[\protect\citeauthoryear{Lodieu et al.}{2012}]{lodi2012}
Lodieu, N., Deacon, N. R., \& Hambly, N. C. 2012, MNRAS, 422, 1495
\bibitem[\protect\citeauthoryear{Lopes-Morales}{2007}]{lope2007}
Lopes-Morales, M. 2007, ApJ, 660, 732
\bibitem[\protect\citeauthoryear{Luhman et al.}{2003}]{luhm2003}
Luhman, K. L., Stauffer, J. R., et al. 2003, ApJ, 593, 1093
\bibitem[\protect\citeauthoryear{Luo et al.}{2015}]{luo+2015}
Luo, A.-L., Zhao, Y.-H., Zhao, G., et al. 2015, RAA (Research in Astronomy and Astrophysics), 15, 1095
\bibitem[\protect\citeauthoryear{Mann et al.}{2012}]{mann2012}
Mann, A. W., Gaidos, E. G., et al. 2012, ApJ, 753, 90
\bibitem[\protect\citeauthoryear{Melis et al.}{2014}]{meli2014}
Melis, C., Reid, M. J., et al. 2014, Sci, 345, 1029
\bibitem[\protect\citeauthoryear{Mermilliod et al.}{2009}]{merm2009}
Mermilliod, J.-C., Mayor, M., \& Udry, S. 2009, A\&A, 498, 949
\bibitem[\protect\citeauthoryear{Meunier \& Lagrange}{2013}]{meun2013}
Meunier, N., \& Lagrange, A. M. 2013, Astron. Nachr., 334, 141
\bibitem[\protect\citeauthoryear{Meynet et al.}{1993}]{meyn1993}
Meynet, G., Mermilliod, J.-C., \& Maeder, A. 1993, A\&AS, 98, 477
\bibitem[\protect\citeauthoryear{Morales et al.}{2008}]{mora2008}
Morales, J. C., Ribas, I., \& Jordi, C. 2008, A\&A, 478, 507
\bibitem[\protect\citeauthoryear{Mullan \& MacDonald}{2001}]{mull2001}
Mullan, D. J., \& MacDonald, J. 2001, ApJ, 559, 353
\bibitem[\protect\citeauthoryear{Neff et al.}{1995}]{neff1995}
Neff, J. E., O'Neal, D., \& Saar, S. H., 1995, ApJ, 452, 879
\bibitem[\protect\citeauthoryear{O'Donnel}{1994}]{odon1994}
O'Donnell J. E. 1994, ApJ, 422, 158
\bibitem[\protect\citeauthoryear{O'Neal et al.}{1996}]{onea1996}
O'Neal, D., Saar, S. H., \& Neff, J. E. 1996, ApJ, 463, 766
\bibitem[\protect\citeauthoryear{O'Neal et al.}{1998}]{onea1998}
O'Neal, D., Neff, J. E., \& Saar, S. H. 1998, ApJ, 507, 919
\bibitem[\protect\citeauthoryear{O'Neal et al.}{2004}]{onea2004}
O'Neal, D., Neff, J. E., Saar, S. H., \& Cuntz, M. 2004, AJ, 128, 1802
\bibitem[\protect\citeauthoryear{Pinsonneault et al.}{1998}]{pins1998}
Pinsonneault, S. H., Stauffer, J., et al. 1998, ApJ, 504, 170
\bibitem[\protect\citeauthoryear{Rajpurohit et al.}{2013}]{rajp2013}
Rajpurohit, A. S., Reyle, C., et al. 2013, A\&A, 556, A15
\bibitem[\protect\citeauthoryear{Ramsey \& Nations}{1980}]{rams1980}
Ramsey, L. W., \& Nations, H. L. 1980, ApJ, 239, 121
\bibitem[\protect\citeauthoryear{Reid et al.}{1995}]{reid1995}
Reid, I. N., Hawley, S. L., Gizis, J. E. 1995, AJ, 110, 1838
\bibitem[\protect\citeauthoryear{Ribas}{2006}]{riba2006}
Ribas, I. 2006, Ap\&SS, 304, 89
\bibitem[\protect\citeauthoryear{Rojas-Ayala et al.}{2012}]{roja2012}
Rojas-Ayala, B., Covey, K. R., et al. 2012, ApJ, 748, 93
\bibitem[\protect\citeauthoryear{Rosvick et al.}{1992}]{rosv1992}
Rosvick, J. M., Mermilliod, J.-C., \& Mayor, M. 1992, A\&A, 255, 130
\bibitem[\protect\citeauthoryear{Savanov \& Strassmeier}{2008}]{sava2008}
Savanov, I. S., \& Strassmeier, K. G. 2008, Astron. Nachr., 329, 364
\bibitem[\protect\citeauthoryear{Smolinski et al.}{2011}]{smol2011}
Smolinski, J. P., Lee, Y. S., et al. 2011, AJ, 141, 89

\bibitem[\protect\citeauthoryear{Soderblom et al.}{1993}]{sode1993}
Soderblom, D. R., Stauffer, J. R., et al. 1993, ApJS, 85, 315

\bibitem[\protect\citeauthoryear{Soderblom et al.}{2005}]{sode2005}
Soderblom, D. R., Nelan, E., et al. 2005, AJ, 129, 1616
\bibitem[\protect\citeauthoryear{Soderblom et al.}{2009}]{sode2009}
Soderblom, D. R., Laskar, T. et al. 2009, AJ, 138, 1292
\bibitem[\protect\citeauthoryear{Strassmeier}{2002}]{stra2002}
Strassmeier, K. C. 2002, Astron. Nachr., 323, 309
\bibitem[\protect\citeauthoryear{Strassmeier}{2009}]{stra2009}
Strassmeier, K. C. 2009, Astron. Astrophys Rev, 17, 251
\bibitem[\protect\citeauthoryear{Stauffer et al.}{1984}]{stau1984}
Stauffer, J. R., \& Hartmann, L., et al. 1984, ApJ, 280, 202
\bibitem[\protect\citeauthoryear{Stauffer \& Hartmann}{1987}]{stau1987}
Stauffer, J. R., \& Hartmann, L. W. 1987, ApJ, 318, 337
\bibitem[\protect\citeauthoryear{Stauffer et al.}{1998}]{stau1998}
Stauffer, J. R., Schultz, G., \& Kirkpatrick, J. D. 1998, ApJ, 499, 199
\bibitem[\protect\citeauthoryear{Stauffer et al.}{2003}]{stau2003}
Stauffer, J. R., Jones, B. F., Backman, D., et al. 2003, AJ, 126, 833
\bibitem[\protect\citeauthoryear{Stauffer et al.}{2007}]{stau2007}
Stauffer, J. R., Hartmann, L. W., et al. 2007, ApJS, 172, 663
\bibitem[\protect\citeauthoryear{Torres \& Ribas}{2002}]{torr2002}
Torres, G. \& Ribas, I. 2002, ApJ, 567, 1140
\bibitem[\protect\citeauthoryear{Torres}{2013}]{torr2013}
Torres, G. 2013, Astron. Nachr., 334, 4
\bibitem[\protect\citeauthoryear{van Leeuwen}{2009}]{vanl2009}
van Leeuwen, F. 2009, A\&A, 497, 209
\bibitem[\protect\citeauthoryear{Vogt}{1979}]{vogt1979}
Vogt, S. S. 1979, PASP, 91, 616
\bibitem[\protect\citeauthoryear{Vogt}{1981}]{vogt1981}
Vogt, S. S. 1981, ApJ, 247, 975
\bibitem[\protect\citeauthoryear{West et al.}{2004}]{west2004}
West, A. A., Hawley, S. L., Walkowicz, L. M., et al. 2004, AJ, 128, 426
\bibitem[\protect\citeauthoryear{Wright et al.}{2011}]{wrig2011}
Wright, N. J., Darke, J. J., et al. 2011, ApJ, 743, 48
\bibitem[\protect\citeauthoryear{Woolf \& Vallerstein}{2006}]{wool2006}
Woolf, V. M., \& Vallerstein, G. 2006, PASP, 118, 218
\bibitem[\protect\citeauthoryear{Wu et al.}{2011}]{wu++2011}
Wu, Y., Luo, A.-L., Li, H.-N., et al. 2011, RAA (Research in Astronomy and Astrophysics), 11, 924
\bibitem[\protect\citeauthoryear{Zhao et al.}{2012}]{zhao2012}
Zhao, G., Zhao, Y.-H., Chu, Y.-C., et al. 2012, RAA (Research in Astronomy and Astrophysics), 12, 723
\end{thebibliography}

% Alternatively you could enter them by hand, like this:
% This method is tedious and prone to error if you have lots of references

%%%%%%%%%%%%%%%%% APPENDICES %%%%%%%%%%%%%%%%%%%%%
%See appendix~\ref{sec:advanced}
\appendix
\section{Radial velocities for M-type stars}
\label{sec:rv_m}
\begin{figure}
\centering
\includegraphics[width=\columnwidth]{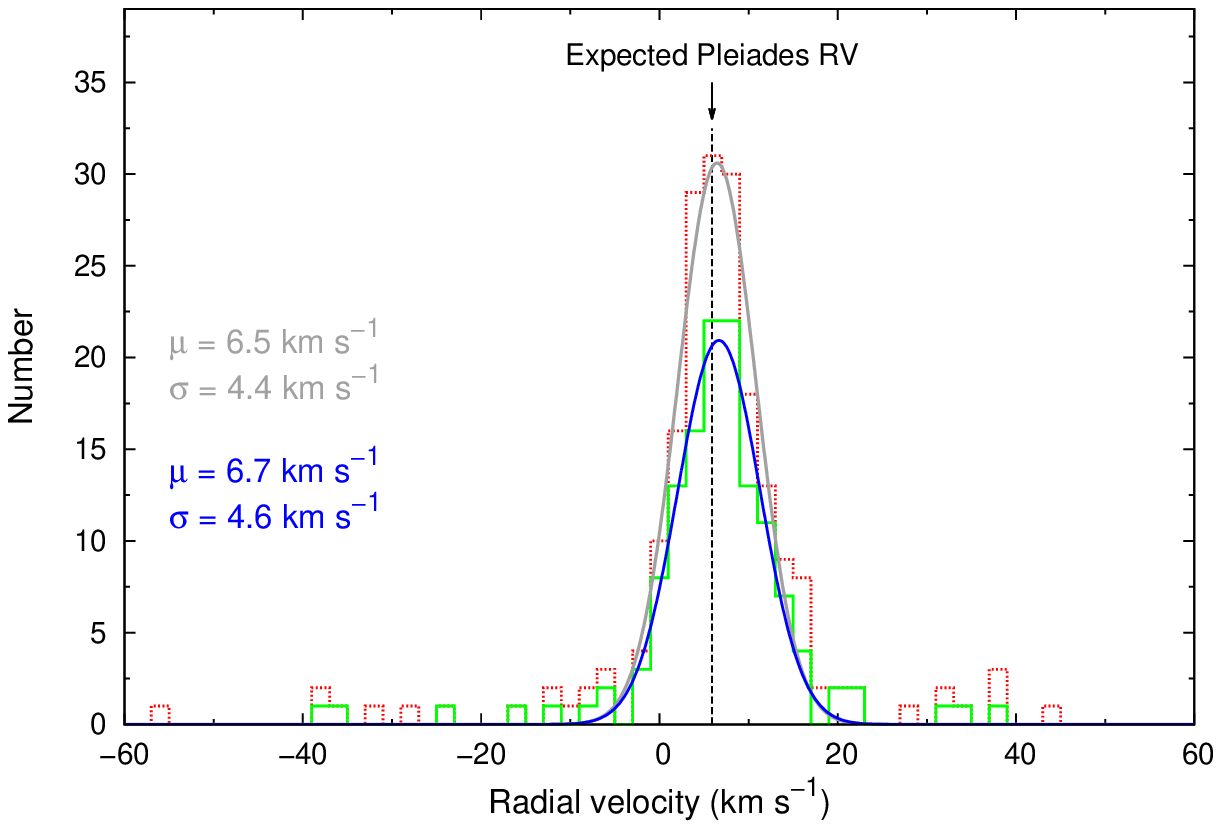}
\caption{Distribution of RVs for 200 Pleiades candidates of spectral type later than K7 is shown in red dashed line.
Gaussian fit to this distribution is shown in grey solid line. 
Green solid line represents the distribution for 135 probable single members. Corresponding  Gaussian fit is shown in blue solid line.
Mean ($\mu$) and standard deviation ($\sigma$) of respective fits are given.
Expected Pleiades RV value at 5.9 $km\,s^{-1}$ is marked as black dashed line.}
\label{fig:rv_m22}
\end{figure}

Radial velocities (RVs) of M-type stars 
are measured by cross-correlating each spectrum with best matched subtype M dwarf templates of \citet{boch2007}. 
To dilute the impact of telluric lines, for each target, 
we selected four spectral regions in the spectra that includes strong atomic and molecular bandhead features  
with relatively free of strong telluric lines (See Table~\ref{tab:bandrv}). 
Note different regions are used for different M sub-type stars. 
For each region RV is considered as corresponding peak of the cross-correlation function, which was obtained through parabolical fit to correlation values.  
For each target, among four RVs, we removed the one with largest deviation and taken average of remaining measurements as final RV. 
Fig.~\ref{fig:rv_m22} shows the measured RV distribution of Pleiades members with spectral type later than K7. 
 Our derived RV values for Pleiades members agree well with 
the literature value of Pleiades cluster \citep{rosv1992,merm2009}, 
indicating the validation of such method to estimate RV for M-type stars.

%%%%%%%%%%%%%%%%%%%%%%%%%%%%%%%%%%
\begin{table}
\centering
\caption{Spectral regions used in RV measurements for M-type stars.}
\label{tab:bandrv}
\begin{tabular}{lcccccc}
   \hline
Wavelength &  Strong line/band & For sub-types  \\
 region (\AA) &  features &  \\
   \hline
   5860-5930  &  Na D lines         & M0-M1  \\
   6090-6230  &  TiO band           & M0-M7 \\
   6540-6590  &  $H_{\alpha}$ line  & > M7 \\
   7040-7180  &  TiO band           & > M1\\
   7560-7740  &  TiO band           & > M7\\
   8160-8230  &  \ion{Na}{i} doublet lines   & $\geq$ M0\\
   8510-8700  &  \ion{Ca}{ii} IRT lines         & M0-M7 \\
   \hline
\end{tabular}
\end{table}

%%%%%%%%%%%%%%%%
\section{Corrections for measured spectral indices}
\label{sec:correct_m}
\begin{figure}
\centering
\includegraphics[width=\columnwidth]{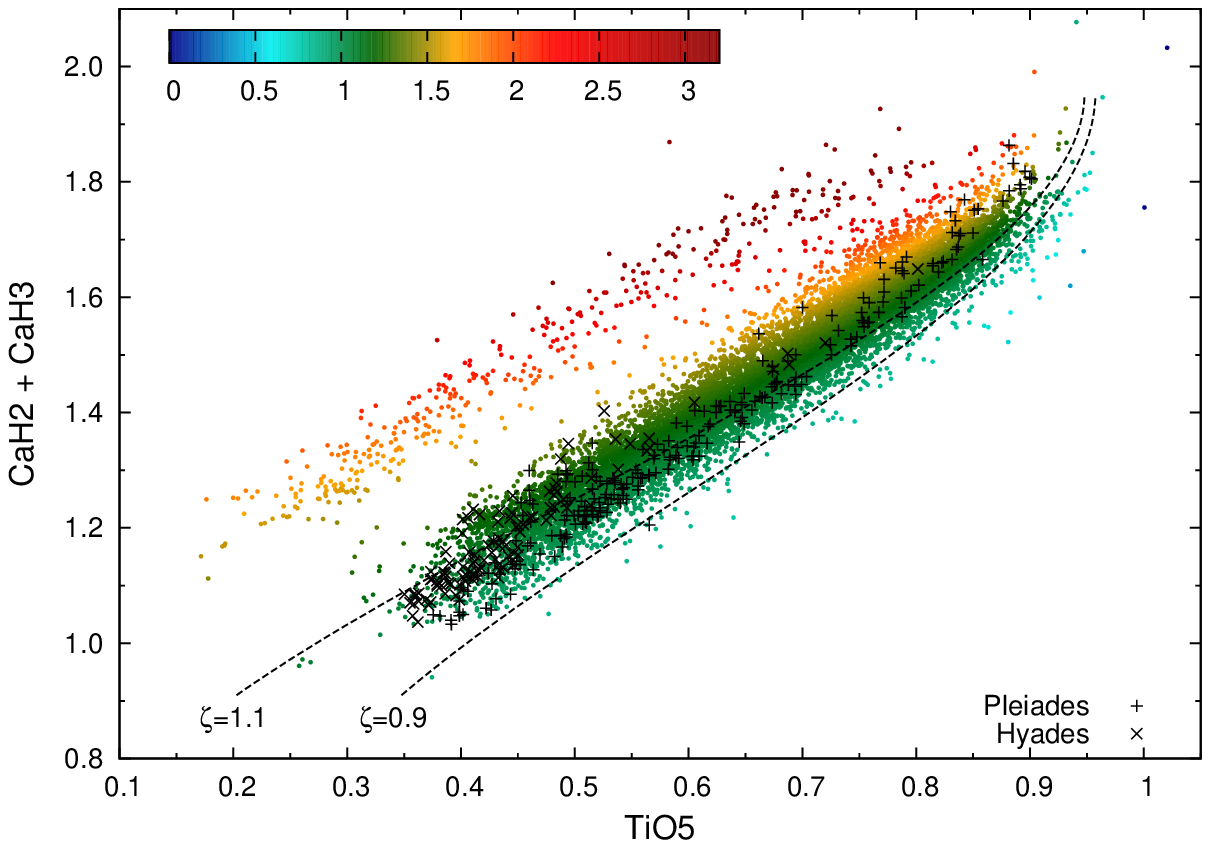}
\caption{Same as Figure~2, but for [CaH] and TiO5.}
\label{fig:zeta_mo}
\end{figure}

\begin{figure*}
\centering
\includegraphics[width=164mm,height=62mm]{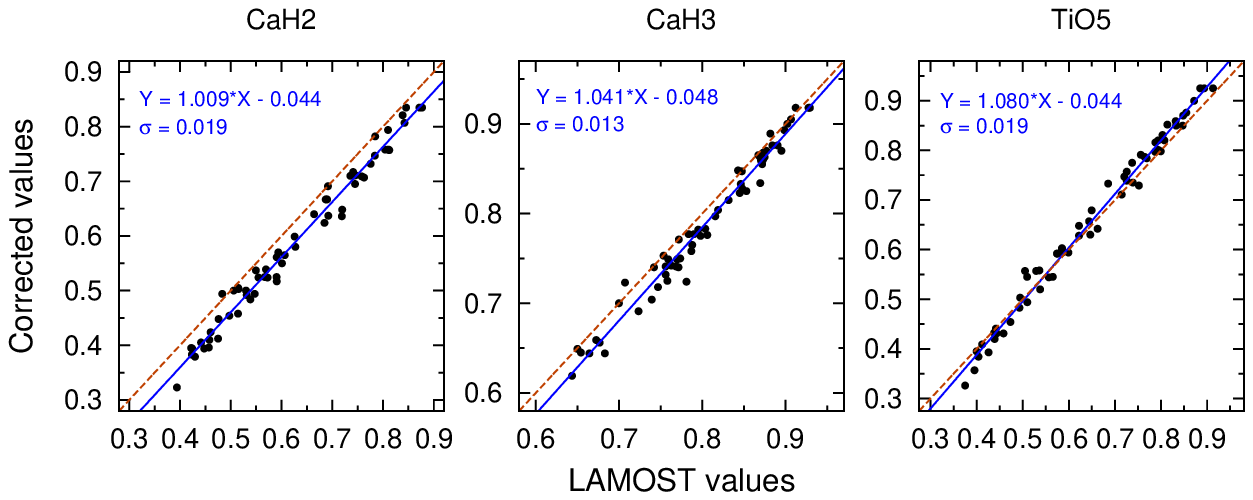}
\caption{Relations of CaH2, CaH3 and TiO5 between our measurements from LAMOST spectra
and corrected (standard) values reported by \citet{lepi2013}. 
Blue solid lines are the linear fits to the offsets and orange dashed lines represent 1:1 relations. 
Relations and corresponding rms ($\sigma$) of residuals of fit are labelled in respective panels.}
\label{fig:caho_cahc}
\end{figure*}

Our measurements of [CaH] and TiO5 for initially selected M-type stars and probable 
members of two open clusters with solar metallicities, Pleiades and Hyades, 
are shown in Fig.~\ref{fig:zeta_mo}. Also, marked the predicted lines corresponding to $\zeta=0.9$ and $\zeta=1.1$ \citep[the stars with solar 
metallicities have $\zeta\sim1.0$, see][]{lepi2013}. Note there exists a systematic offset in $\zeta$s based on our measurements, 
e.g., larger in case of early M sub-type stars. In fact, as pointed out by \citet{lepi2013}, 
since these spectral indices are susceptible to spectral resolution and spectrophotometric calibration, 
the $\zeta$ calibration may vary from instrument to instrument in different observatories. 
In order to check the systematic offsets in our measurements of TiO5, CaH2, and CaH3 indices with 
respect to standard values of the stars reported by \citet{lepi2013}, 
we cross-matched LAMOST DR2 M-type stars catalogue with the sample of \citet{lepi2013}, and 
picked good quality spectra of 55 common stars. 
We found a systematic offset in these spectral indices, 
shown in Fig.~\ref{fig:caho_cahc}, and calculated corrections for our measurements relative 
to standard scales \citep[the values reported by][]{lepi2013} using linear fitting. 
The relations with corrected indices are following,
\begin{equation}
 CaH2_{c} =  1.009 \times CaH2_{L} - 0.044,\\
\end{equation}
\begin{equation}
 CaH3_{c} =  1.041 \times CaH3_{L} - 0.048,
\end{equation}
\begin{equation}
 TiO5_{c} =  1.080 \times TiO5_{L} - 0.044,
\end{equation}
where `L' denotes the original measurement from LAMOST spectra and `c' denotes the corrected value. 

%%%%%%%%%%%%%%
\section{Determination of effective temperature for M dwarf reference stars}
\label{sec:teff_m}
Due to the difficulty in measuring stellar bolometric fluxes and lack of accurate molecular line opacities, 
there has been a challenge in the determination of effective temperature for M-type stars. 
Thanks to the availability of improved model atmospheres with new molecular line lists and implementation of new techniques, 
recently different $T_{eff}$ scales of M dwarfs are derived. 
For instance, \citet{boya2012} empirically determined effective temperatures for 21 nearby K and M dwarfs 
based on the interferometric angular diameter measurements; \citet{casa2008} semi-empirically 
determined effective temperatures using a modified infrared flux method (IRFM) for more than 340 M dwarfs;
\citet{roja2012} derived effective temperatures for 133 nearby M dwarfs from their H$_{2}$O-K2 index 
using the solar BT-Settl models; \citet{rajp2013} obtained a revised effective temperature scales for more than 150 thin disc M dwarfs 
by fitting their overall shape of the optical spectra using BT-Settl models with solar metallicity.
However, the effective temperature scale remains model and method dependent, 
e.g., \citet{rajp2013} found that temperatures derived from model spectra are 
systematically $\sim300$ K warmer than that empirically obtained by \citet{casa2008}.  

\begin{figure*}
\centering
\includegraphics[width=\columnwidth]{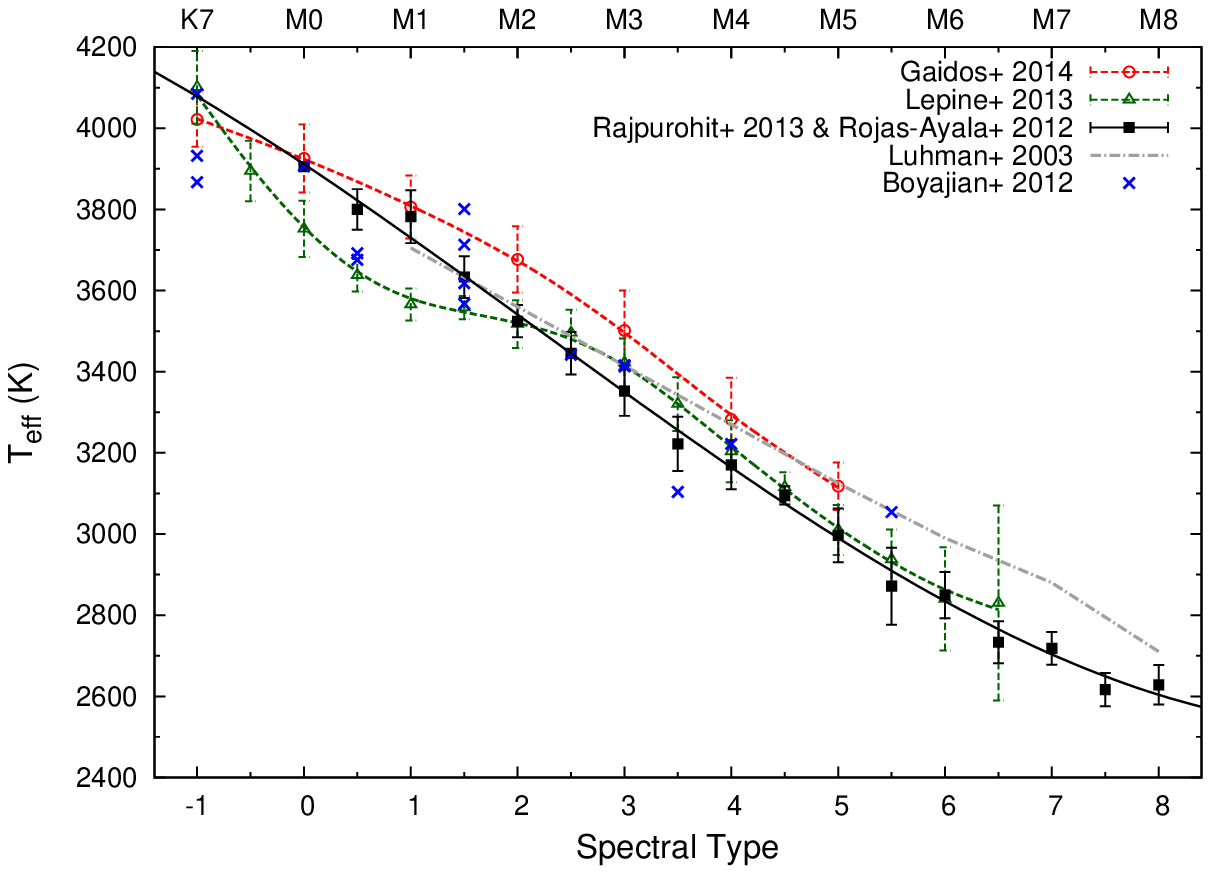}
 \includegraphics[width=\columnwidth]{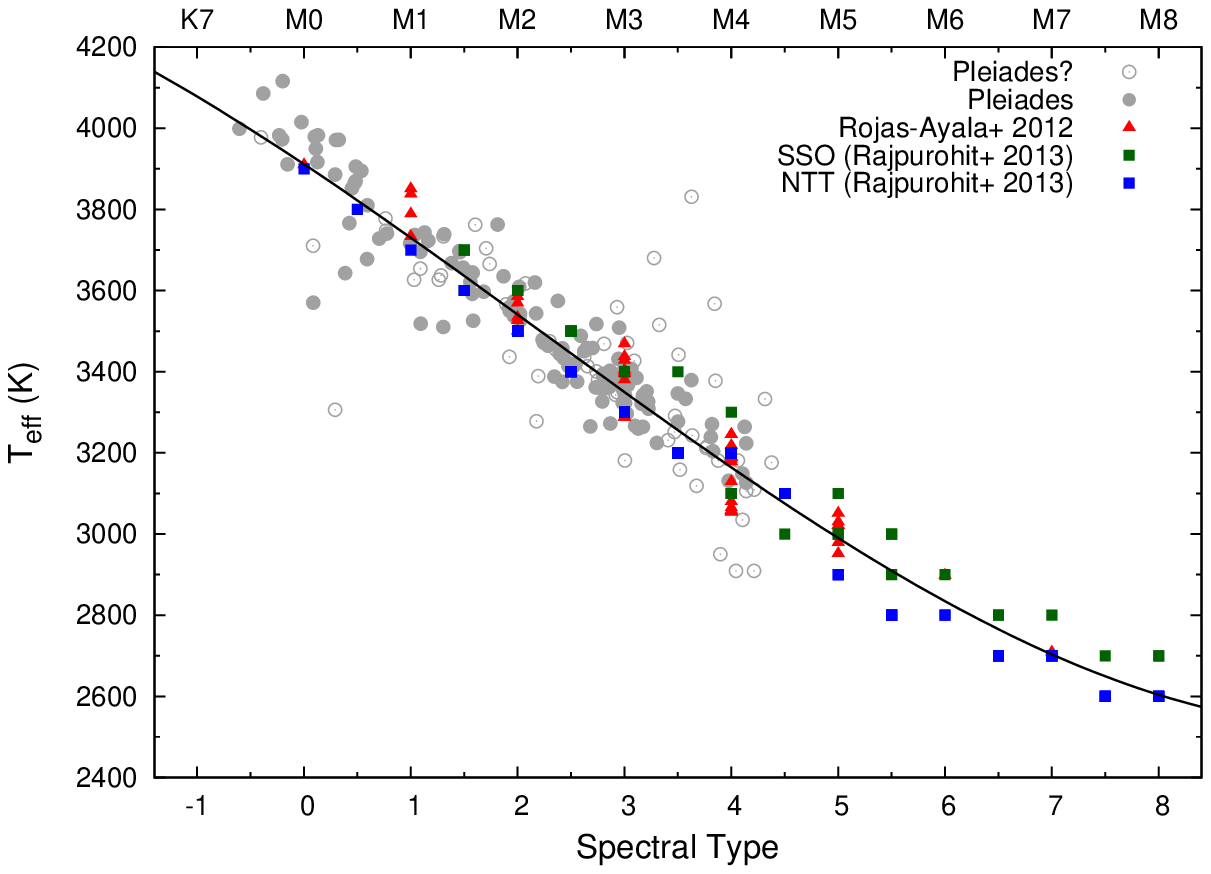}
\caption{Left: Relations between $T_{eff}$ and spectral subtype from K7 to M8 are shown  for respective sample from literature. 
The error bars indicate $T_{eff}$ scatter in each sub-type.  Dashed lines denote smooth trend of $T_{eff}$ to spectral subtype. 
Right: Relation between $T_{eff}$ and spectral subtype for probable members of Pleiades along with sample stars from literature. 
Filled circles represent probable single members and open circles with dot denote probable binary members or non-members.} 
\label{fig:teff_spt}
\end{figure*}

We have collected several subsample of K7 and M dwarfs with solar metallicities 
whose effective temperatures and spectral types have been scaled in previous studies: 
42 M dwarfs ($-0.1\leq[Fe/H]\leq0.1$), 1104 K7 and M dwarfs ($0.9\leq\zeta\leq1.1$), 
144 thin disc M0-M8 dwarfs and 725 K7 and M dwarfs ($-0.2<[Fe/H]<0.2$), are from \citet{roja2012}, 
\citet{lepi2013}, \citet{rajp2013}, and \citet{gaid2014}, respectively. 
$T_{eff}$ values and spectral types of the above sample are taken from the respective references. Mean $T_{eff}$ values in each spectral sub-type for these late-K and M dwarf samples are estimated, and shown in Fig.~\ref{fig:teff_spt}. 
Also, the relation of $T_{eff}$ versus spectral sub-type adopted by \citet{luhm2003} is plotted for comparison. 
The effective temperatures of 16 M dwarfs (with [Fe/H] from -0.4 to +0.4) derived by \citet{boya2012} 
from interferometric angular diameter measurements are also shown in Fig.~\ref{fig:teff_spt}. 
One can notice the complicated systematic offsets between different $T_{eff}$ scales. 
For Pleiades M-type members, we measured $T_{eff}$ based on broad band colors ($T_{eff}$ from $V-I_{c}$, and/or corrected $T_{eff}$ from other colors like $V-K_{s}$, see Appendix~\ref{sec:teffq}) and spectral types from TiO5$_{c}$, and compared with different $T_{eff}$ scaling relations, and found that the $T_{eff}$ scale provided by \citet{rajp2013} agrees well with $T_{eff}$ derived from broad band colors (See right panel of Fig.~\ref{fig:teff_spt}).
However, among the stars from \citet{rajp2013}, there still exist a little deviation between the scales of 55 
stars whose spectra were obtained from Siding Spring Observatory (SSO subsample) 
and 89 stars whose spectra were collected from New Technology Telescope (NTT subsample) 
\citep[see the right panel of Fig.~\ref{fig:teff_spt}, also see][]{rajp2013}. We noticed that the $T_{eff}$ scale of 42 M 
dwarfs in \citet{roja2012} agree well with M dwarfs in SSO subsample of \citet{rajp2013}, 
thus we added 42 M dwarfs to SSO subsample (55)  
to make two set of subsamples (combined SSO subsample and NTT subsample) with similar volume.
Finally, using these two subsamples, we obtained the correlation between $T_{eff}$ and spectral sub-type by 
fitting to mean $T_{eff}$ value for each sub-spectral type with a third order polynomial,
\begin{equation}
T_{eff}=1.096(SpT)^3 -7.315(SpT)^2 -175.072(SpT) + 3911.43 K,
\end{equation}
which is shown with black solid lines in Fig.~\ref{fig:teff_spt}. 
The above relation is used to estimate the $T_{eff}$ for our reference M dwarfs sample based on their spectral types. 
The spectral type was adopted from mean value of two spectral types for each object, 
derived based on TiO5$_{c}$ and CaH2$_{c}$, respectively, 
using the relations of spectral type as a function of TiO5$_{c}$ and CaH2$_{c}$ reported by \citet{lepi2013} (See equation 6 \& 8~in their paper). 

\begin{figure}
\centering
\includegraphics[width=\columnwidth]{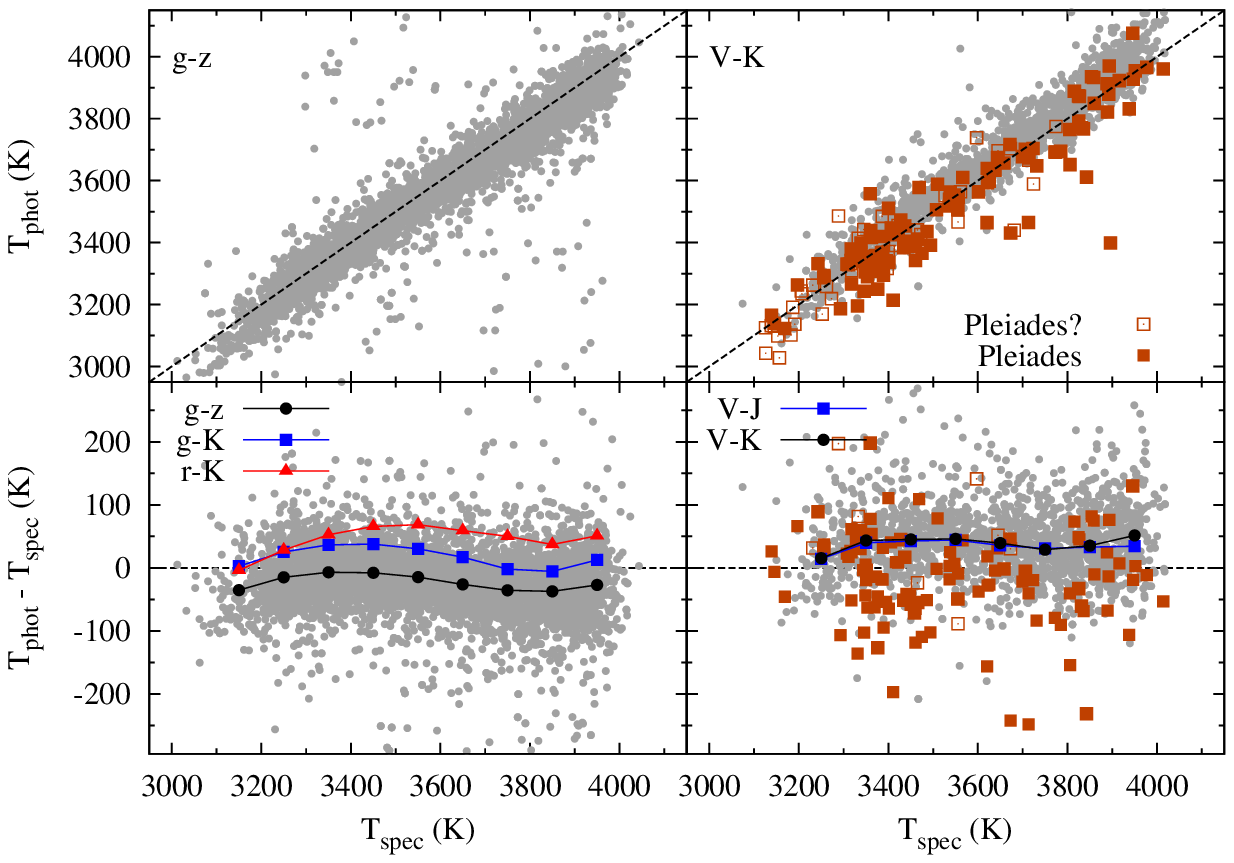}
\caption{Comparison between $T_{phot}$ from broad band colors, $g-z$ (left) and $V-K_{s}$ (right), 
and $T_{spec}$ are in upper panels. Dashed lines corresponds to 1:1 correlation. 
Corresponding differences ($T_{phot}$-$T_{spec}$) are shown for various broad band colors, binned mean values as solid black circles. 
Mean values of differences with $T_{phot}$ from colors of $g-K_{s}$, $r-K_{s}$ and $V-J$ are displayed as well. 
Pleiades probable single members and binary/non-member stars are shown as filled and open squares with dot, respectively, in right panels.}
\label{fig:teffph_sp2}
\end{figure}

To further check any potential systematic deviation of effective temperatures from spectral 
indices (hereafter $T_{spec}$) with respect to the effective temperatures from broad-band colors (hereafter $T_{phot}$), 
we selected a subsample from our M-type reference stars, 
located in northern hemisphere with Galactic latitude $b>40^{\circ}$ (reddening could be negligible), 
consists of $\sim$5300 objects with good Sloan $grz$ magnitudes \citep[SDSS DR9,][]{ahn+2012} 
and $\sim$1900 objects with APASS $V$ magnitudes \citep[APASS DR9,][]{hend2016}. 
We then derived their $T_{phot}$ values from $g-z$, $g-K_{s}$, $r-K_{s}$ and $V-K_{s}$ colors using 
color-$T_{eff}$ relations from PARSEC models \citep{chen2014} with 5 Gyr and solar metallicity. 
$K_{s}$ magnitudes for entire sample are retrieved from 2MASS archive \citep{cutr2003}. 
We compared their $T_{phot}$ values with $T_{spec}$ values, shown in Fig.~\ref{fig:teffph_sp2} and noticed 
that estimated $T_{phot}$ values  based on color-$T_{eff}$ relation are color-dependent. 
On average, the difference between these two temperature scales ($T_{phot}$-$T_{spec}$) 
approximately vary from -30 K to 50 K for respective colors from $g-z$ to $r-K_{s}$.  
In addition, by comparing $T_{phot}$ from $V-K_{s}$ with $T_{spec}$ from TiO5$_{c}$ for Pleiades candidate members, 
we found that these two temperature scales are consistent with each other. However, considering $V-K_{s}$ color is susceptible to spottedness, 
$T_{eff}$ from this color could be underestimated by 20-50 K for M-type Pleiades stars (see Appendix~\ref{sec:teffq}). 
In this case, the effective temperatures of M dwarfs in current work may be systematically underestimated by about 50 K.  
Such a systematic deviation of about 50 K in M dwarfs temperature would not affect the estimation of 
spot coverage severely for active G/K-type stars, but would result in large uncertainties for M-type Pleiades members, as discussed in the text.

%%%%%%%%
\section{Estimation of quiescent photosphere temperature}
\label{sec:teffq} 
The cool spots on the stellar surface contribute light that affect spectral features and broad-band colors,  
thus derived effective temperature $T_{eff}$ becomes the ``average'' surface temperature (flux-weighted mean temperature) in spotted stars. 
In other words, for a star with surface covered with large cool spots, the $T_{q}$ must be larger than $T_{eff}$.
Moreover, \citet{stau2003} showed that Pleiades K and M dwarfs have anomalous spectral energy distributions (SEDs) that differ 
from inactive counterparts. The striking SED feature is that these cool dwarfs have blue excesses 
 also slight red excesses, which account for the observed blueward displacement in $B-V$ color  
and redward displacement in $V-K$ color, respectively. Recently, \citet{kama2014} confirmed these findings using new $BVI_{c}$ photometry.
\citet{stau2003} proposed that this SED anomaly may be due to presence of more than one photospheric temperature in Pleiades cool members, 
e.g., the ``normal'' quiescent photosphere with both cool (spot) and warm region (plage), 
or simply warmer quiescent photosphere (warmer than it would be if it were not spotted) with cool spotted region. 
Therefore, it is a challenge to  determine definite quiescent photosphere temperature for a spotted star. 
Fortunately, unlike the anomalous $B-V$ and $V-K$ colors, the $V-I$ is less influenced by the spot \citep{stau2003}, 
which may be a result of balance of the warm and cool components on the stellar surface.  
Therefore, in this work, we estimated $T_{q}$ using $V-I_{c}$ for most of the Pleiades candidates (223 members have $V-I_{c}$ colors);
If $V-I_{c}$ is unavailable, we then estimated $T_{q}$ from a corrected effective temperature based on other colors (see the following text). 
Thus, we used $V-K_{s}$ as second choice (e.g., 48 members with no $V-I_{c}$ have $V-K_{s}$ colors), 
considering the fact that $V-K_{s}$ is a good indicator of $T_{eff}$ for low mass stars \citep{bess2005}. 
If $V-K_{s}$ is also not available, $r-K_{s}$ was used as third choice and $I_{c}-K_{s}$ as the last choice.

\begin{figure*}
\centering
\includegraphics[width=\columnwidth]{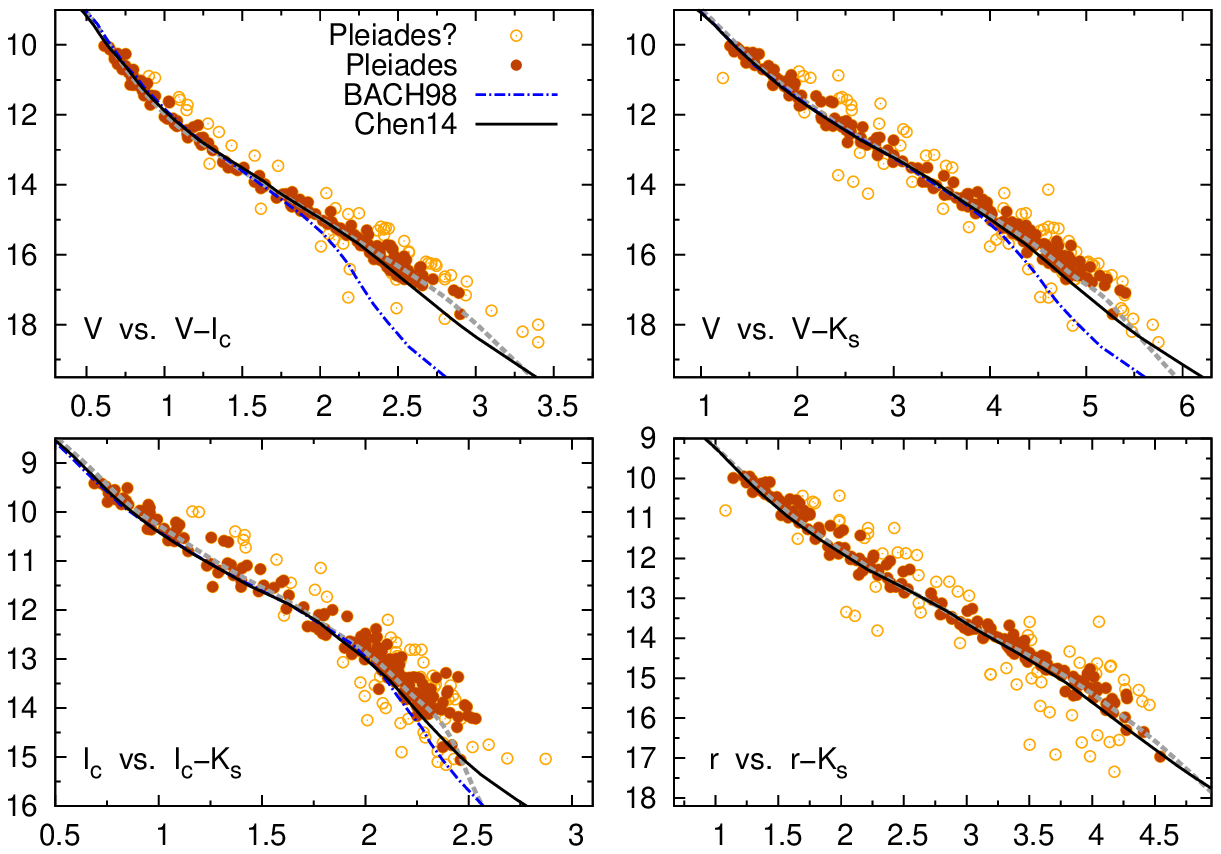}
 \includegraphics[width=\columnwidth]{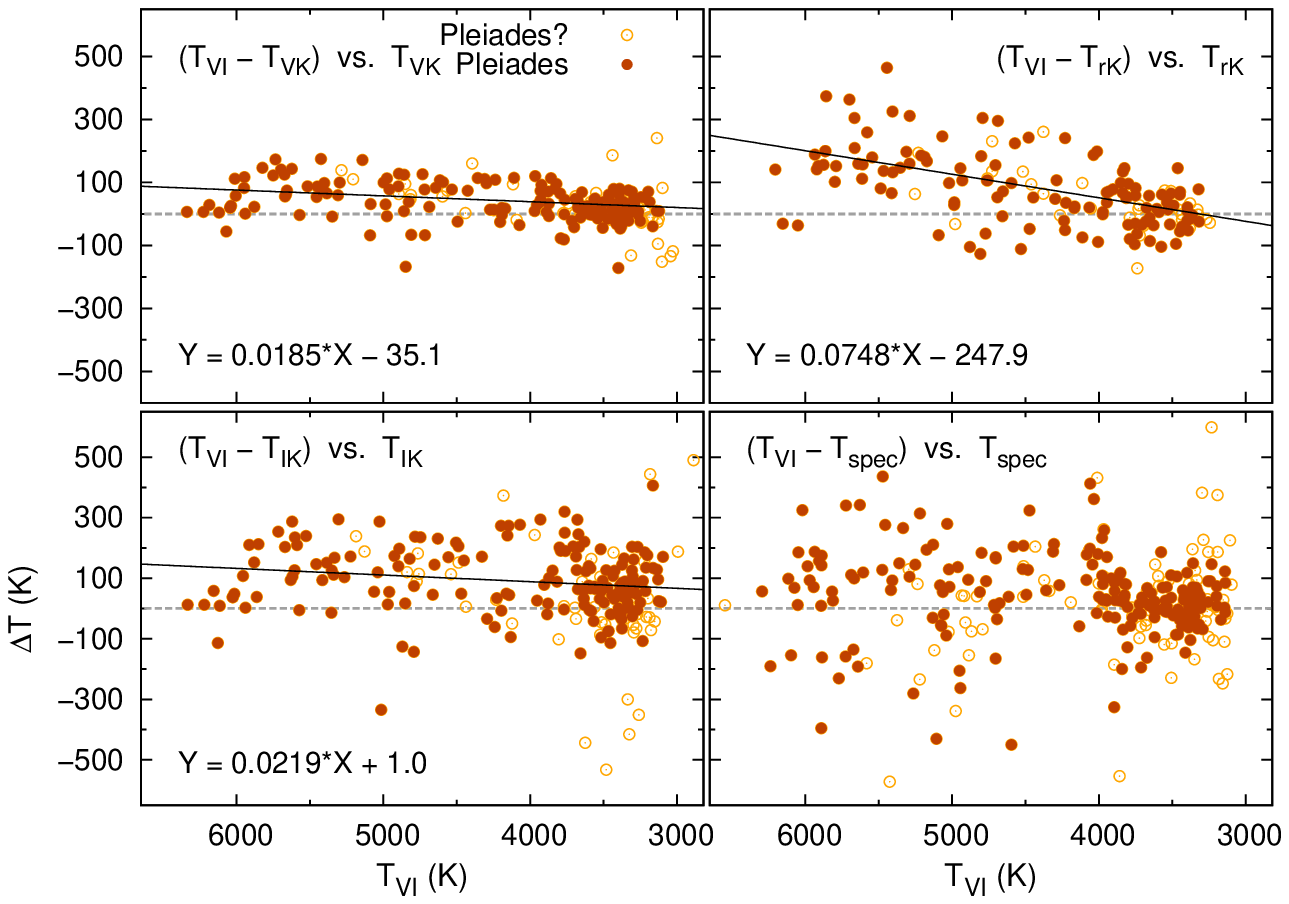}
\caption{Left: CMDs of Pleiades members. Black solid and blue dot-dashed lines represents isochrones with 120 Myr and 125 Mr 
from Chen14 \citep{chen2014} and BACH98 \citep{bara1998}, respectively.  
Grey dashed lines represent empirical Pleiades single-loci (see text). 
Right: Comparison of $T_{VK}$, $T_{rK}$, and $T_{IK}$ with $T_{VI}$ for Pleiades members. 
Black solid lines indicate linear fits to the offsets, resultant relations are labelled. 
$T_{VI}$ (\&$T^{c}_{VI}$) with $T_{spec}$ are compared for all 304 Pleiades candidate members in lower right panel.}
\label{fig:teffph_color}
\end{figure*}

For members in sample-1, we first adopted $V$ and $I_{c}$ magnitudes from catalogue of \citet{stau2007}, 
and updated with new homogeneous $V$ and $I_{c}$ magnitudes from \citet{kama2014}.
 For members in sample-2, we adopted Sloan $r$ magnitudes from \citet{bouy2015}.
We added APASS Johnson V and Sloan $r$ magnitudes for members when required. 
The $K_{s}$ magnitudes for all the members are taken from 2MASS database. 
All these magnitudes are dereddened using the extinction of Pleiades $A_{V} = 0^{m}.12$ (correspondingly, $A_{r}$=0.10, 
$A_{I}$=0.07, $A_{K}$=0.01 mag, based on the extinction 
law of \citet{card1989} \& \citet{odon1994} with $R_{V}=3.1$). 
We noticed that five stars (HII 476, HII 625, HII 676, HII 1039, HII 1136) in our sample have been known to have large reddening \citep{sode1993}, 
which lie in or near the small region of the Merope CO cloud running into the Pleiades \citep{stau1987}.
Therefore, for these five stars, we dereddened their colors using the reddening value provided by \citet{sode1993}. 
The corresponding color-$T_{eff}$ relations were estimated from the PARSEC models of \citet{chen2014} 
with an age of 120 Myr and solar metallicity.
The left panels of Fig.~\ref{fig:teffph_color} show four color-magnitude diagrams (CMDs) 
of these member stars having corresponding colors, adopting a distance modulus for Pleiades of 5.62 (equivalently a distance of about 133~pc). 
Isochrones from \citet{bara1998} and \citet{chen2014} are plotted. 
Note the CIT system K-band magnitudes from \citet{bara1998} have been transformed to 2MASS K-band magnitudes using the transformation 
equation reported by \citet{carp2001}.
 
To check for any offsets among different color-resultant $T_{eff}$ values, e.g., the difference due to the influence of spottedness,
and any potential intrinsic systematic errors due to the theoretical color-$T_{eff}$ relations,  
we compared $T_{eff}$ values from $V-K_{s}$, $r-K_{s}$ and $I_{c}-K_{s}$ (hereafter $T_{VK}$, $T_{rK}$, $T_{IK}$ respectively) 
with $T_{eff}$ values from $V-I_{c}$ color ($T_{VI}$) for Pleiades stars (See right panels of Fig.~\ref{fig:teffph_color}).
Note there exist differences between these $T_{eff}$ values.
For instance, on average, $T_{VK}$ are 20-100 K lower than $T_{VI}$ through M to GK spectral types.
We estimated offsets from $T_{VK}$, $T_{rK}$, $T_{IK}$ to $T_{VI}$ values by linear fitting and 
applied to the members with no $V-I_{c}$ color and got their $T_{VI}$-like values (hereafter $T^{c}_{VI}$).
In the lower right panel, we showed the comparison between derived $T_{VI}$ \& $T^{c}_{VI}$ values and $T_{spec}$ 
values (for members with spectral type earlier than K7, $T_{spec}$ 
were derived by LASP; for K7 and M-type stars, 
$T_{spec}$ were estimated from their $TiO5_{c}$).
We can see that $T_{VI}$ (\&$T^{c}_{VI}$) are averagely hotter than $T_{spec}$, as expected. 
However, broad-band colors based $T_{eff}$ of some members are largely underestimated, 
e.g., for HHJ 435 $T_{VI}$ $\sim$ 3570 K, 
while its spectrum indicates that it should be a late-K or early M type star, being consistent with $T_{spec}\sim3900$ K estimated from $TiO5_{c}$.   
In fact, HHJ 435 lies in the area of the Merope CO cloud, indicating this star is probably effected by large reddening.  
For the stars outside the CO cloud region, the large deviation in $T_{phot}$ indicate that the colors maybe affected by heavy spottedness, 
or simply indicate they are binaries/non-member stars. In this paper, 
for these 39 stars  with $T_{VI}$ (\&$T^{c}_{VI}$) of 100 K lower than $T_{spec}$, 
we simply adopted the $T_{spec}$ as $T_{q}$, instead of $T_{VI}$ or $T^{c}_{VI}$. 
The adopted $T_{q}$ along with $T_{eff}$ from different colors are listed in Table~\ref{tab:objects_results} and Table~\ref{tab:objects_list}.  

To identify potential binaries and non-members, we first preferred $V$ versus $V-I_{c}$ CMD, 
further $V$ versus $V-K_{s}$ as second choice, 
then $r$ versus $r-K_{s}$ and $I_{c}$ versus $I_{c}-K_{s}$ are used as third and last choices.
Based on empirical single-star loci of Pleiades in these CMDs shown in Fig.~\ref{fig:teffph_color} 
(if available, the average single-star loci of values provided by \citet{stau2007} and \citet{kama2014} were used, 
and the single-star loci in $r$ vs. $r-K_{s}$ CMD were derived 
based on the sample of \citet{bouy2015}),    
brighter stars with $-1.^{m}0 < \Delta mag < -0^{m}.5$ in one CMD (e.g., $\Delta V$ in $V$ vs. $V-I_{c}$ CMD) 
were classified as probable binary members, 
while the fainter stars with $\Delta mag > 0^{m}.3$ or very much brighter stars with $\Delta mag < -1^{m}.0$ were identified 
as probable non-member stars, as flagged in Table~\ref{tab:objects_list}. 

%%%
\section{Which index is the best indicator of cool spots?}
\label{sec:best_tio}
\begin{figure}
\centering\includegraphics[width=\columnwidth]{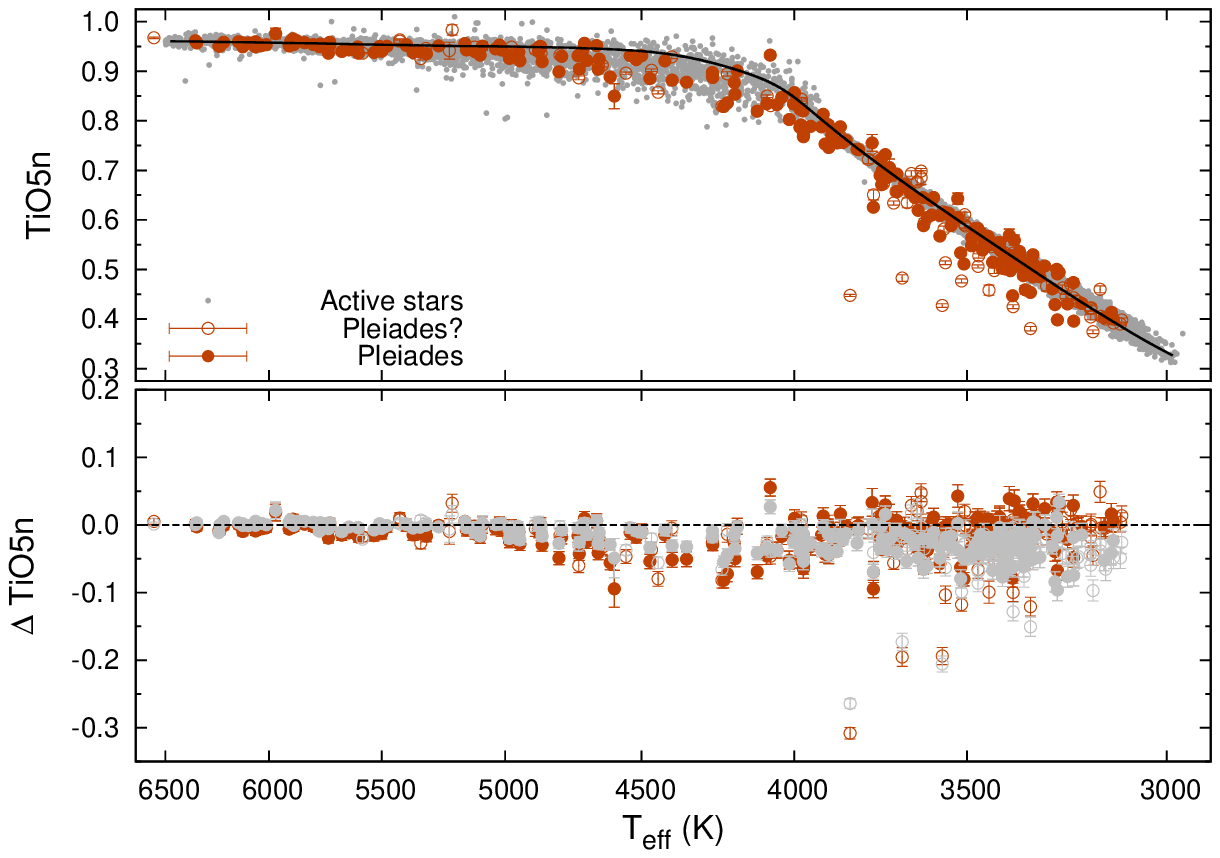}
\caption{Same as right panel of Fig.~\ref{fig:ewha_tio2_active}, but for TiO5n. 
 Grey symbols in lower panel indicates $\Delta$TiO2n values.}
\label{fig:results_tio5n}
\end{figure}

\begin{figure}
\centering
\includegraphics[width=\columnwidth]{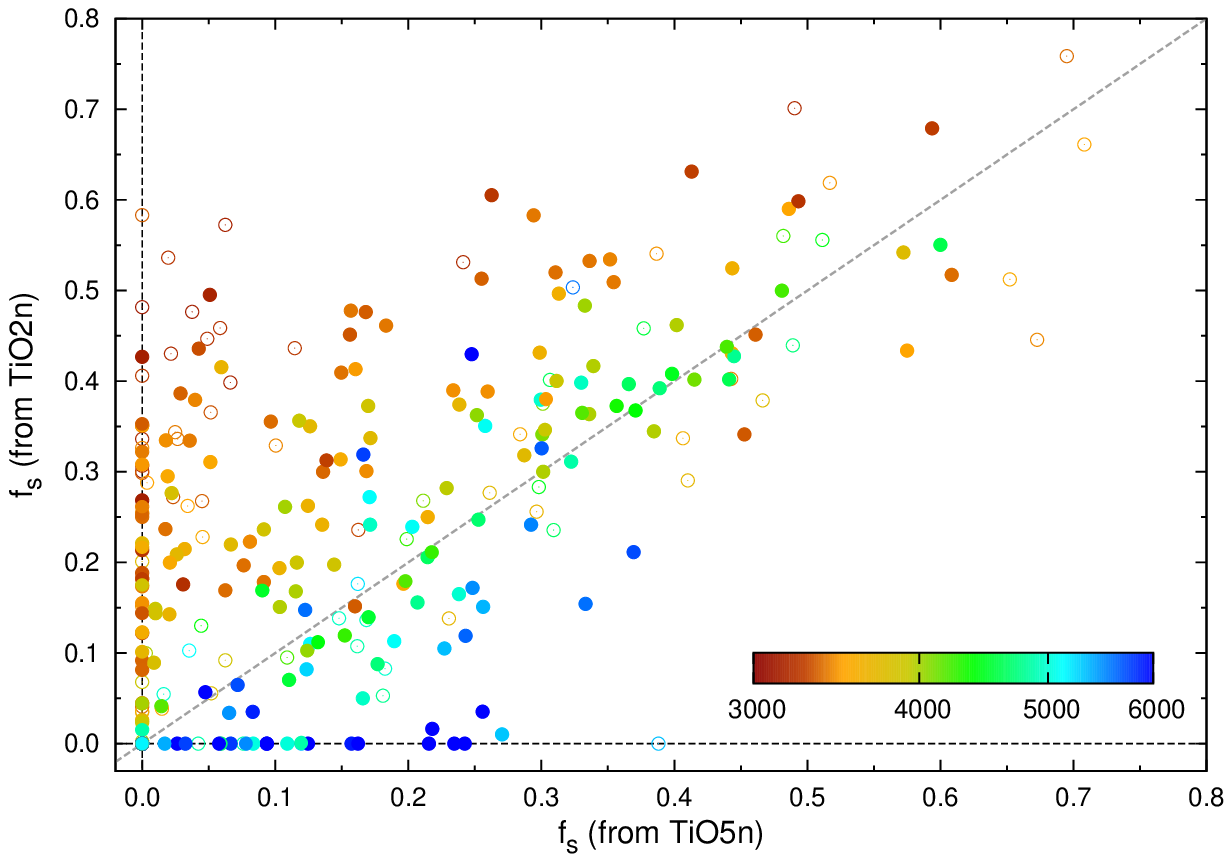}
\caption{Spot filling factors from TiO2n and TiO5n are compared. Colors are scaled as per quiescent photosphere temperature. 
Dashed line corresponds to 1:1 correlation.}
\label{fig:results_tio2n_tio5n}
\end{figure}

Equation~\ref{equ:Ftotal} shows that the observed relative flux of any molecular band, $F_{t}$, depends on both the intrinsic strength of spot, $F_{s}$, 
and the contrast of nearby continuum surface flux between the spotted and quiescent photosphere, $R_{\lambda_{c}}$.
In other words, just as noted by \citet{onea1998}, to be a better diagnostic for cool spot, 
a molecular band should not only be intrinsically strong, 
but the contrast of continuum surface flux should be large enough to produce a measurable signal in the observed spectra. 
Therefore, these intrinsically strong molecular bands in the red spectral region are more useful for measuring spot parameters. 
In fact, Late-K and M-type dwarfs have numerous strong molecular bands in the red spectral region, 
like the CaH bands near 6380~\AA~(CaH1) and 6800~\AA~(CaH2), 
TiO bands near 7050~\AA~(TiO2, TiO5), 7600~\AA~and 8860~\AA~(hereafter TiO7600 and TiO8860). 
However, considering their strong temperature-sensitivity over a wide temperature range but weak gravity-sensitivity, 
TiO bands are ideal diagnostics of cool spots. 
Moreover, comparing with other TiO band systems like TiO7600 and TiO8860 in the red spectral region, 
TiO bands near 7050~\AA~have strong temperature-sensitivity 
over very wide temperature range (e.g., from $\sim$K5 to M7) and marginally susceptible to telluric lines or strong metal lines. 
Therefore, we took the TiO bandhead of 7050~\AA~as the ``best'' band for measuring cool spot parameters for Pleiades cool members. 
The bandhead of 7050~\AA~consists of three strong absorption subbands, namely, TiO2, TiO3 and TiO4. 
Comparing with TiO2, TiO3 and TiO4 are slight weaker, 
particularly, the pseudo-continua of these two bands are susceptible to bluer bands (e.g., TiO2 eats the pseudo-continuum of TiO3).
TiO5, measuring the full depth of this TiO band system, is thus stronger than any of those three bands, 
e.g., the depth ratios of TiO2n to TiO5n are in the range of  0.5 to 0.7 for M dwarfs with temperature from $\sim4000$ K to 3000 K. 
In other words, it is easier to detect signals of cool spot from the TiO5 index, as illustrated in Fig.~\ref{fig:results_tio5n}. 
One can see more extra absorption from TiO5 than TiO2 for G/K- and early M-type active stars and Pleiades members. 
For cooler active stars and Pleiades members, TiO5 is similar to TiO2 (or even weaker), 
which maybe partly due to weaker temperature-sensitivity of TiO5, also partly due to the 
systematic underestimate of temperatures for these active stars (since we estimated their temperatures using CaH2$_{c}$ and TiO5$_{c}$). 
In this study, we have estimated spot coverages using TiO2 and TiO5 bands for Pleiades members, 
and are listed in Table~\ref{tab:objects_results}. 
We compared the resultant spot filling factors from TiO2n and TiO5n, and shown in Fig.~\ref{fig:results_tio2n_tio5n}.
Similar spot coverages are seen, from these two indices, for stars of late-G to early-M Pleiades members. 
For hotter or cooler member stars, as expected, results from TiO5n and TiO2n show opposite behaviours. However, in the paper, 
we discussed the results from TiO2n, since the results from TiO5n provide not much extra information.

%%%%%%%%%%%%%%%%%%%%%%%%%%%%%%%%%%%%%%%%%%%%%%%%%%

% Don't change these lines
\bsp	% typesetting comment
\label{lastpage}
\end{document}